\documentclass[a4paper,10pt]{article}

\usepackage{amsmath,amssymb,amsthm,ulem,array,mathtools}
\usepackage[a4paper]{geometry}
\usepackage{graphicx}
\usepackage[font={footnotesize}]{caption}
\usepackage{enumerate}
\usepackage{tikz}

\def\be{\begin{equation}}
\def\ee{\end{equation}}

\theoremstyle{plain}
\newtheorem{ex}{Exercise}
\newtheorem*{np}{Numerical project}

\theoremstyle{definition}

\newenvironment{recap}
{
\begin{samepage}
\vspace{0.2cm}
\hrule width \hsize height 1pt \kern .5mm \hrule width \hsize
\vspace{0.2cm}
\noindent{\bfseries Intermediate recap}\\
\noindent
}{
\hfill$\blacksquare$
\vspace{0.2cm}
\hrule width \hsize \kern .5mm \hrule width \hsize height 1pt
\vspace{0.2cm}
\end{samepage}
}

\newenvironment{beyond}
{
\footnotesize
\noindent{\sc To go further}\\
\footnotesize
}
{
}

\newenvironment{intro}
{
\noindent
}{
}

\begin{document}

\title{\bf 
Lecture notes on\\
``Quantum chromodynamics and statistical physics''%
\footnote{Lectures given at the {\it ``Huada school on QCD''},
Central China Normal University, Wuhan, China, June 2-13, 2014.}
}

\author{St\'ephane Munier
\\
\small\it Centre de physique th\'eorique, \'Ecole Polytechnique, 
CNRS, Palaiseau, France.
}
\date{}
\maketitle

\begin{abstract}
The concepts and methods used for the study of disordered systems
have proven useful in the analysis of the evolution equations 
of quantum chromodynamics in the high-energy regime:
Indeed, parton branching in the semi-classical approximation 
relevant at high energies
is a peculiar branching-diffusion process, 
and parton branching supplemented by
saturation effects (such as gluon recombination)
is a reaction-diffusion process.
In these lectures, we first introduce the basic concepts in the context of simple
toy models, we study the properties of the latter, 
and show how the results obtained for the simple models 
may be taken over to quantum chromodynamics.
\end{abstract}

\tableofcontents


\clearpage


\section{Branching random walks and the Fisher-Kolmogorov-Petrovsky-Piscounov equation}

\begin{intro} 
In this section, we shall introduce branching random walks, which are
a class of stochastic processes appearing in many different branches of
science, and in particular in particle physics.
We show how a nonlinear diffusion equation called the
FKPP equation characterizes some properties of the realizations
of branching random walks. We start by recalling some elementary facts
on ordinary Brownian motion, before adding in the branching process.
\end{intro}

\subsection{Brownian motion}

Consider a one-dimensional lattice indexed by the variable $x$, with lattice spacing
$\Delta x$.
We start with a single particle at site 0.
We take the following rule for the evolution of the system from time $t$ to
time $t+\Delta t$, which consists in two elementary processes: 
The particle has the probability $\mu$ to jump on the lattice site to the left,
and the probability 
$\mu$ to jump to the right. Hence by probability conservation, the particle 
has the probability $1-2\mu$ to stay at its current position (from which we see that $\mu$ has
to be chosen less than $\frac12$).
\begin{figure}[h]
\begin{center}
\includegraphics[width=.6\textwidth,angle=0]{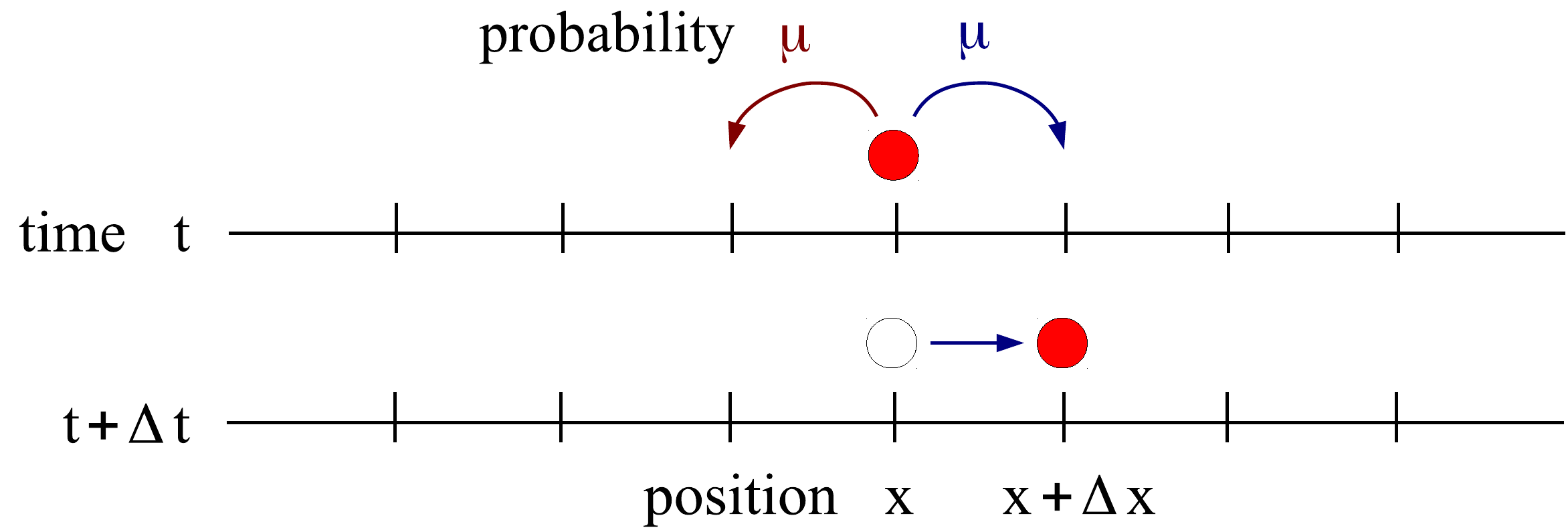}
\end{center}
\caption{\label{fig:brownian1}
Definition of the elementary processes and illustration of one step 
of the Brownian motion on the lattice.
The particle sits at position $x$ at time $t$.
According to the evolution rules, it may jump left with probability $\mu$,
right also with probability $\mu$, and it may stay at $x$ with probability $1-2\mu$.
In the particular realization shown in the figure, the particle jumps right, in
such a way that its position be $x+\Delta x$ at time $t+\Delta t$.
}
\end{figure}

\begin{figure}
\begin{center}
\includegraphics[width=.7\textwidth,angle=0]{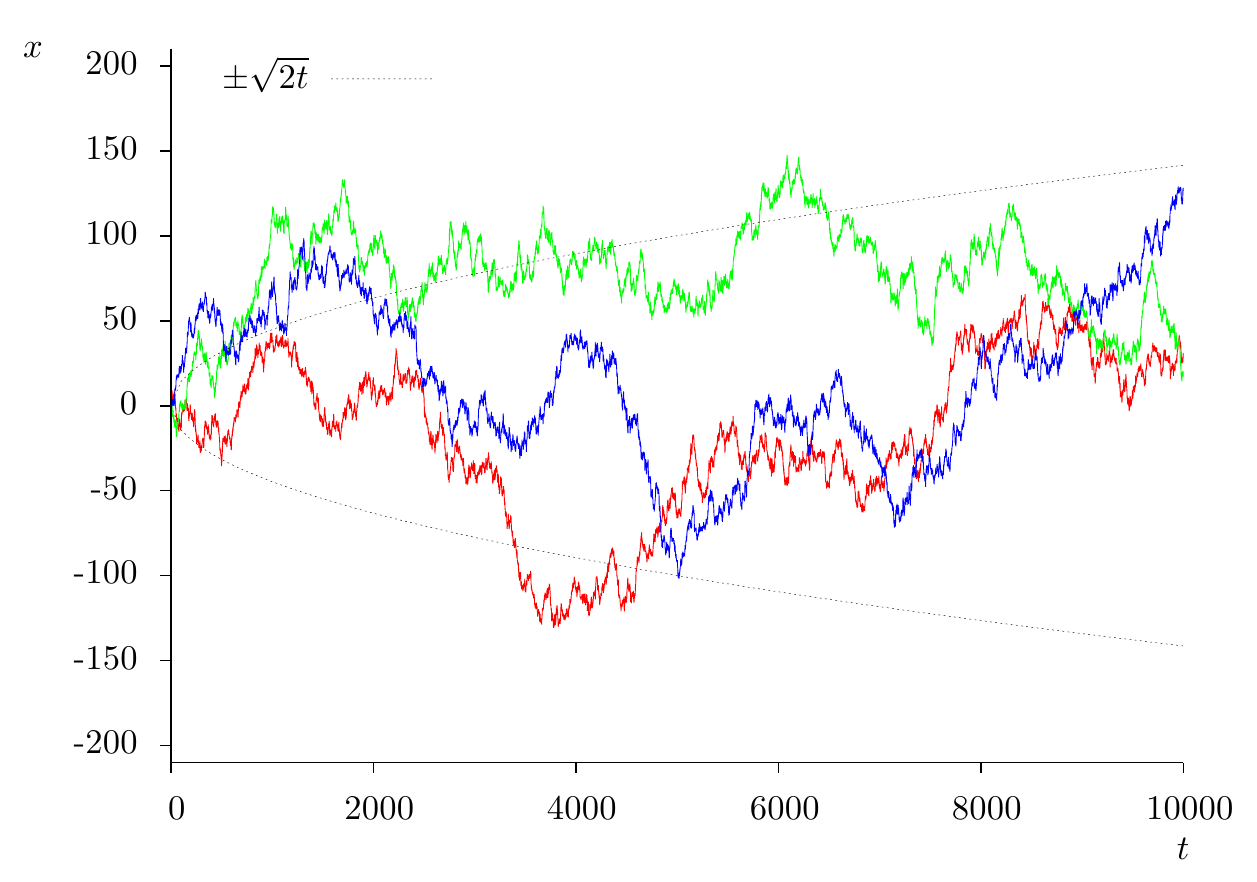}
\end{center}
\caption{\label{fig:plotRW}
Three realizations of the continuous Brownian up to $t=10000$.
The parabola (dotted lines) represents the standard deviation 
$\pm\sqrt{\langle x^2\rangle-\langle x\rangle^2}$.
}
\end{figure}

With this rule, it is straightforward to
establish an equation for the probability $P(x,t)$ that
the particle be on site $x$ at time $t$:
We simply relate $P$ at time $t+\Delta t$ to $P$ at time $t$
with the help of the probabilities of the elementary processes.
The different terms which contribute to $P(x,t+\Delta t)$ stem from
the following cases:
\begin{enumerate}[(i)]
\item The particle is at site $x-\Delta x$ at time $t$
and makes a right jump. This generates the term $P(x-\Delta x,t)\times\mu$;
\item the particle is at $x+\Delta x$ at $t$ and makes a left jump.
The corresponding term reads $P(x+\Delta x,t)\times\mu$;
\item the particle is already at $x$ at time $t$ and does not move.
The term which describes this case is  $P(x,t)\times(1-2\mu)$.
\end{enumerate}
Summing all contributions, one arrives at
\be
P(x,t+\Delta t)=P(x-\Delta x,t)\mu+P(x+\Delta x,t)\mu
+P(x,t)(1-2\mu)
\ee
from which, after a trivial rearrangement, 
we get the finite difference evolution equation
\be
P(x,t+\Delta t)-P(x,t)=\mu\left[
P(x-\Delta x,t)+P(x+\Delta x,t)-2P(x,t)
\right].
\label{eq:finitediff}
\ee
It is often easier to deal analytically with differential equations
rather than difference equations.
If we let the lattice spacing $\Delta x$ and the time step $\Delta t$ go
to zero, the above finite-difference equation becomes a partial differential
equation. We must be careful however to keep the ratio
\be
D\equiv \mu\frac{(\Delta x)^2}{\Delta t}
\ee
finite and fixed when taking this limit in such a way that no relevant
terms in Eq.~(\ref{eq:finitediff}) vanish.
We get
\be
\frac{\partial P}{\partial t}=D\frac{\partial^2 P}{\partial x^2}.
\label{eq:diffusion}
\ee
This equation is the so-called Fokker-Planck equation for our process, and we
recognize that it is the simple diffusion equation.
To set up a well-posed problem, we need to specify the initial condition 
and the boundary conditions.
The initial condition is a single particle at site $x=0$ at time $t=0$, hence
in the continuous limit
\be
P(x,t=0)=\delta(x).
\ee
As for the boundary conditions, we shall first opt for free ones, and second impose a fixed
absorptive boundary.
In the following, we shall set $D=1$ for simplicity. (From dimensional analysis, one may always
re-establish a general diffusion constant). Realizations of this model are shown in Fig.~\ref{fig:plotRW}.

\begin{figure}
\begin{center}
\includegraphics[width=.7\textwidth,angle=0]{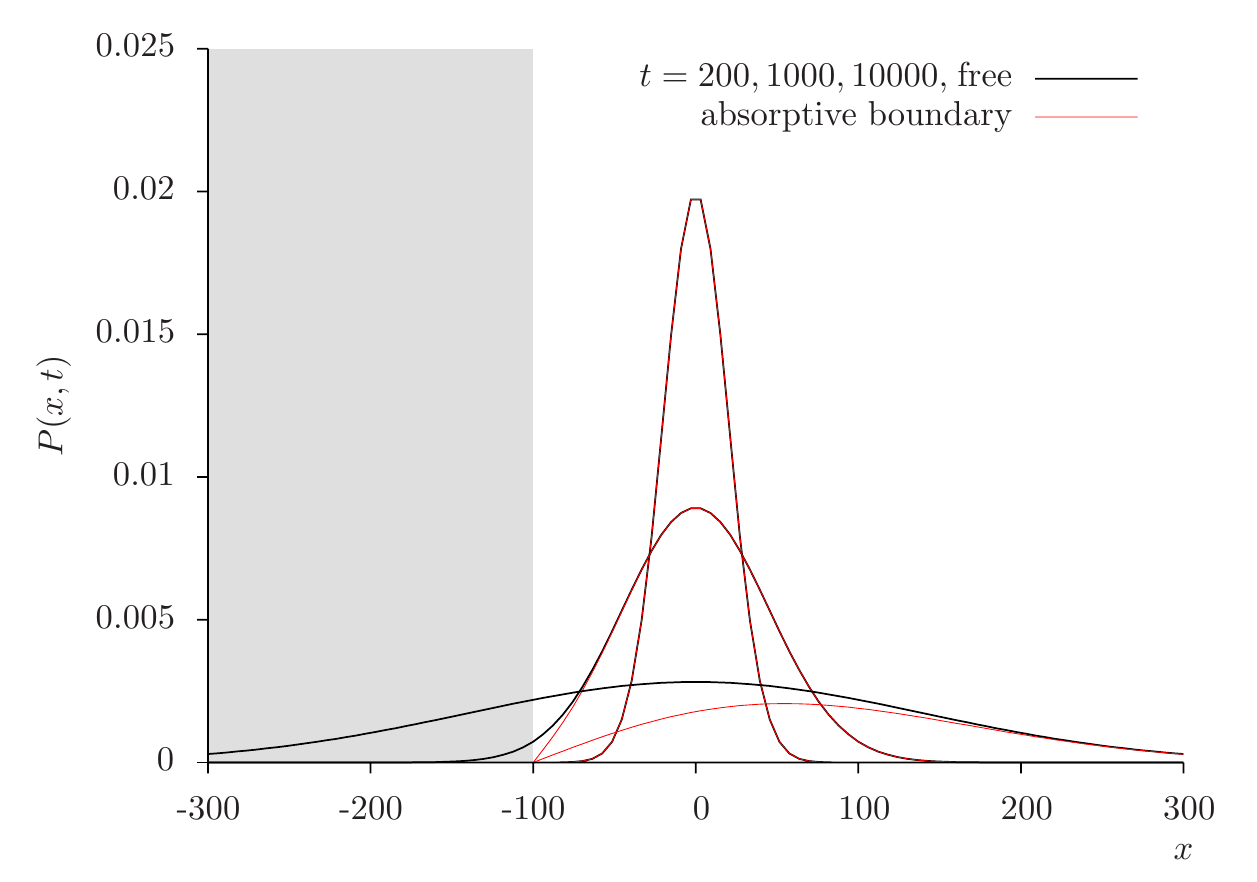}
\end{center}
\caption{\label{fig:plotgaussian}
Probability density $P(x,t)$ to find the particle at position $x$ at time $t$ as a function of $x$
for three different times: $t=200$ (most peaked curve), $t=1000$ and $t=10000$ (flattest curve). 
We consider two different boundary conditions:
free boundary conditions (black curves, Eq.~(\ref{eq:solutiondiffusion})) 
and absorptive boundary condition 
at $X=-100$ (red curves, Eq.~(\ref{eq:solimages})).
The shaded area represents the forbidden region in the case of an absorptive boundary condition.
}
\end{figure}

First, we choose free boundary conditions.
A general solution to Eq.~(\ref{eq:diffusion}) is easily obtained as a superposition
of exponentials,
\be
P(x,t)=\int_{\gamma_0-i\infty}^{\gamma_0+i\infty}\frac{d\gamma}{2i\pi}
e^{-\gamma x}\tilde P(\gamma,t).
\label{eq:mellin}
\ee
From Eq.~(\ref{eq:diffusion}), 
$\tilde P$ obeys the ordinary first-order differential equation
\be
\frac{d\tilde P(\gamma,t)}{dt}=\gamma^2 \tilde P(\gamma,t),
\ee
with the initial condition $\tilde P(\gamma,t=0)=1$.
The solution is trivial:
\be
\tilde P(\gamma,t)=e^{\gamma^2 t}.
\ee
One then inserts this expression in 
Eq.~(\ref{eq:mellin}) in order to compute $P(x,t)$:
\be
P(x,t)=\int_{\gamma_0-i\infty}^{\gamma_0+i\infty}\frac{d\gamma}{2i\pi}
e^{-\gamma x+\gamma^2 t}
=e^{-\frac{x^2}{4t}}\int_{\gamma_0-i\infty}^{\gamma_0+i\infty}\frac{d\gamma}{2i\pi}
e^{t\left(\gamma-\frac{x}{2t}\right)^2}.
\ee
Performing the change of variable $\gamma=\frac{x}{2t}+i\frac{\nu}{\sqrt{t}}$
and sliding the integration contour in such a way that
$\gamma_0=\frac{x}{2t}$, we are left with a standard Gaussian integral
\be
P(x,t)=\frac{e^{-\frac{x^2}{4t}}}{\sqrt{t}}
\int_{-\infty}^{+\infty}\frac{d\nu}{2\pi}e^{-\nu^2}.
\ee
Finally,
\be
P(x,t)=\frac{1}{\sqrt{4\pi t}}e^{-\frac{x^2}{4t}}.
\label{eq:solutiondiffusion}
\ee
This function (of $x$) is represented in Fig.~\ref{fig:plotgaussian}
(at different times~$t$).

In order to characterize such a probability distribution, it is useful to introduce
the generating function of its moments:
\be
G(\lambda,t)=\left\langle e^{\lambda x}\right\rangle
\equiv\int_{-\infty}^{+\infty}dx\,P(x,t)\, e^{\lambda x},
\ee
whose expansion in powers of $\lambda$ has the moments of $x$ as
coefficients:
\be
G(\lambda,t)=\sum_{n=0}^{\infty}\frac{\lambda^n}{n!}
\left\langle
x^n
\right\rangle.
\label{eq:expansionG}
\ee
We note that $G(\lambda,t)=\tilde P(\lambda,t)=e^{\lambda^2 t}$. 
Expanding the latter function in powers of $\lambda$
and identifying the result
to Eq.~(\ref{eq:expansionG}), one gets the following expression for the moments:
\be
\left\langle x^{2n}\right\rangle=(2n-1)!!(2t)^n,\
\left\langle x^{2n+1}\right\rangle=0.
\ee
Another useful tool to characterize a probability distribution is the set of its
cumulants.
We shall denote by $\langle x^n\rangle_c$ the cumulant of order $n$.
The generating function for the cumulants is just the logarithm of $G$, namely
\be
W(\lambda,t)=\ln G(\lambda,t)=\sum_{n=1}^{+\infty}\frac{\lambda^n}{n!}
\left\langle x^n\right\rangle_c.
\ee
In the case of Brownian motion, $W(\lambda,t)=\lambda^2 t$, and thus
all cumulants except the second order one (the variance) are zero:
\be
\langle x^2\rangle_c=2t,\ 
\langle x^{n}\rangle_c=0\ \ \text{for $n\neq 2$}.
\ee
The value of the variance means that the random walk explores a region
of typical size $\sqrt{2t}$ around the origin.
(Of course, this is just the width of the Gaussian 
in Eq.~(\ref{eq:solutiondiffusion})
in our simple case).

\begin{ex}
Prove the following relations between the cumulants and the moments:
\be
\langle x\rangle_c=\langle x \rangle,\
\langle x^2\rangle_c=\langle x^2 \rangle-\langle x \rangle^2,\
\langle x^3\rangle_c=\langle x^3 \rangle-3\langle x^2 \rangle \langle x\rangle
+2\langle x\rangle^3.
\ee
\end{ex}

So far, we have solved the diffusion problem in the case of free boundary conditions
(The particle could diffuse on the whole lattice, without any restriction).
Let us now put an absorptive boundary at position $X<0$.
If the particle hits the position $x=X$, it is lost, and the random walk stops.
{\it The solution to this problem will be used later to address the
branching random walk.}
\begin{figure}
\begin{center}
\includegraphics[width=.7\textwidth,angle=0]{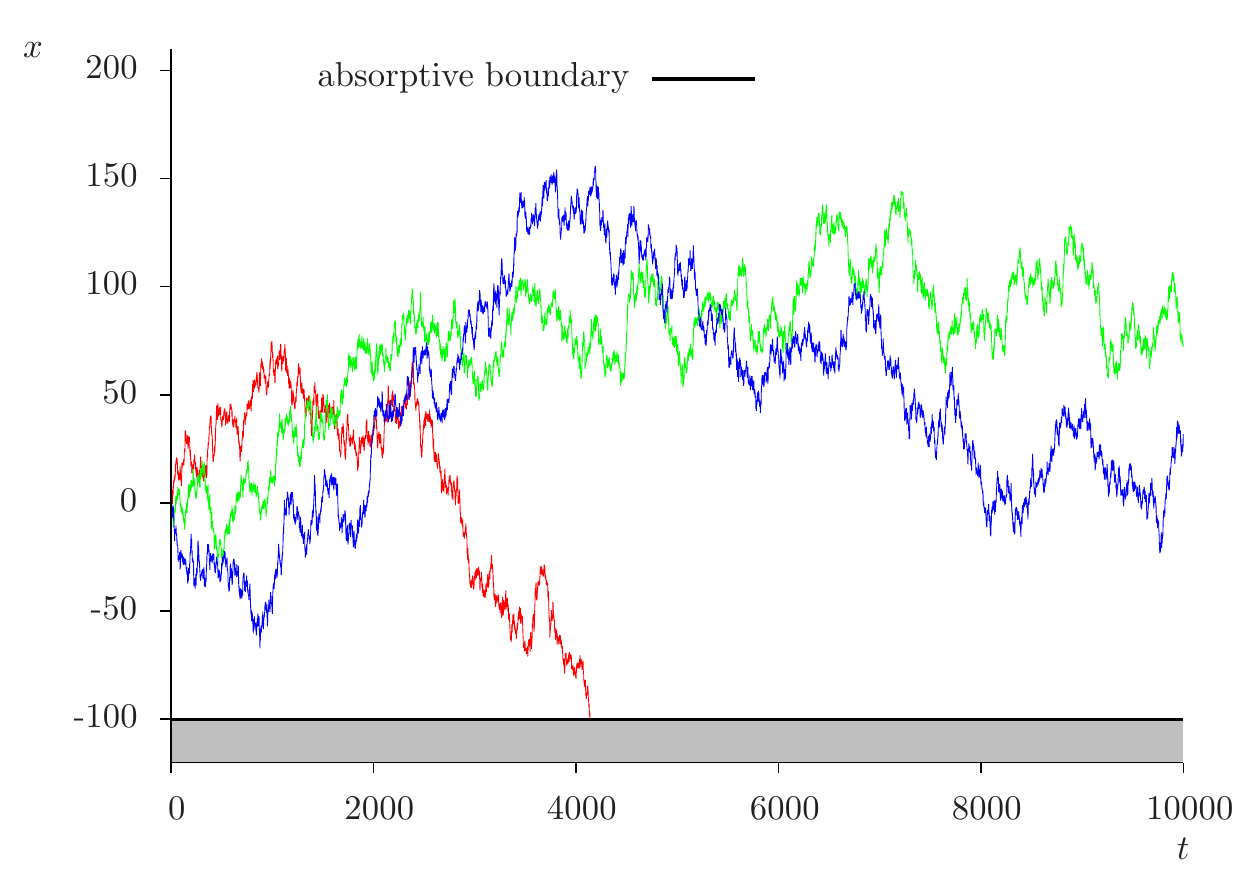}
\end{center}
\caption{\label{fig:plotRWboundary}
Three realizations of the continuous Brownian with an absorptive boundary at $X=-100$
up to $t=10000$.
One of the paths hits the boundary.
}
\end{figure}

Let us state mathematically the problem: We need to solve the diffusion equation
$\partial_t P=\partial_x^2P$ for $x>X$,
with the initial condition $P(x,0)=\delta(x)$,
and the boundary condition $P(X,t)=0$.
It is actually possible to replace this boundary problem by an initial-value problem,
taking advantage of the linearity of the diffusion equation.
It is easy to check that the initial-value problem
\be
\partial_t P=\partial_x^2 P,\
P(x,0)=\delta(x)-\delta(x-2X)
\ee
is equivalent to the boundary problem as long as $x\geq X$.
This is the so-called {\it method of images}.
The solution is then just the difference of Eq.~(\ref{eq:solutiondiffusion})
and of the latter translated by $2X$:
\be
P(x,t)=\frac{1}{\sqrt{4\pi t}}\left[
e^{-\frac{x^2}{4t}}-e^{-\frac{(x-2X)^2}{4t}}
\right].
\label{eq:solimages}
\ee
This function is represented in Fig.~\ref{fig:plotgaussian}.
Let us take the large-time limit of this expression.
We assume that $X$ be of order 1.
We may then write
\be
\begin{split}
P(x,t)&=\frac{1}{\sqrt{4\pi t}}
\left[e^{-\frac{[(x-X)+X]^2}{4t}}-e^{-\frac{[(x-X)-X]^2}{4t}}\right]\\
&=e^{-\frac{(x-X)^2}{4t}}\left[
e^{-\frac{X(x-X)}{2t}}-e^{\frac{X(x-X)}{2t}}
\right]
e^{\frac{X^2}{4t}}\\
&\underset{t\gg 1}{\simeq} \frac{1}{\sqrt{\pi t}}e^{-\frac{(x-X)^2}{4t}}
\sinh\frac{(-X)(x-X)}{2t}.
\end{split}
\ee
The remaining Gaussian factor is significant only in the range
$x-X\ll \sqrt{t}\ll t$. The second inequality is trivial for large $t$.
When the first inequality is satisfied, one may expand the $\sinh$ factor, and
one gets
\be
P(x,t)=\frac{(-X)}{2\sqrt{\pi}}
\frac{(x-X)}{t^{3/2}}
e^{-\frac{(x-X)^2}{4t}}.
\label{eq:imagesres}
\ee


\begin{ex}
Show that Eq.(\ref{eq:imagesres}) is actually an exact solution to the diffusion equation.
(Do not use the method of images.)
\end{ex}

\begin{ex}
Perform the integral
\be
\int_X^{+\infty}dx\,P(x,t)
\ee
and comment on the result. 
Then, compute the mean value of the position of the particle at time~$t$.
\end{ex}


\subsection{Branching random walk}

\begin{figure}[h]
\begin{center}
\begin{tabular}{cc|c}
\multicolumn{2}{c}{Ordinary Brownian motion} & 
\multicolumn{1}{c}{Branching}\\
\\
\includegraphics[width=.25\textwidth,angle=0]{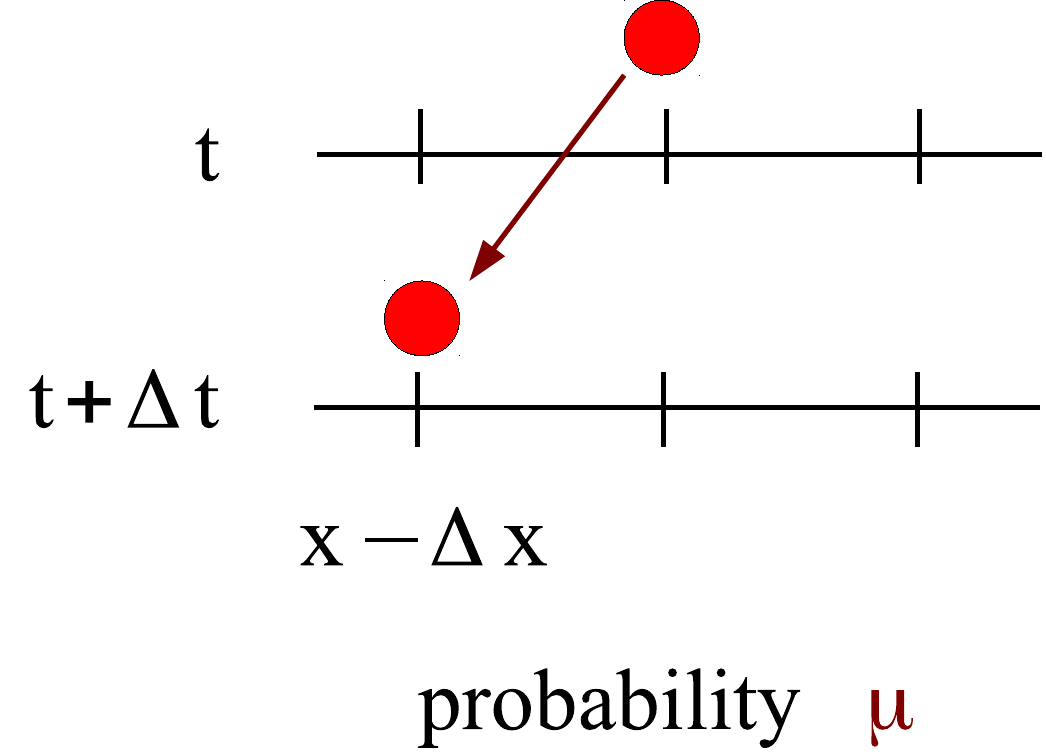}\ \ \ \ &
\includegraphics[width=.25\textwidth,angle=0]{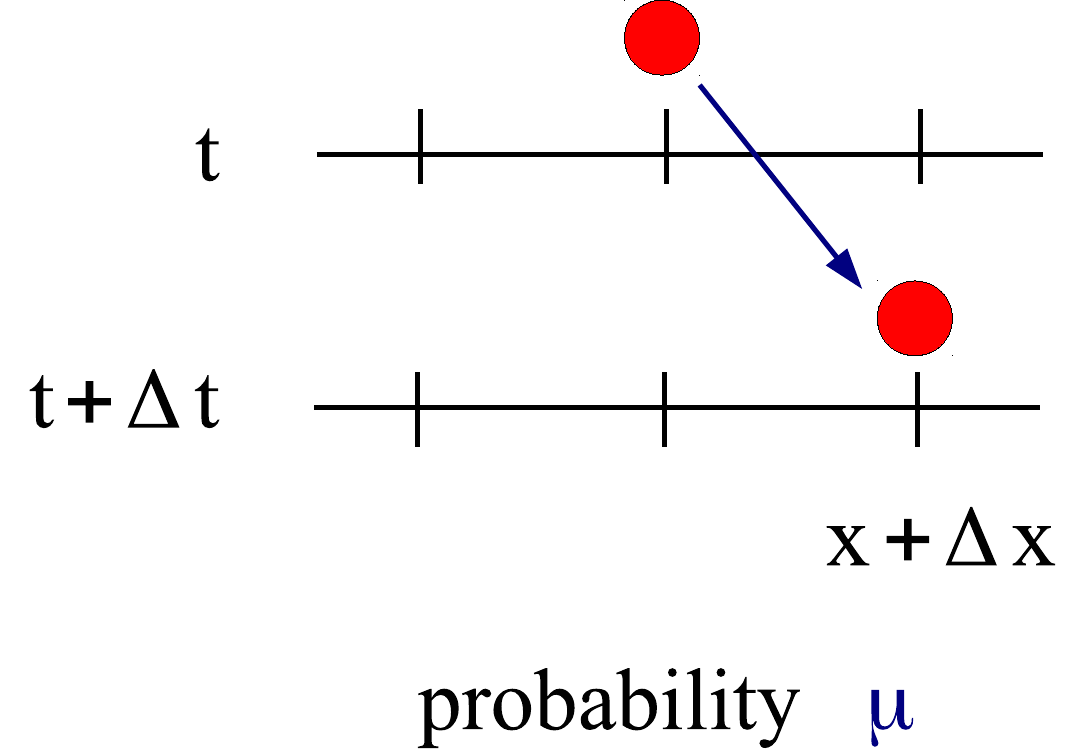}\ \ \ \ &
\includegraphics[width=.25\textwidth,angle=0]{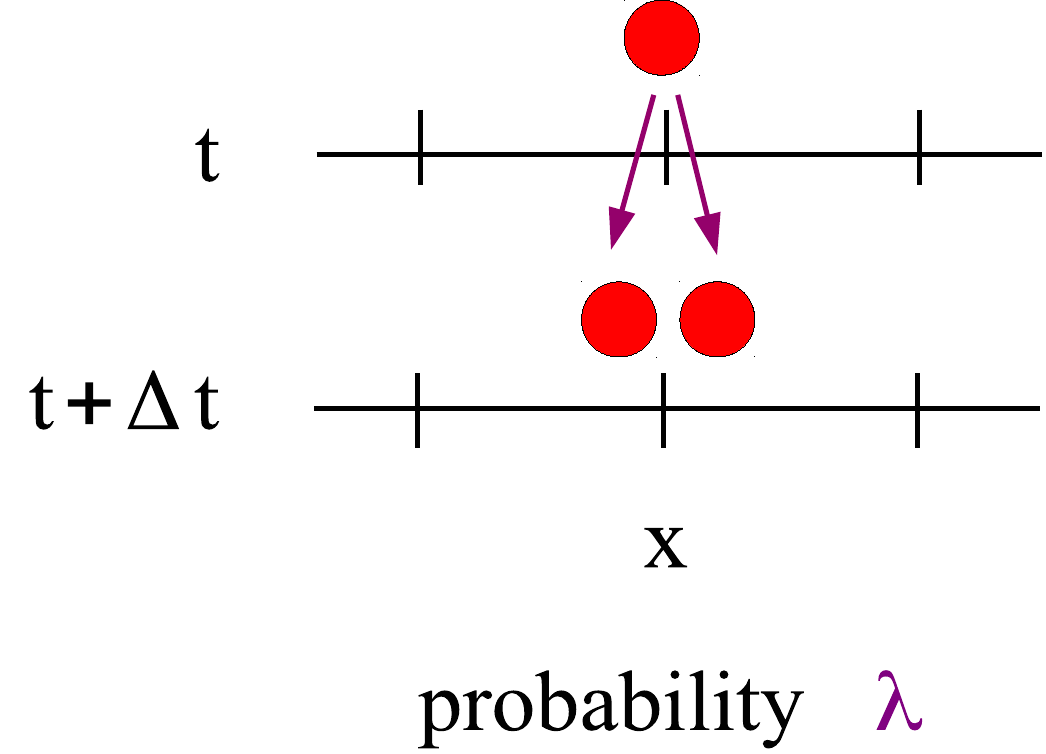}
\end{tabular}
\end{center}
\caption{\label{fig:elemBRW}
Elementary processes defining the branching random walk on a lattice.
}
\end{figure}

\begin{figure}[ht]
\begin{center}
\begin{tabular}{c|c|c}
\includegraphics[width=.3\textwidth,angle=0]{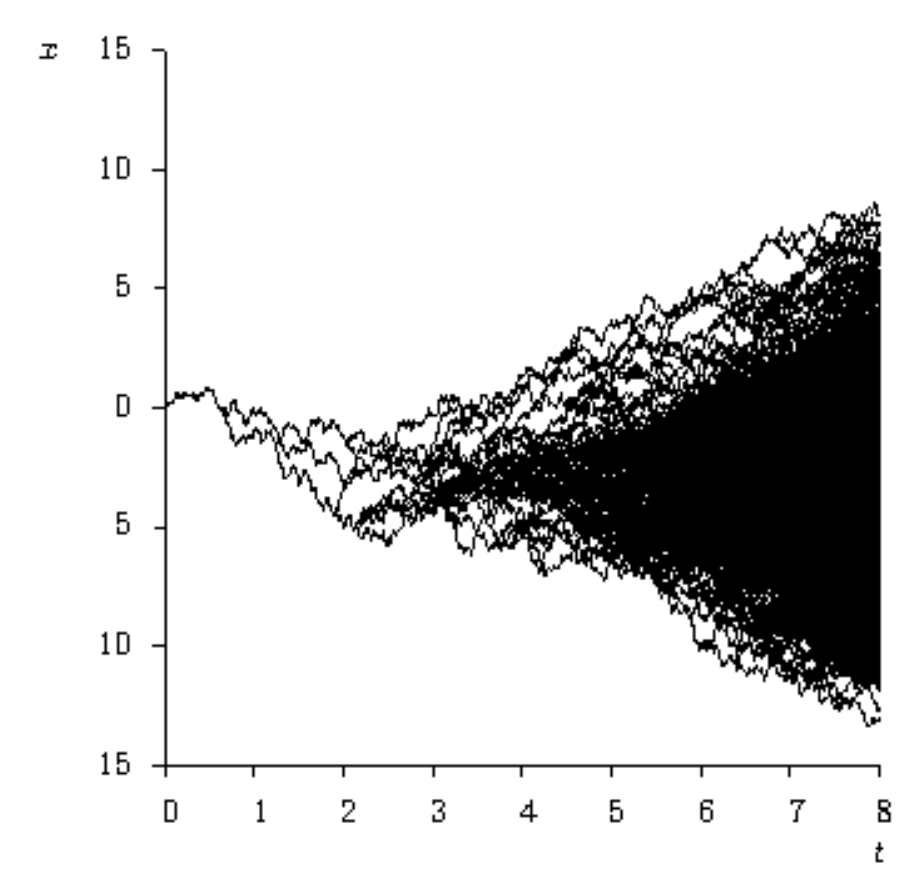}&
\includegraphics[width=.3\textwidth,angle=0]{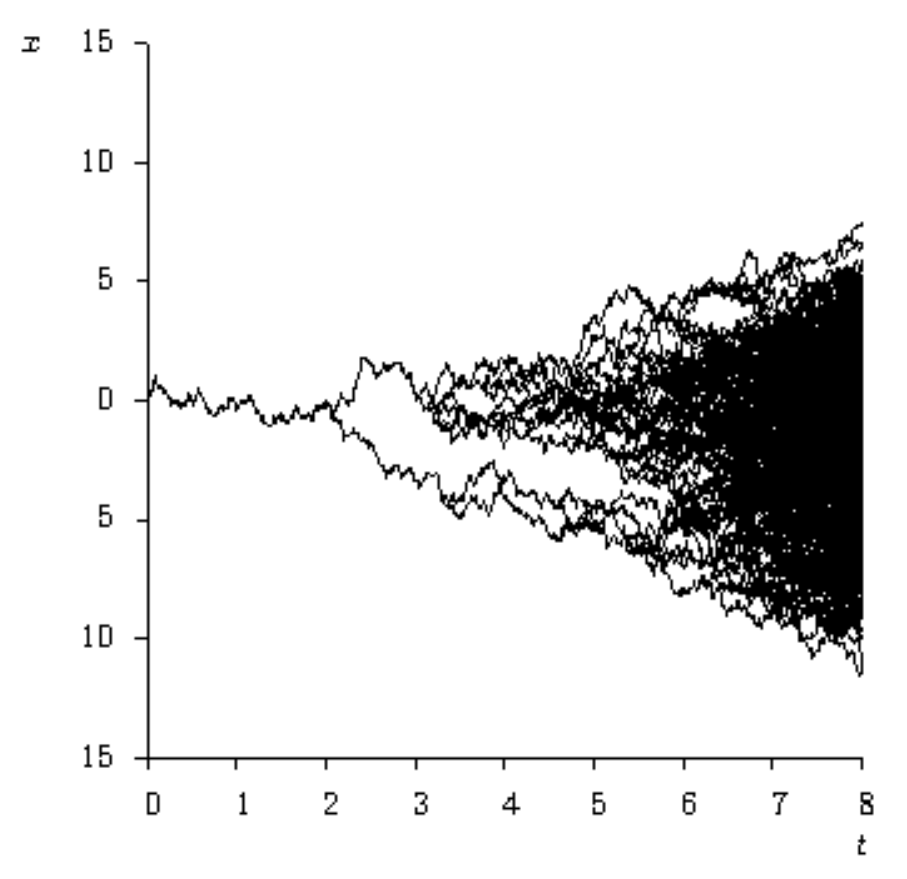}&
\includegraphics[width=.3\textwidth,angle=0]{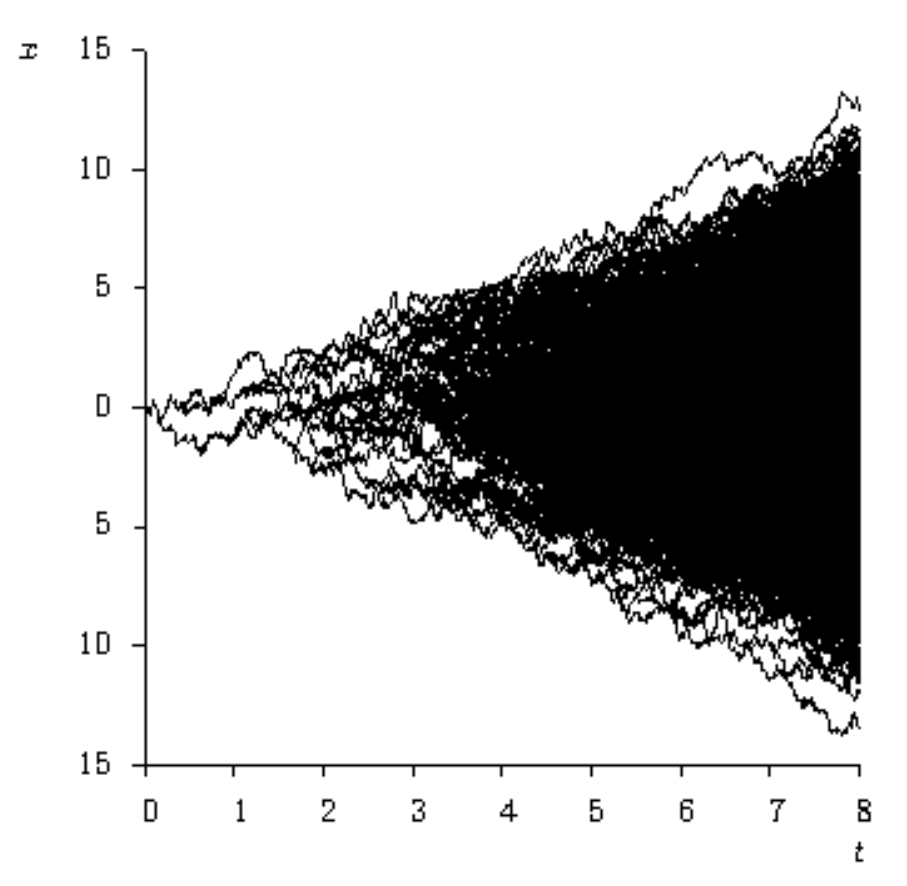}\\
\includegraphics[width=.3\textwidth,angle=0]{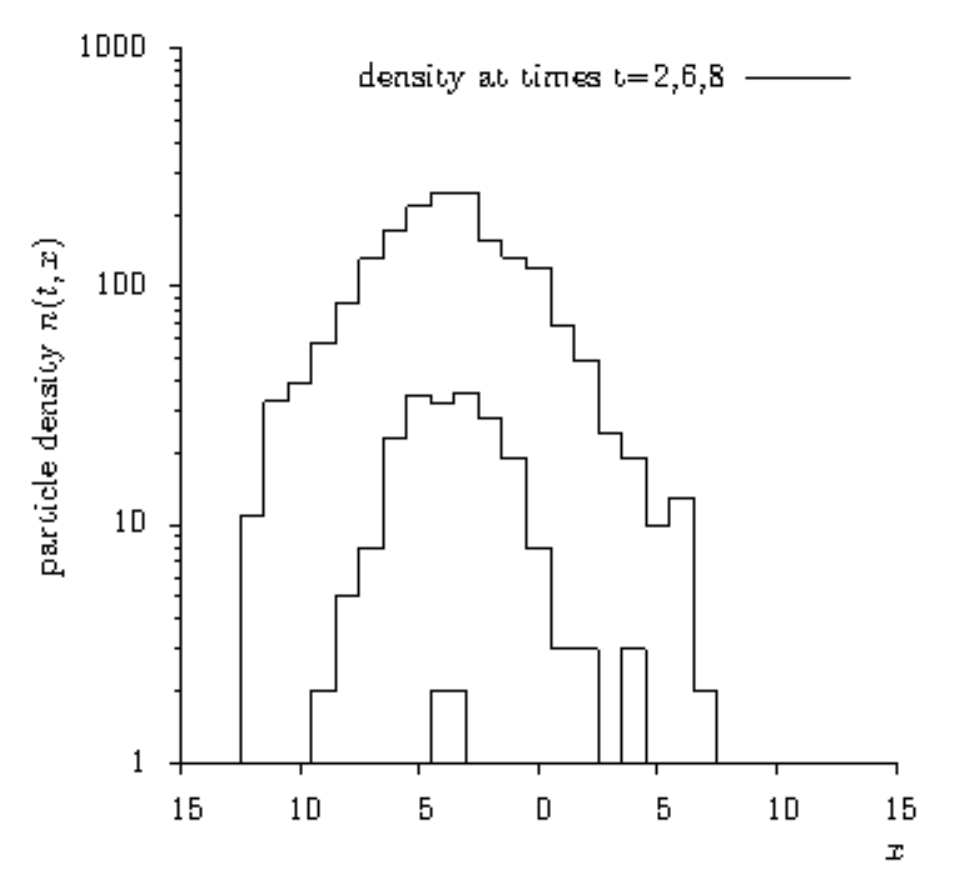}&
\includegraphics[width=.3\textwidth,angle=0]{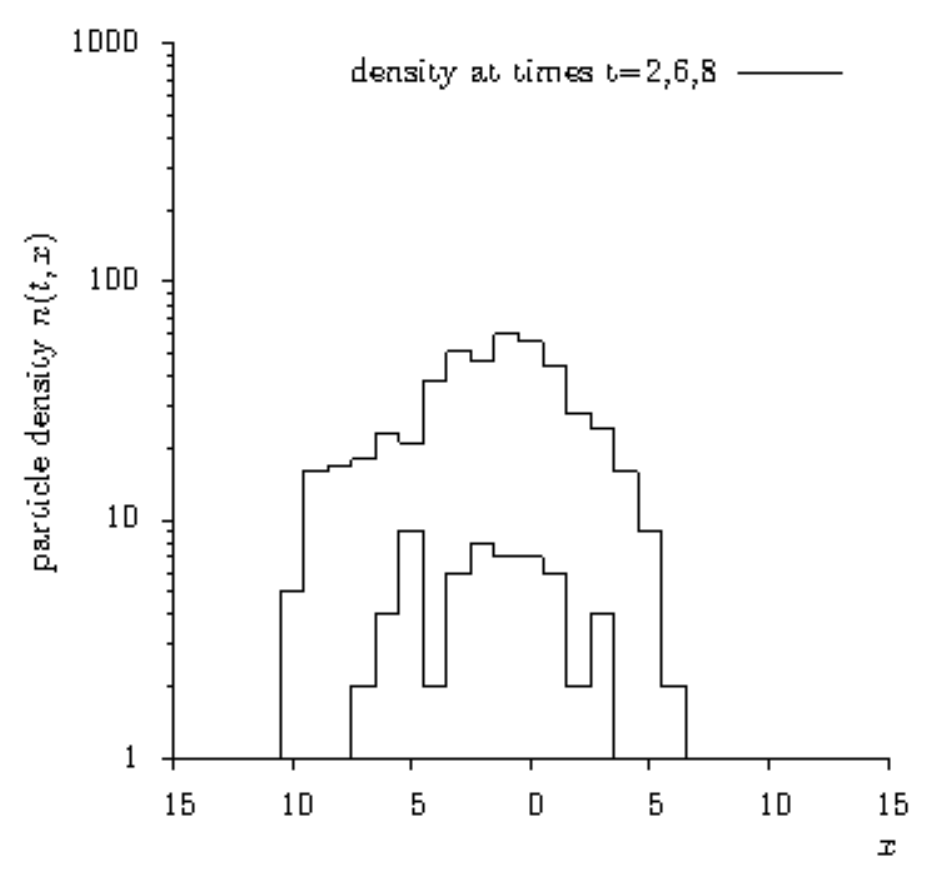}&
\includegraphics[width=.3\textwidth,angle=0]{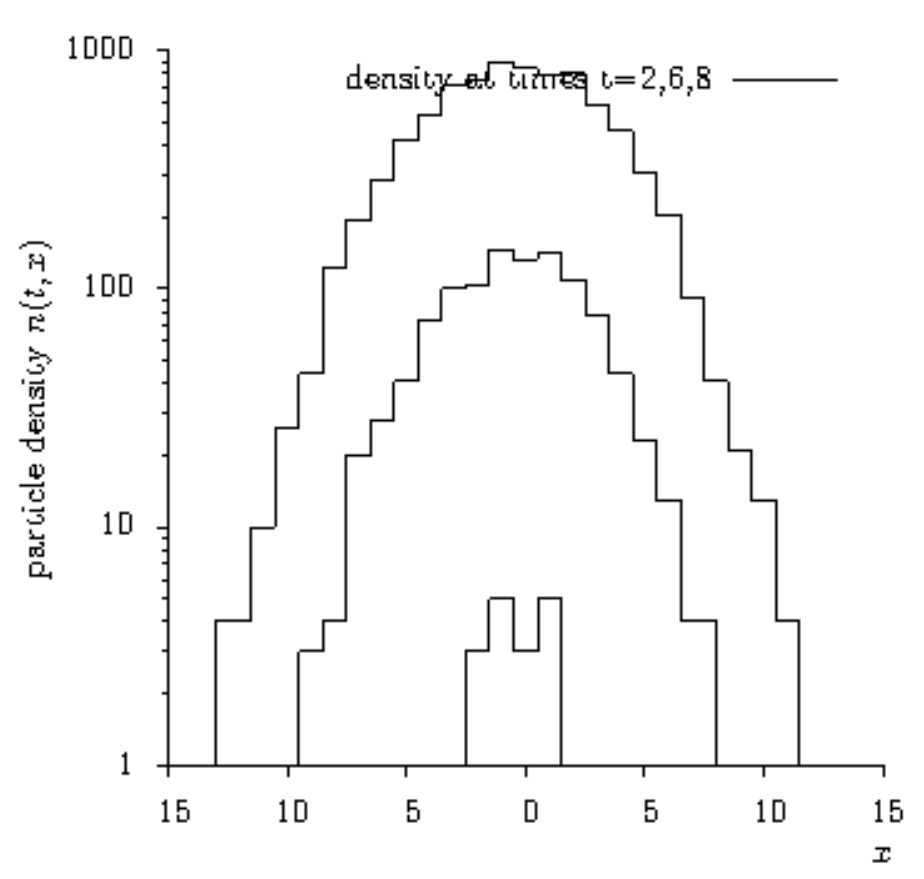}\\
(a) & (b) & (c)
\end{tabular}
\end{center}
\caption{\label{fig:brw}Three realizations of a branching random walk. {\it Top:} particle positions
as a function of time. {\it Bottom:} Corresponding particle number densities in bins
of size 1 (logarithmic scale on the $y$-axis) at three different times: $t=2,6$ and~8.
In realization~(a) and especially~(b), the first splittings happen quite late, leading to a total
number of particles at later time
which is quite low, while in~(c), the first splittings are very fast.
It is during the low density phase, at the beginning of the evolution, that diffusion is very
effective to shift the particle distributions. It is clear in these figures that the
effect of the early-time fluctuations persist to very late times.
}
\end{figure}

We add a process to the Brownian motion defined in Fig.~\ref{fig:brownian1}:
During the time interval $\Delta t$, each particle may split to two particles
on the same site with probability $\lambda$ (see Fig.~\ref{fig:elemBRW}).
Numerical simulations of realizations of a branching random walk in the continuum limit 
are shown in Fig.~\ref{fig:brw}.
Now at time $t$, one has a distribution of particles, whose number and set of
positions are random variables.

Let us establish an equation for the average number of particles on site $x$
at time $t+\Delta t$, given the full distribution of particles at time $t$.
On the average, a fraction $1-2\mu$ of the $n(x,t)$ particles already present
at $x$ at time $t$ does not move and hence contributes to 
$\langle n(x,t+\Delta t)\rangle_{[t,t+\Delta t]}$
(the subscript ${[t,t+\Delta t]}$ means that the average is
taken in the corresponding time interval only),
the fraction $\mu$ of the particles at $x-\Delta x$ add to the latter,
as well as the same fraction $\mu$ of the particles on site $x+\Delta x$.
Finally, a fraction $\lambda$ of the $n(x,t)$ split, which
adds $\lambda n(x,t)$ to $\langle n(x,t+\Delta t)\rangle_{[t,t+\Delta t]}$.
This leads to the equation
\be
\langle n(x,t+\Delta t)\rangle_{[t,t+\Delta t]}=n(x,t)(1-2\mu)
+n(x-\Delta x,t)\mu
+n(x+\Delta x,t)\mu
+\lambda n(x,t).
\ee
Now in order to get a closed equation, we may average over the whole
history between time $0$ and time $t$:
\be
\langle n(x,t+\Delta t)\rangle-\langle n(x,t)\rangle
=\mu
\left[
\langle n(x+\Delta x,t)\rangle+\langle n(x-\Delta x,t)\rangle
-2\langle n(x,t)\rangle
\right]
+\lambda
\langle n(x,t)\rangle.
\ee
We can set $\lambda=\Delta t$
and $\mu(\Delta x)^2=\Delta t$,
and take to limits $\Delta x,\Delta t\rightarrow 0$
to arrive at a partial differential equation:
\be
{\partial_t}\langle n\rangle
=\partial_x^2\langle n\rangle+\langle n\rangle.
\ee
The first term in the right-hand side of this equation is a diffusion term,
while the second term represents the branchings.
Using the integral transform~(\ref{eq:mellin}) (we call $\tilde n(\gamma,t)$
the transform of $\langle n(x,t)\rangle$), we obtain an equation which can
be viewed as an ordinary differential equation
\be
\frac{d \tilde n(\gamma,t)}{dt}=(\gamma^2+1) \tilde n(\gamma,t),
\ee
with the initial condition $\tilde n(\gamma,t=0)=1$.
The solution is again trivial:
\be
\tilde n(\gamma,t)=e^{(\gamma^2+1) t},
\ee
and transforming back to $\langle n\rangle$ using the inverse Mellin 
transform~(\ref{eq:mellin}):
\be
\langle n(x,t)\rangle=
\int\frac{d\gamma}{2i\pi}e^{-\gamma x+(\gamma^2+1)t}
=\frac{1}{\sqrt{4\pi t}}
{\exp\left({t-\frac{x^2}{4t}}\right)}.
\label{eq:meann}
\ee
This is an exact result, and was obtained very simply.

\begin{figure}
\begin{center}
\includegraphics[width=.7\textwidth,angle=0]{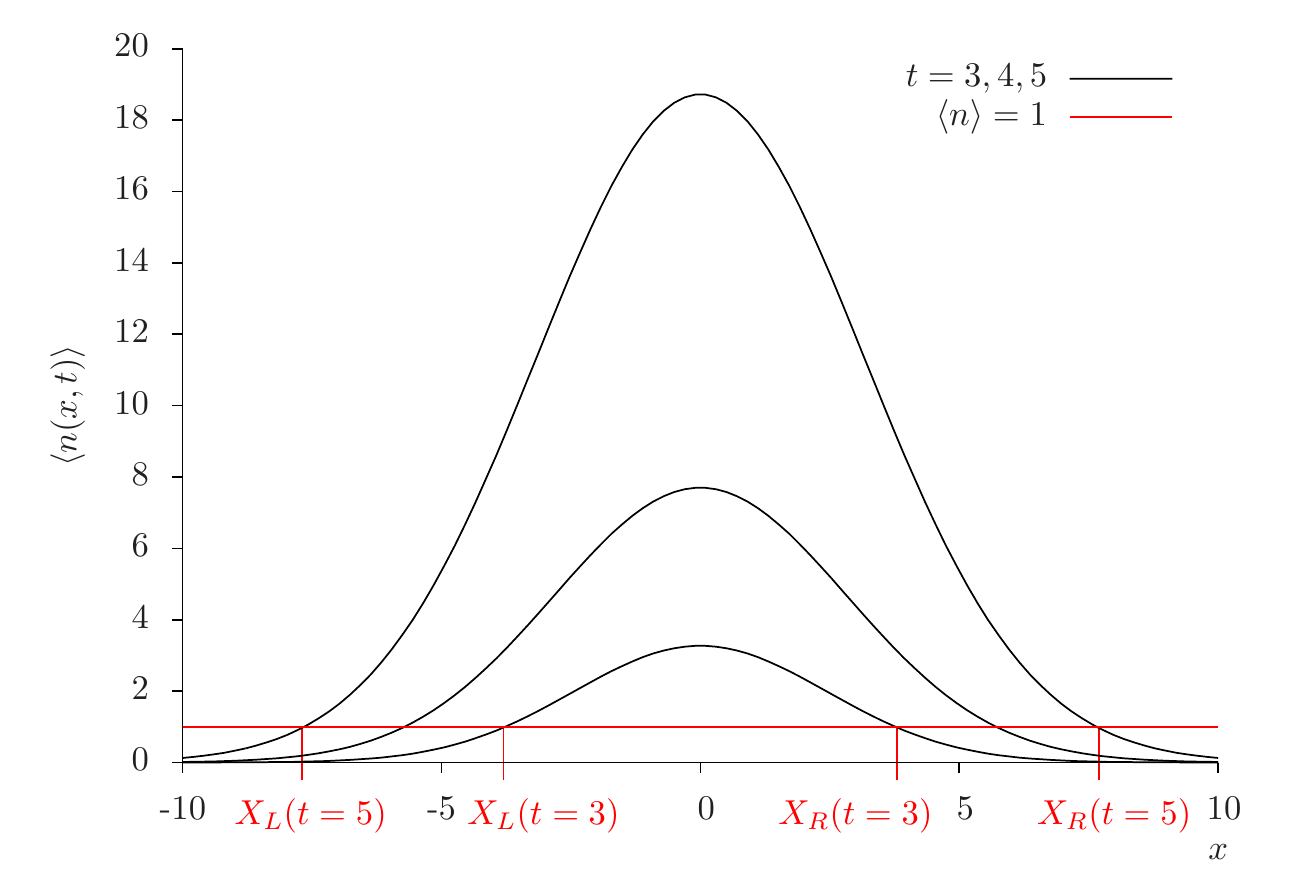}
\end{center}
\caption{\label{fig:plotbranchinggaussian}
Mean particle number $\langle n(x,t)\rangle$ (Eq.~(\ref{eq:meann})) 
generated by a branching random walk
for three different times: $t=3,4,5$.
The values of $x$ for which $\langle n(x,t)\rangle=1$ are also
represented for $t=3$ and $t=5$.
}
\end{figure}

The function~(\ref{eq:meann}) is represented in Fig.~\ref{fig:plotbranchinggaussian}.

There are other quantities related to the branching random walk for which
an analytical expression is much less easy to get.
One of them is the mean position of the rightmost (or leftmost) particle 
in the branching random
walk, as a function of time.

\begin{figure}[ht]
\begin{center}
\begin{tabular}{c|c}
\includegraphics[width=.4\textwidth,angle=0]{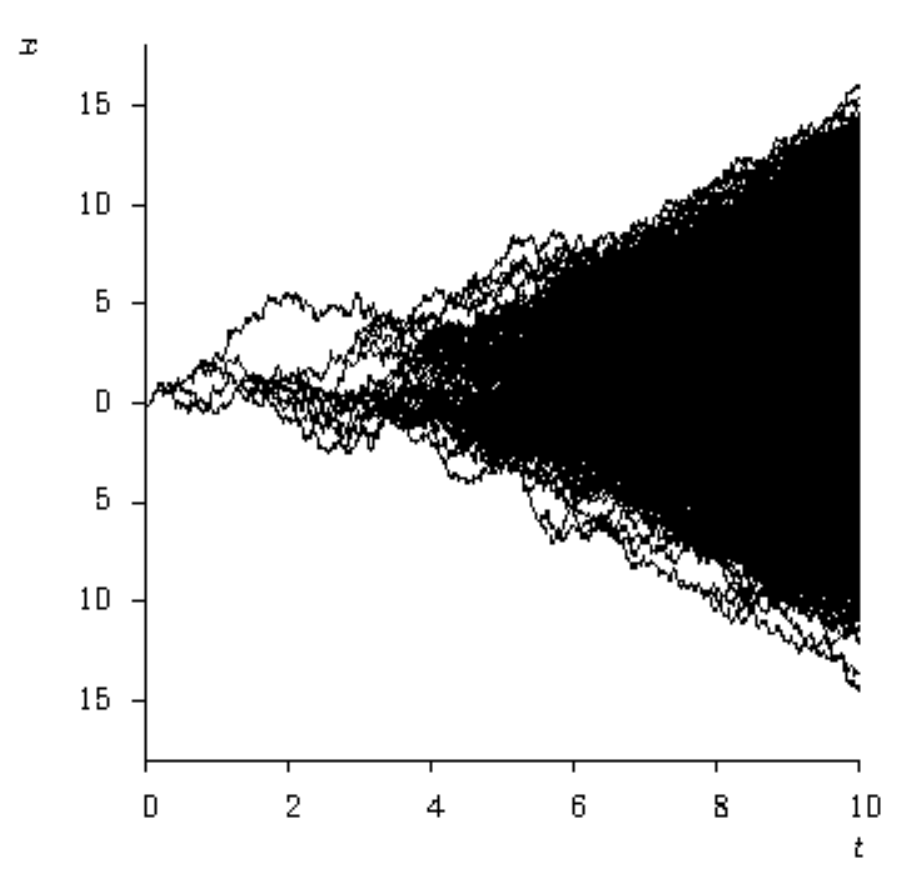}&
\includegraphics[width=.4\textwidth,angle=0]{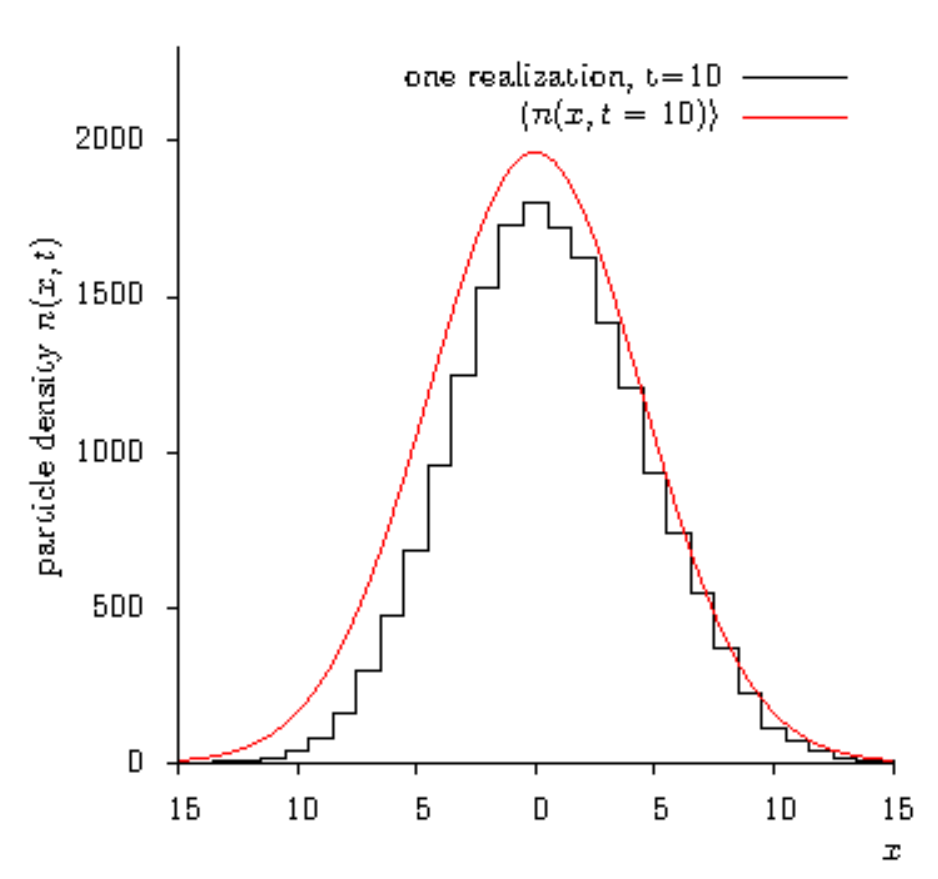}\\
(a) & (b)
\end{tabular}
\end{center}
\caption{\label{fig:brw_bis}Another realization of a branching random walk
until $t=10$.
(a) Trajectories of the particles as a function of time.
(b) Comparison between the particle density in the realization at time $t=10$ 
and the mean particle density (red continuous line). We selected a particular
realization for which the density 
is not very different in
shape from the latter. There are actually large event-by-event
variations with respect to the mean.
One important feature of the realization is that there is of course a leftmost
and a rightmost occupied bin, while $\langle n(x,t>0)\rangle$ is nonzero for all $x$.
}
\end{figure}

Let us first try the most naive approach.
We assume that the mean particle density $\langle n\rangle$
reflects the particle distribution in each realization. Then, the position
$X_R(t)$, $X_L(t)$ of the rightmost and leftmost particles respectively
would be the values of $x$ for which $\langle n(x,t)\rangle$ is say 1 
(see Fig.~\ref{fig:plotbranchinggaussian}).
To determine $X_{R,L}(t)$, we just need to solve $\langle n(X_{R,L}(t),t)\rangle=1$.
From Eq.~(\ref{eq:meann}), we find, at large $t$,
\be
X_R(t)=2t-\frac12\ln t+\text{const},\
X_L(t)=-2t+\frac12\ln t+\text{const}.
\label{eq:XRLnaive}
\ee
This result is actually not fully correct, which is not so suprising given that there
are large fluctuations between realizations in the particle number densities 
(see Fig.~\ref{fig:plotbranchinggaussian}, and Fig.~\ref{fig:brw_bis}
for a comparison between one realization and the mean density).
It turns out however that the first terms $\pm 2t$ are the correct ones. The fact that the
subleading terms are logarithmic is also correct, but the coefficients
of these logs are wrong.
In order to obtain the correct result, we need to establish an exact equation
for the probability distribution
$p(X,t)$ of the position ($X$) of say the rightmost particle in the branching random walk.

To this aim, let us introduce the probability $Q(x,t)$
that at time $t$, all particles
be on the left of position $x$, starting from one single particle at position 0.
We establish an evolution equation for $Q(x,t)$ in the same way as for $P$ in the case of the
Brownian motion or of $\langle n\rangle$ above, that is by trying to relate $Q$ at time
$t+\Delta t$
to $Q$ at time $t$. However, in the present case, it is better to divide
the time interval as $[0,t+\Delta t]=[0,\Delta t]\cup[\Delta t,t+\Delta t]$,
namely to add the small interval $\Delta t$ at the beginning when the system
still consists in a single particle.

After the first time step of size $\Delta t$, 
the system consists either (i) in a single particle at position $+\Delta x$ (this
happens with probability $\mu$), or (ii) in a single particle
at position $-\Delta x$ (with the same probability), 
or (iii) of two particles at
position $0$ (with probability $\lambda$), or finally (iv) 
of one single particle at position 0
if nothing happens in the first time step (probability $1-2\mu-\lambda$).
In case~(i), the probability $Q(x,t+\Delta t)$ that all particles be on the left
of position $x$ at time $t+\Delta t$ is the probability that all particles
be on the left of $x$ after evolution of a particle initially at $\Delta x$
over a time interval $t$, namely $Q(x-\Delta x,t)$.
In case~(ii), the same line of reasoning leads to $Q(x+\Delta x,t)$.
In the third case,
at time $\Delta t$, we have two particles at position~0, which evolve independently
of each other over $t$ additional steps of time. Hence the probability that all particles
be to the left of $x$ at time $t+\Delta t$ is the probability that all particles of both
{\it independent} branching random walks be to the left of $x$, namely $[Q(x,t)]^2$.
The assumption that the particles have independent evolutions is of course crucial here
to obtain this term as a simple product. Case~(iv) is trivial.

Translating this discussion into a mathematical expression,
we get the equation
\be
Q(x,t+\Delta t)=\mu\left[Q(x-\Delta x,t)+Q(x+\Delta x,t)\right]
+\lambda Q^2(x,t)+(1-2\mu-\lambda)Q(x,t),
\ee
which can be recast as a finite-difference evolution equation:
\be
Q(x,t+\Delta t)=Q(x,t)+\mu\left[Q(x-\Delta x,t)+Q(x+\Delta x,t)-2Q(x,t)\right]
+\lambda \left[Q^2(x,t)-Q(x,t)\right].
\ee
Taking the usual continuous limit
($\Delta t, \Delta x\rightarrow 0$ with $\mu(\Delta x)^2=\Delta t$ and $\lambda=\Delta t$), 
we arrive at a nonlinear partial differential
equation called the Fisher-Kolmogorov-Petrovsky-Piscounov (FKPP)
equation
\be
\frac{\partial Q}{\partial t}=\frac{\partial^2 Q}{\partial x^2}-Q+Q^2.
\ee
At $t=0$, if one starts with a single particle at position 0, then
obviously the probability $Q(x,0)$ is 1 for $x> 0$ and $0$ for $x\leq 0$, namely
\be
Q(x,0)=\theta(x).
\ee
Recalling the definition of $Q$,
we see immediately that the probability distribution $p(x,t)$
of the position of the rightmost particle
is just the $x$-derivative of $Q$:
\be
p(X,t)=\frac{\partial}{\partial x}Q(x,t)|_{x=X}
\ee
and hence the average position of the rightmost particle reads
\be
X_R(t)=\langle x\rangle_t=\int_{-\infty}^{+\infty}dx\,x\,p(x,t)=
\int_{-\infty}^{+\infty}dx\,x\,\frac{\partial}{\partial x}Q(x,t).
\ee


\begin{ex}
We introduce the number $N(t)$ of particles at time $t$ in a given realization,
the set $\{x_i(t)\}$ of their positions,
and a function $f(x)$.
Prove that 
\begin{equation}
F(x,t)\equiv\left\langle\prod_{i=1}^{N(t)}f(x-x_i(t))\right\rangle
\label{eq:F}
\end{equation}
obeys the FKPP equation with $f(x)$ as initial condition.
\end{ex}

In these lectures, we shall mainly use an alternative form of the FKPP equation,
which is obeyed by the function $u\equiv 1-Q$:
\be
\partial_t u=\partial_x^2 u+u-u^2.
\label{eq:FKPPu}
\ee
Of course, $u(x,t)$ is simply the probability that at least one particle be
located to the right of $x$ at time $t$.


\begin{np}
Branching random walks on a spacetime lattice are relatively easy to implement
numerically. It is useful to write a code which generates realizations of such a model,
in order to be able to ``play'' with the model and build up an intuition of its
behavior.

Consider the model described at the begining of this section.
(We may set, for example, $\mu=\lambda=\Delta t=10^{-2}$ and $\Delta x=1$).

The most straightforward method would be to simulate the behavior of each 
individual particle as one increases time from
$t$ to $t+\Delta t$, namely to ``decide'' for each particle 
whether it moves
right, left, duplicates, or stays as is in this time interval. 
However, the complexity of this
method is linear in the number of particles, that is to say exponential in time,
and thus becomes unpractical after a few times steps.

However, since we have a spacetime lattice and since the particles are
indistinguishable, we can instead decide for each site how many particles
move right, how many move left and so on.

The first step of our project is to prove
that given a number $n$ of particles on a particular
 site at time $t$, the joint distribution
of the number $n_L$ of particles
that move left, $n_R$ that move right, and $n_+$ that duplicate is given by the
multinomial law
\be
P(n_L,n_R,n_+)=\binom{n}{n_L,n_R,n_+,n-n_L-n_R-n_+}\mu^{n_L+n_R}\lambda^{n_+}
(1-2\mu-\lambda)^{n-n_L-n_R-n_+},
\ee
where the multinomial coefficient is a generalization of the binomial coefficient:
\be
\binom{n}{k_1,k_2,\cdots,k_j}=\frac{n!}{k_1!k_2!\cdots k_j!}
\ \ \text{with}\ \
n=\sum_{i=1}^j k_i.
\ee
Since according to our naive estimate, the number of sites which are occupied at time
$t$ grows linearly with $t$, the complexity also depends linearly on $t$.

It is now an easy programming exercise to implement this evolution rule.
The only practical issue may be with the bookkeeping of the moves of the particles.

\end{np}

\begin{recap}
We have introduced branching random walks in one space dimension.
It is a class of stochastic models with basically two elementary processes
which determine the dynamics: diffusion in space, and branching.
We have seen that the mean density of particles obeys a simple linear
partial differential equation.
Other ``observables'' on this branching random walk such as
the mean position of the boundaries (namely of the rightmost/leftmost particles)
are derived from nonlinear
partial differential equations instead, 
such as the FKPP equation~(\ref{eq:FKPPu}) in the
simplest case of branching Brownian motion (continuous space and time) with diffusion constant and
branching rate both set to unity.
\end{recap}

\clearpage


\section{Solving the FKPP equation}

\begin{intro}
This section is dedicated to finding solutions, or rather, properties of the solutions
to the FKPP equation
\begin{equation*}
\partial_t u=\partial_x^2 u+u-u^2.
\tag{\ref{eq:FKPPu}$'$}
\end{equation*}
Our approach will essentially be heuristic; We will nevertheless
state a fundamental mathematical theorem on the convergence of the solutions to traveling waves
at large times. Then, 
we shall generalize the obtained properties
to a wider class of equations.
\end{intro}

\subsection{Heuristic analysis of the equation}

\begin{figure}[h]
\begin{center}
\includegraphics[width=.8\textwidth,angle=0]{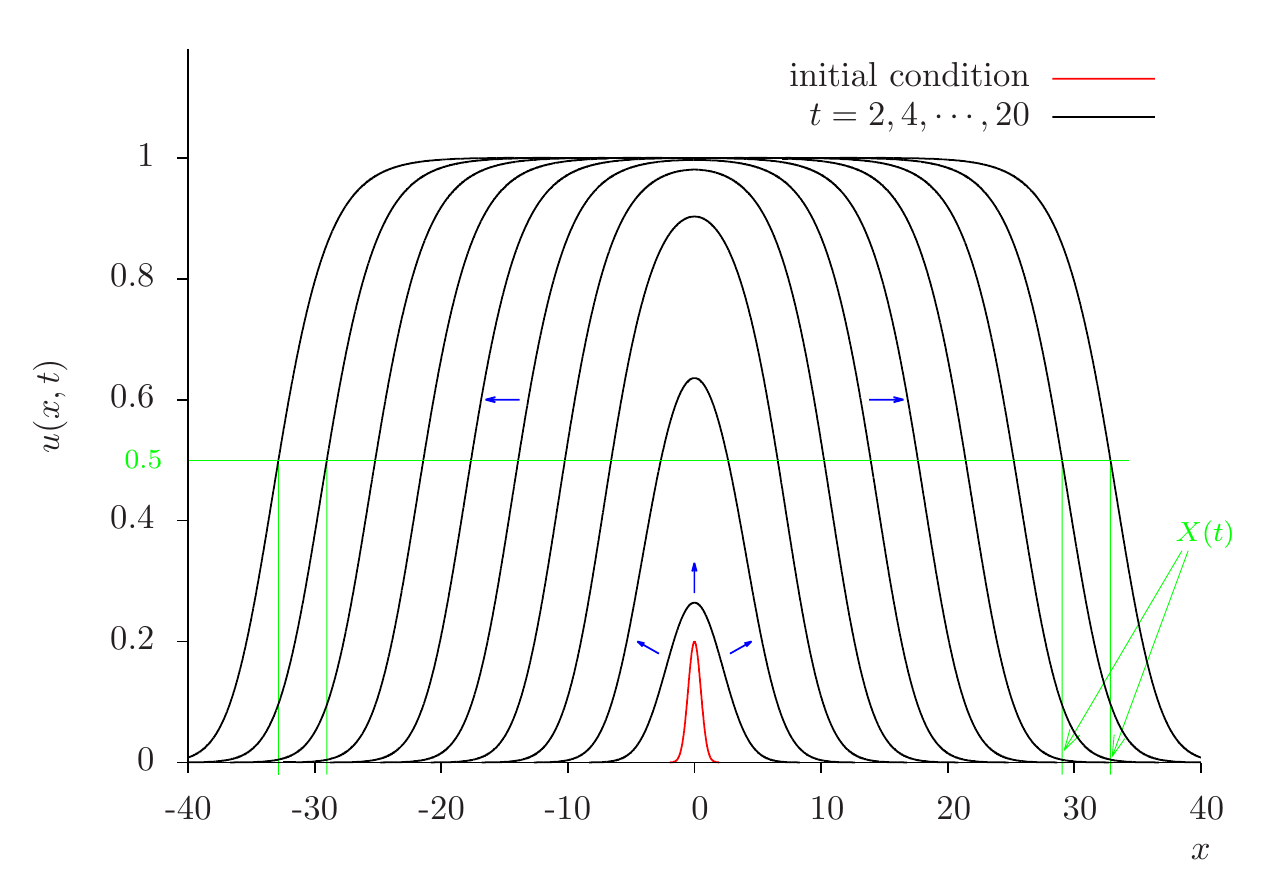}
\end{center}
\caption{\label{fig:plotFKPP}
Numerical solution of the FKPP equation~(\ref{eq:FKPPu}) at different times 
(black curves),
starting from a localized initial condition (red curve which
represents Eq.~(\ref{eq:initialfkpp}) for
$\varepsilon=0.3$).
The blue arrows indicate the sense of the evolution from one time to the
next one.
}
\end{figure}
We first look for spatially homogeneous solutions $u(x,t)=U(t)$.
Then Eq.~(\ref{eq:FKPPu}) reduces to the simple first-order equation
\be
U'(t)=U(t)-U^2(t),
\ee
whose solution is trivial.
We shall however limit ourselves to analyze the two fixed points $U=0$ and $U=1$.
The latter is stable, while the former is unstable.
In order to see these facts, we consider infinitesimal perturbations of
these fixed points, and follow their $t$ evolution.

If $U(t=0)=\varepsilon\ll 1$, 
then $U(t)\simeq \varepsilon e^t$ (as long as $t\ll\ln 1/\varepsilon$).
A small perturbation grows exponentially with time, which means that $U=0$ is indeed an
unstable fixed point.
If one perturbs instead the other fixed point by setting the initial condition
$U(t=0)=1-\varepsilon$, 
then $U(t)\simeq 1-\varepsilon e^{-2t}$, and thus $U$ goes back 
to the fixed point $U=1$, which
means that it is stable.

We go back to the full equation~(\ref{eq:FKPPu}), and
we start the evolution with a localized, small initial condition, say
\be
u(x,t=0)=\frac{\varepsilon}{\sqrt{2\pi}} 
e^{-\frac{x^2}{2\varepsilon^2}}\ \text{with}\ \varepsilon\ll 1.
\label{eq:initialfkpp}
\ee
This is a perturbation to the unstable fixed point, and thus
we know that it should grow exponentially
with time. At small times ($t\ll \ln 1/\varepsilon$), $u\ll 1$ and the nonlinear term 
in Eq.~(\ref{eq:FKPPu}) can be neglected compared to the linear growth term.
Thus the FKPP equation may be replaced by its linearized part
\be
\partial_t u=\partial_x^2 u+u,
\label{eq:FKPPulinearized}
\ee
which encodes an exponential growth in time and a diffusion in space.
At large enough $t$, there are regions in $x$
in which $u$ reaches 1, and where the nonlinear term
is no longer negligible. Actually, it starts to compensate the linear
growth term, and tames the exponential growth, bringing $u$ to its stable
fixed point.
The growth may continue only at larger values of $|x|$. Thus, wave fronts form,
and move to larger values of $|x|$ as time elapses. These fronts are called ``traveling waves'',
and are characterized by their position $X(t)$ and their shape in the comoving frame.
We show a numerical solution of the FKPP equation in Fig.~\ref{fig:plotFKPP} in order to
illustrate the dynamics just described, and the formation of the traveling
wave starting from a small localized initial condition.

This intuitive discussion is actually backed by a rigorous mathematical theorem,
which we are going to state in the following section.

\subsection{Bramson's theorem: traveling waves}

We are going to put the theorem in a general form that will be easy to take over
to different kinds of branching random walks later.
To this aim, we introduce notations that may seem arbitrary at this stage, but whose
meaning will become transparent later on.

Let us define the function
\be
v(\gamma)=\gamma+\frac{1}{\gamma},
\label{eq:v}
\ee
which is determined by the linearized part of the FKPP 
equation (Eq.~(\ref{eq:FKPPulinearized})), 
see below.
Let us also introduce
$\gamma_0$, the solution of $v'(\gamma_0)=0$.

The theorem states that if one chooses an initial condition such that 
$u(x,0)$ decreases smoothly from 1 to 0 as $x$ goes from $-\infty$ to $+\infty$, with 
the asymptotic behavior 
\be
u(x,0)\underset{x\rightarrow+\infty}{\sim}
e^{-\beta x}\ \text{with $\beta\neq\gamma_0$}\
\ \text{or}\ \ u(x,0)\underset{x\rightarrow+\infty}{\sim}x^{\nu}e^{-\gamma_0 x},
\ee
then, at large time, $u$ becomes a function of a single variable:
\be
u(x,t)\underset{t\rightarrow +\infty}{\sim}{\cal U}(x-X^{(\beta)}(t)),
\label{eq:travelingwaveproperty}
\ee
where
\be
X^{(\beta)}(t)=\begin{cases}
v(\beta) t+{\cal O}(1) &\text{for $\beta<\gamma_0$ or ($\beta=\gamma_0$ and $\nu<-2$)}\\
v(\gamma_0) t+\frac{\nu-1}{2\gamma_0} \ln t+{\cal O}(1)  
&\text{for $\beta=\gamma_0$ and $\nu>-2$}\\
v(\gamma_0) t-\frac{3}{2\gamma_0} \ln t+{\cal O}(1)  &\text{for $\beta>\gamma_0$}
\end{cases}
\ee
Here, $\gamma_0=1$ and $v(\gamma_0)=2$, but we shall deal with
variants of the FKPP equation, for which it will be enough to
replace the function $v$ and hence the parameters $\gamma_0$ and $v(\gamma_0)$ for
the above formulae to apply.

This is (our reformulation of) a theorem which was proved rigorously.
In the next sections, we shall motivate these formulae through heuristic arguments.

Note that for applications to
QCD, only the last case of Bramson's theorem will be relevant to us.


\subsection{Heuristic derivation of the properties of the traveling waves}


\subsubsection{Asymptotic shape and velocity}

Let us go far to the right of the position $X(t)$ of the wave front.
There, the linearized equation
\be
\partial_t u=\partial_x^2 u+u
\tag{\ref{eq:FKPPulinearized}$'$}
\ee
is a good approximation to the full FKPP equation.
We have seen that the solution of such an equation reads
(see Eq.~(\ref{eq:meann}) with the substitution $\langle n\rangle\rightarrow u$)
\be
u(x,t)=\int\frac{d\gamma}{2i\pi}\tilde u(\gamma,0)
e^{-\gamma x+(\gamma^2+1)t}
=\int\frac{d\gamma}{2i\pi}\tilde u(\gamma,0)
e^{-\gamma[ x-v(\gamma)t]},
\ee
where $v(\gamma)$ is the function
given by Eq.~(\ref{eq:v}) and
is the velocity of the wave of ``wave number'' $\gamma$.

We now go to the moving frame defined by the change of coordinates
$x=\xi+V t$, where $V$ is a constant representing the velocity of this new frame
with respect to the original one.
We choose the initial condition $u(x,t=0)=e^{-\beta x}$ for $x>0$ and
$u(x,t=0)=1$ for $x\leq 0$. Then, obviously,
\be
\tilde u(\gamma,0)=\frac{\beta}{\gamma(\beta-\gamma)}.
\ee
In the new frame and with this initial condition,
\be
u(\xi,t)=\int\frac{d\gamma}{2i\pi}\left[
\frac{\beta}{\gamma(\beta-\gamma)}
\right]
e^{-\gamma[\xi-(v(\gamma)-V)t]}.
\ee
The integration goes over say a straight line parallel to the imaginary axis
in the complex $\gamma$ plane, and intersects the real axis between $0$ and $\beta$.
If $t$ is a large parameter, we may try and evaluate the integral over $\gamma$ using
the saddle-point method. We recall that generically, this method
consists in the following approximation:
\be
I=\int\frac{d\gamma}{2i\pi}g(\gamma)e^{f(\gamma)t}\underset{\text{$t$ large}}{\sim}
\sum_{\gamma_s}g(\gamma_s)e^{f(\gamma_s)t},
\ee
where $\gamma_s$ represents the extrema of $f$, namely the solution(s) of
the saddle-point equation $f^\prime(\gamma_s)=0$.
In our case, the following identifications are in order:
\be
I=u(\xi,t),\ 
g(\gamma)=\frac{\beta}{\gamma(\beta-\gamma)}e^{-\gamma\xi}
\ \ \text{and}\ \
f(\gamma)=\gamma[v(\gamma)-V].
\ee
The saddle-point equation reads $(\gamma_s v(\gamma_s))^\prime=V$.
With $v$ given by Eq.~(\ref{eq:v}), the latter obviously
has a single solution.
Hence
\be
u(\xi,t)=\frac{\beta}{\gamma_s(\beta-\gamma_s)}e^{-\gamma_s\xi+\gamma_s(v(\gamma_s)-V)t},
\ee
which is an acceptable solution so long as one can move the contour
in such a way that it does not hit the singularity at $\gamma=\beta$,
so for $\beta>\gamma_s$.

\begin{figure}
\begin{center}
\includegraphics[width=.8\textwidth,angle=0]{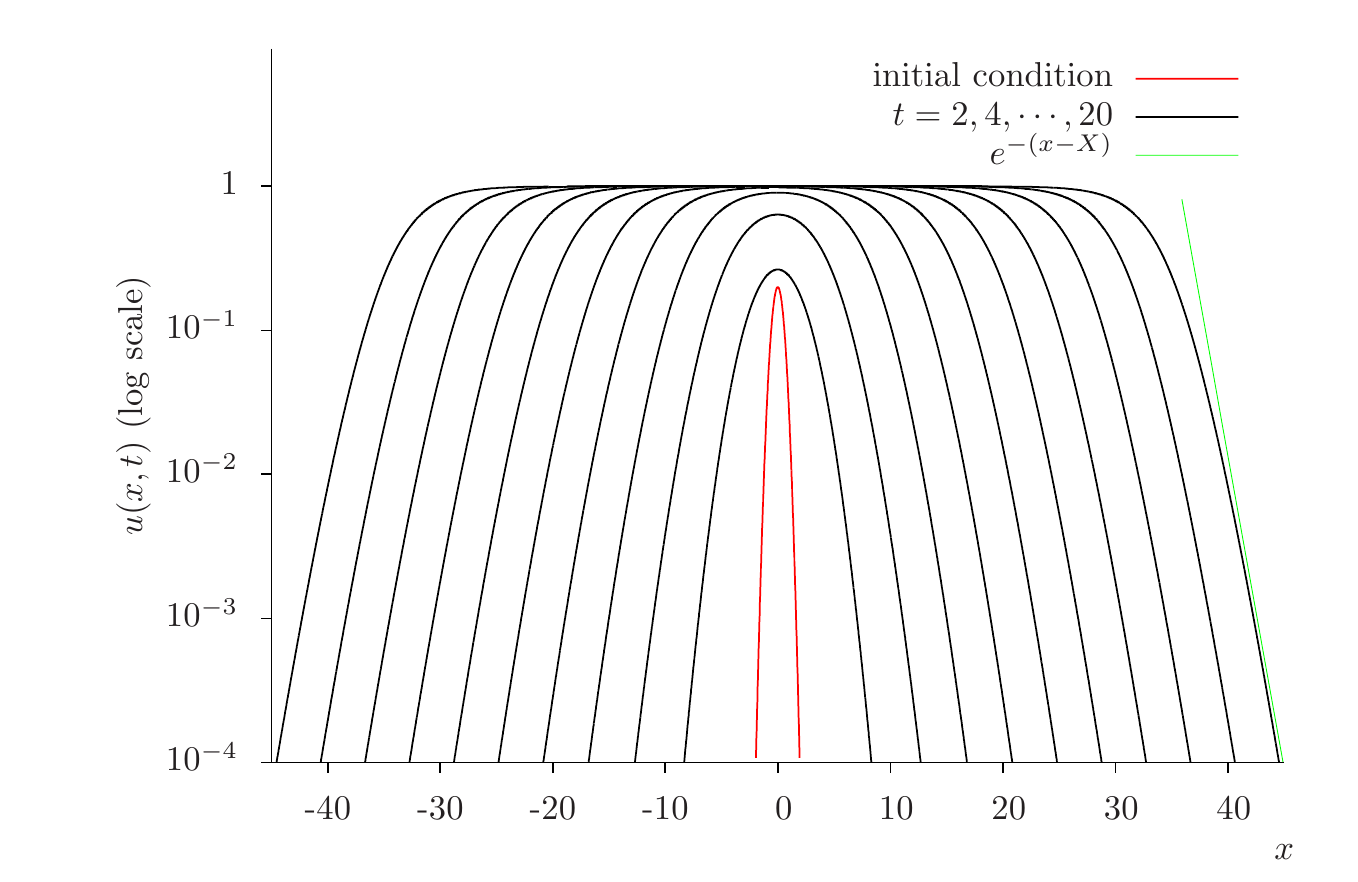}
\end{center}
\caption{\label{fig:plotFKPPlog}
The same as Fig.~\ref{fig:plotFKPP} but in logarithmic scale in the $y$-axis
in order to see the exponential shape $e^{-x}$ setting in at large $t$ and $x$,
and progressively replacing the initial Gaussian.
}
\end{figure}

Next, one chooses the velocity of the frame $V$ in such a way that the solution
be stationary. One easily sees that $V=v(\gamma_s)$. Together with the saddle-point equation,
these equations define $\gamma_s$ to be the value $\gamma_0$ of $\gamma$ which minimizes
$v(\gamma)$, i.e. $v^\prime(\gamma_0)=0$.
Hence the actual front velocity at large time is the minimum possible velocity allowed
by the dispersion relation, and the shape of the
front is $u\sim e^{-\gamma_0\xi}$. The convergence towards this shape is seen in 
Fig.~\ref{fig:plotFKPPlog}.

We mention only briefly
the case $\beta<\gamma_s$, since as we shall see later, is not of interest for QCD.
In this case, the dominant contribution to the integral is the pole at $\gamma=\beta$, and
thus
\be
u(\xi,t)=e^{-\beta\xi+\beta(v(\beta)-V)t}
\ee
which can be made stationary by setting $V=v(\beta)$.
So in this case, the front velocity at large times is the velocity of the tail of the initial
state, whose shape $u\propto e^{-\beta\xi}$ is preserved through the evolution.

From now on, we shall consider a steep enough initial condition, 
such as $Q(x,t=0)=1-u(x,t=0)=\theta(x)$, or a localized one like
Eq.~(\ref{eq:initialfkpp}).


\subsubsection{Finite-time corrections}

The full nonlinear problem is of course too difficult to solve.
Let us try and replace it by a simpler problem.

From our earlier heuristic analysis, we convinced ourselves 
that the effect of the nonlinearity is just to tame the exponential growth
of $u$ which results from the (linear) branching term in the evolution equation.
So it is natural to expect that the wave velocity be determined by the linear part
of the latter.
The easiest way to represent the effect of the nonlinearity is to replace it with a 
moving absorptive boundary set at a fixed distance of the position of the front.

We solve the linear equation in the frame moving at velocity~2: $\xi=x-2t$.
We define
\be
u(x,t)=e^{-\xi}h(\xi,t).
\ee
The linear equation~(\ref{eq:FKPPulinearized}) on $u$ translates 
into an equation for $h$.
Indeed,
\be
\begin{split}
\partial_t u&=(2h+\dot\xi \partial_\xi h+\partial_t h)e^{-\xi}\\
\partial_x u&=(-h+\partial_\xi h)e^{-\xi}\\
\partial^2_x u&=(h-2\partial_\xi h+\partial_\xi^2 h)e^{-\xi},
\end{split}
\ee
with $\dot\xi=-2$.
Therefore,
Eq.~(\ref{eq:FKPPulinearized}) reduces to
\be
\partial_t h=\partial_\xi^2 h,
\ee
which is the simple diffusion equation.

We try and put an absorptive boundary in the moving frame at $\xi=0$. 
Then according to the discussion which led to
Eq.~(\ref{eq:imagesres}),
the solution of the diffusion equation
reads
\be
h(\xi,t)\propto \frac{\xi}{t^{3/2}}e^{-\frac{\xi^2}{4t}},
\ee
namely
\be
u(x,t)\propto\xi e^{-\xi-\frac32\ln t}e^{-\frac{\xi^2}{4t}}.
\label{eq:sol_u_it1}
\ee
The lines of constant $u$
correspond to the trajectory of the front.
Thus we define the position $\xi_t$
of the front in the moving frame as
$u(x=2t+\xi_t,t)=\text{const}$.
From Eq.~(\ref{eq:sol_u_it1}), it is clear that at large $t$, $\xi_t=-\frac32\ln t+\cdots$
hence the position of the front in the original frame reads
\be
X(t)=2t+\xi_t=2t-\frac32\ln t+\cdots
\label{eq:XtFKPP}
\ee
which is Bramson's result.

\begin{figure}
\begin{center}
\includegraphics[width=.8\textwidth,angle=0]{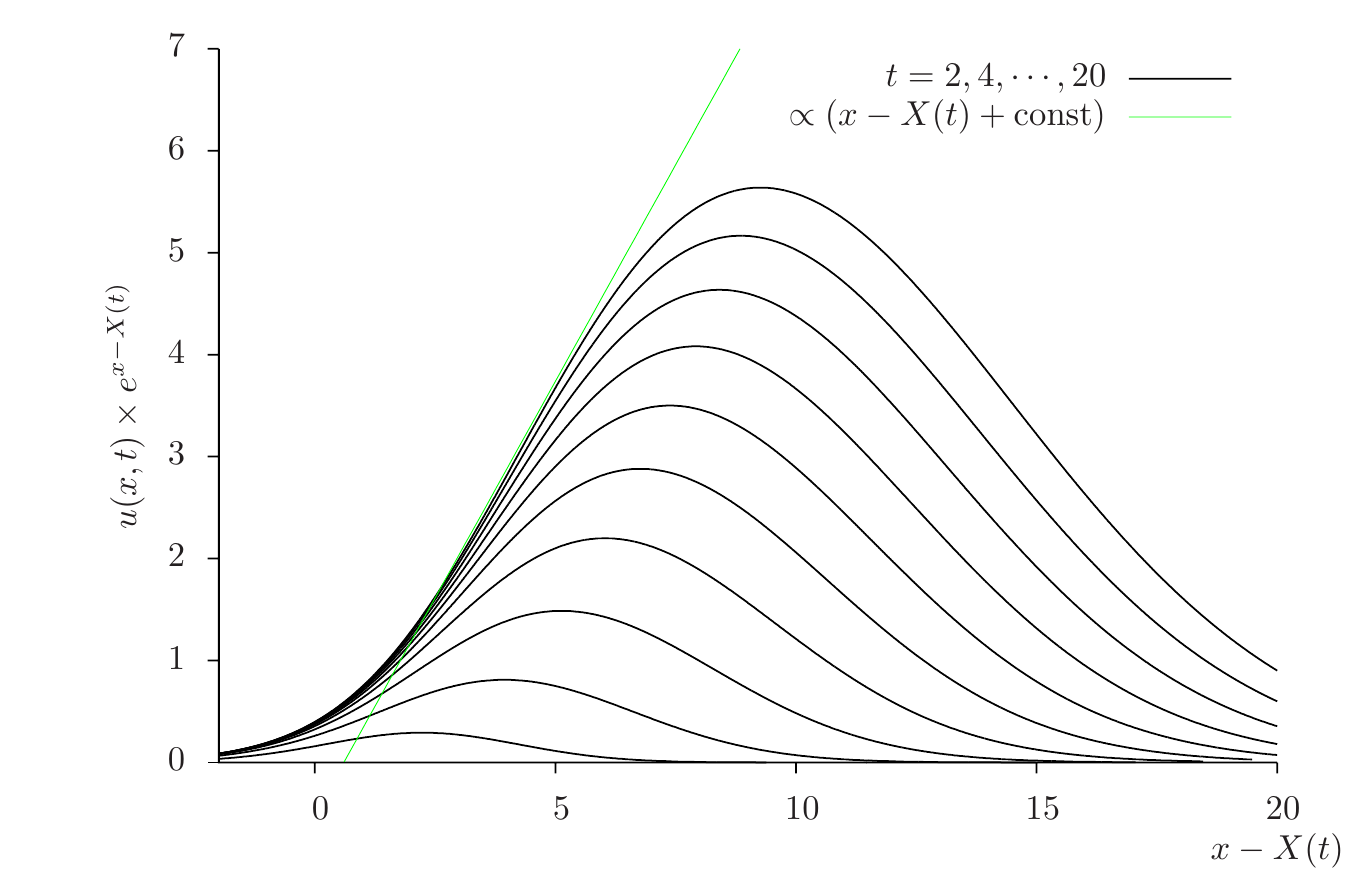}
\end{center}
\caption{\label{fig:plotFKPPrescale}
Numerical evaluation (as in Fig.~\ref{fig:plotFKPP}) 
of $u(x,t)\times e^{(x-X(t))}$. One sees the convergence to a
straight line $\text{const}\times (x-X(t)+\text{const})$ (green line).
}
\end{figure}

Now, we should adjust the position of the boundary in such a way that it matches a line of constant
$u$. We would then get for the shape of $u$:
\be
u(x,t)\propto(x-X(t)+\text{const})
e^{-(x-X(t))}
e^{-\frac{(x-X(t))^2}{4t}}.
\label{eq:shapeFKPP}
\ee
The shape is best seen if one plots $u(x,t)\times e^{x-X(t)}$ against $x$, which is sometimes
called the ``reduced front'' (see Fig.~\ref{fig:plotFKPPrescale}).
This function converges to the ``scaling'' function
$\text{const}\times (x-X(t)+\text{const})$ at large times.

The initial problem was to find the probability distribution $p(X,t)$ of the 
position $X$ of the rightmost particle. It was related to the function $Q$ satisfying 
the FKPP equation,
and thus to $u$ through
\be
p(X,t)=\frac{\partial Q(x,t)}{\partial x}|_{x=X}=-\frac{\partial u(x,t)}{\partial x}|_{x=X}.
\ee
We leave as an exercise to find
the relation between $X(t)$ computed above and
the average value of the rightmost particle in the branching random walk:
\begin{ex}
Prove that $\langle X\rangle_t$ is related to an integral of $u$:
\be
U=\int^{+\infty}dx\, u(x,t)+\text{const}.
\ee
How should the lower bound of this integral be set? How should the additive 
constant be chosen?
Relate $U$ to the expectation value of the random variable $X$.
Finally, show that, given the fact that $u$ is a traveling wave,
$U$ is a way to define the position of the front, 
namely that at large time, $U=X(t)$.
\end{ex}

Recall that the naive estimate of the average position of the rightmost particle gave
\be
X_R(t)=2t-\frac12\ln t+\text{const}
\tag{\ref{eq:XRLnaive}$'$}
\ee
(see also Fig.~(\ref{fig:plotbranchinggaussian})).
Comparing to Eq.~(\ref{eq:XtFKPP}),
we see that the leading term ($2t$) was correct, 
the form of the subleading term ($\propto \ln t$) also,
but the coefficient was not the correct one.
The nonlinearity/absorptive boundary just changed this coefficient $\frac12$ to a $\frac32$.


\subsection{\label{sec:dual}``Dual'' interpretation of the solution to the FKPP equation}

The main manifestation 
of the discreteness of the number of particles is obviously
to bring the particle density $n$ in each event to~0 to the
right of the rightmost occupied site. This fact is of course neglected in the
``mean-field'' approximation to branching diffusion
in which one replaces $n$ by its expectation value $\langle n\rangle$.

One may try to model discreteness by
an absorptive boundary on the linear equation which gives the evolution of the mean number
of particles $\langle n\rangle$.
In this case, we would solve again a diffusion equation with an absorptive boundary condition.
The result would be very similar to the one obtained in the case of the FKPP equation,
where the absorptive boundary represented the nonlinearity which forced~$u$ 
to keep less than~1.
We would get the following expression, near the right discreteness boundary:
\be
\langle n(x,t)\rangle\propto
\left(
X(t)-x
\right)
e^{-(x-X(t))-\frac{(x-X(t))^2}{4t}}
\theta(X(t)-x).
\ee

Before, we had a deterministic nonlinear equation~(\ref{eq:FKPPu}), 
which we replaced by its linearized
approximation~(\ref{eq:FKPPulinearized}) 
supplemented with an absorptive boundary.
In the present case, 
the time evolution
of the branching random walk
is a priori represented by a stochastic equation.
We replace the latter by its mean-field approximation also
supplemented with
an absorptive boundary with mimics the main origin of the noise, namely the discreteness of
the number of particles.
The two problems are very similar from the mathematical point of view,
so it is not surprising that the position of the ``discreteness boundary'',
namely of the rightmost particle in the branching random walk,
has the time dependence of the position of the FKPP traveling 
wave~(\ref{eq:XtFKPP}).

Actually, there would be a difference between the two calculations
if we went to the next order in the large-$t$ expansion, a fact
which we shall comment at the end of this chapter.


\subsection{\label{sec:gen}Generalization to other branching-diffusion processes}

We are going to extend the results just obtained in the case of the simple
branching random walk to more general branching-diffusion processes. 
In particular, we will consider models
in discrete space and/or time, which are more suitable for numerical implementation.

This generalization mainly relies on the observation that what we have done only depends on
the linear branching-diffusion kernel.

We had the equation
\be
\partial_t u=\partial_x^2 u+u.
\tag{\ref{eq:FKPPulinearized}$'$}
\ee
The eigenfunctions of the kernel were $e^{-\gamma x}$, and the corresponding eigenvalues
$\chi(\gamma)=\gamma^2+1$.
More precisely, $u(x,t)=e^{-\gamma(x-v t)}$ solves Eq.~(\ref{eq:FKPPulinearized})
provided that $v$ be related to the eigenvalues through 
$v(\gamma)=\chi(\gamma)/\gamma$.
We saw that at large times, the eigenfunction $e^{-\gamma_0 x}$ dominates
when one looks in the vicinity of a given value of $u$, where
$\gamma_0$ solves $v^\prime(\gamma_0)=0$.

All this was very general: One may replace $\chi(\gamma)$ by the eigenvalues of
any branching-diffusion kernel.
Let us go back to the solution to the linearized equation expressed in a more general
form, namely with the help of $\chi(\gamma)$:
\be
u(x,t)=\int\frac{d\gamma}{2i\pi}\tilde u(\gamma,0)
e^{-\gamma x+\chi(\gamma)t}.
\ee
Since we know that large times single out the wave number $\gamma_0$, we 
expand $\chi(\gamma)$ to second order around some $\gamma_0$, 
take the prefactor $\tilde u(\gamma,0)$ at $\gamma_0$,
and eventually go to the frame
moving at velocity $v(\gamma_0)$ by redefining $x$ as $x=\xi+v(\gamma_0)t$:
\be
\begin{split}
u(x,t)&\simeq \tilde u(\gamma_0,0)
\int\frac{d\gamma}{2i\pi}
e^{-\gamma x+[\chi(\gamma_0)+(\gamma-\gamma_0)\chi^\prime(\gamma_0)+\frac12
(\gamma-\gamma_0)^2\chi^{\prime\prime}(\gamma_0)]t}\\
&=\tilde u(\gamma_0,0)
\int\frac{d\gamma}{2i\pi}
e^{-\gamma\xi
+\left[
\chi(\gamma_0)-\gamma v(\gamma_0)+(\gamma-\gamma_0)\chi^\prime(\gamma_0)
+\frac12(\gamma-\gamma_0)^2\chi^{\prime\prime}(\gamma_0)
\right]t
}\\
&=\tilde u(\gamma_0,0)
e^{-\gamma_0\xi}
\int_{c-i\infty}^{c+i\infty}\frac{d\gamma}{2i\pi}
e^{-(\gamma-\gamma_0)\xi+\frac12(\gamma-\gamma_0)^2\chi^{\prime\prime}(\gamma_0)t}\\
&=\tilde u(\gamma_0,0)
e^{-\gamma_0\xi}
\int_{c+\gamma_0-i\infty}^{c+\gamma_0+i\infty}\frac{d\gamma}{2i\pi}
e^{-\gamma\xi+\frac12\gamma^2\chi^{\prime\prime}(\gamma_0)t}\\
&=\tilde u(\gamma_0,0)
e^{-\gamma_0\xi-\frac{\xi^2}{2\chi^{\prime\prime}(\gamma_0)t}}
\int_{\left(c+\gamma_0+\frac{\xi}{\chi^{\prime\prime}(\gamma_0)t}\right)-i\infty}
^{\left(c+\gamma_0+\frac{\xi}{\chi^{\prime\prime}(\gamma_0)t}\right)+i\infty}
\frac{d\gamma}{2i\pi}
e^{\frac12\chi^{\prime\prime}(\gamma_0)t\gamma^2}.
\end{split}
\ee
The next step is to shift the integration contour to 
make it coincide with
the imaginary axis,
and then to write $\gamma=i\nu$. The remaining integral is then just an ordinary Gaussian integral:
\be
\int_{-\infty}^{+\infty}\frac{d\nu}{2\pi}e^{-\frac12\chi^{\prime\prime}(\gamma_0)t\nu^2}
=\frac{1}{\sqrt{2\pi\chi^{\prime\prime}(\gamma_0)t}}.
\ee
The prefactor in $u$ stemming from 
this integral is proportional to $1/\sqrt{t}$ for free boundary
conditions. If we had
a fixed absorptive boundary condition instead, we would just need to replace it
by $\xi/t^{3/2}$.

We see that the procedure to find the shape and position of the front
is the same as in the case of the simple FKPP equation. 
Only a few constants differ.
The solution eventually reads
\be
u(x,t)\propto (x-X(t))e^{-\gamma_0(x-X(t))}e^{-\frac{(x-X(t))^2}{2\chi^{\prime\prime}(\gamma_0)t}},\ \
\text{where}\ \
X(t)=\chi^\prime(\gamma_0)t-\frac{3}{2\gamma_0}\ln t+\text{const}.
\label{eq:gensolFKPPu}
\ee

This formula applies to a variety of stochastic processes.
It is enough to compute the relevant eigenvalue $\chi(\gamma)$.
Let us give a few examples:

\begin{itemize}
\item
{\it Branching diffusion in continuous space and time with diffusion constant $D=1$.}
The equation which gives the time evolution of the probability 
distribution of the position of the rightmost particle
is the FKPP equation
\be
\partial_t u=\partial_x^2 u +u-u^2,
\tag{\ref{eq:FKPPu}$'$}
\ee
and we have seen that $\chi(\gamma)=\gamma^2+1$, hence $v(\gamma)=\gamma+1/\gamma$.
From the saddle-point equation, $\gamma_0=1$, $v(\gamma_0)=2$, 
$\chi^{\prime\prime}(\gamma_0)=2$. Replacing these constants in Eq.~(\ref{eq:gensolFKPPu}),
we check that we get back the results obtained earlier (compare to 
Eq.~(\ref{eq:XtFKPP}) and~(\ref{eq:shapeFKPP})).

\item
{\it Branching random walk on a lattice in space and time.}
The equivalent of the FKPP equation for this process is the following finite-difference equation:
\begin{multline}
u(x,t+\Delta t)=u(x,t)+\mu
\left[
u(x+\Delta x,t)+u(x-\Delta x,t)-2u(x,t)
\right]\\
+\lambda
u(x,t)\left[
1-u(x,t)
\right].
\end{multline}
Looking for solutions of the linearized equation
in the form $u(x,t)=e^{-\gamma(x-v(\gamma)t)}$, we find
\be
v(\gamma)=\frac{\chi(\gamma)}{\gamma}=\frac{1}{\gamma\Delta t}\ln
\left[
1+\lambda
+\mu
\left(
e^{-\gamma\Delta x}+e^{\gamma\Delta x}-2
\right)
\right].
\ee

\item {\it Population evolution model (biological context).}
Consider a population (i.e. a set of individuals). 
Each individual is characterized by a unique real
number $x$ called the ``fitness''.
We define the time evolution by the following rule: 
Each individual with fitness $x$ present in the population at ``generation'' number $t$ is replaced
at $t+1$ by two offspring, which have respective fitnesses $x_1$ and $x_2$ such that
\be
x_1=x+\varepsilon_1,\ x_2=x+\varepsilon_2,
\label{eq:mutations}
\ee
where $\varepsilon_1,\varepsilon_2$ are random numbers distributed according 
to a ``local enough'' probability
distribution $\rho(\varepsilon)$ (for example $\rho(\varepsilon)=e^{-|\varepsilon|}/2$).
Thus the population doubles at each generation, and the individuals diffuse in fitness.

\begin{ex}
Write the expression of $v(\gamma)$ in this case, as a functional of $\rho$.\\
{\em{Hint:}} Start with a population made of a single individual at $t=0$, at position $x=0$.
\end{ex}

\item Last but not least, the {\it evolution of scattering amplitudes} with the rapidity
(i.e. the logarithm of the center-of-mass energy squared)
is given by an equation established in QCD
which has a lot in common with the FKPP equation.
This is basically due to the fact that gluons may branch.
We are going to specialize to QCD in Sec.~\ref{sec:applicationsQCD}.

\end{itemize}

\begin{recap}
We have analyzed the FKPP equation~(\ref{eq:FKPPu}) and understood some properties
of the solutions. Essentially, at least for large times, the FKPP equation admits
traveling wave solutions, namely fronts which just translate in $x$ at a constant velocity.
Starting with appropriate initial conditions, the traveling wave is reached asymptotically, 
and its velocity depends on the ``steepness'' of the initial condition.
Bramson's theorem provides the expression for the velocity and its finite-time
corrections. We rederived it (Eq.~(\ref{eq:XtFKPP}))
in a heuristic approach consisting in replacing the nonlinearity by
a moving absorptive boundary, and we got also the shape of the front~(\ref{eq:shapeFKPP}).
We then explained how equivalent results may be obtained for a more general branching
random walk characterized by the kernel eigenvalues $\chi(\gamma)$:
\be
u(x,t)\propto (x-X(t))e^{-\gamma_0(x-X(t))}e^{-\frac{(x-X(t))^2}{2\chi^{\prime\prime}(\gamma_0)t}},\ \
\text{\normalfont where}\ \
X(t)=\chi^\prime(\gamma_0)t-\frac{3}{2\gamma_0}\ln t+\text{\normalfont const},
\tag{\ref{eq:gensolFKPPu}}
\ee
where $\gamma_0$ solves $\chi^\prime(\gamma_0)=\chi(\gamma_0)/\gamma_0$.
This is the main result of this section, and holds for a steep enough initial condition.
\end{recap}

\begin{beyond}
The derivation of the velocity of the traveling wave and the shape of
the front can be done a bit more rigorously, but
in the same spirit of these lectures: see Ref.~\cite{vanSaarloos200329}.
It is also possible to compute the next term in the expansion of 
the position of the FKPP front
$X(t)$  given
in Eqs.~(\ref{eq:XtFKPP}),(\ref{eq:gensolFKPPu}) 
by refining the ``moving boundary method''. One finds
\be
X(t)=\chi'(\gamma_0)t-\frac{3}{2\gamma_0}\ln t+\text{const}
-\frac{3}{\gamma_0^2}\sqrt{\frac{2\pi}{\chi^{\prime\prime}(\gamma_0)}}\frac{1}{\sqrt{t}}+\cdots
\ee
Since front shape and velocity are related, 
a corresponding correction to the shape of the front is found, see Ref.~\cite{Ebert20001}.
We refer the reader interested in the FKPP equation, its solutions and its applications
to the extensive review
given in Ref.~\cite{vanSaarloos200329}.

The position of a moving absorptive boundary 
put in the tail in such a way as
to mimic discreteness (as was explained in Sec.~\ref{sec:dual})
exhibits a similar correction, but with the opposite sign \cite{Mueller:2014gpa}:
\be
X_{R,\text{cutoff}}(t)=\chi'(\gamma_0)t-\frac{3}{2\gamma_0}\ln t+\text{const}
+\frac{3}{\gamma_0^2}\sqrt{\frac{2\pi}{\chi^{\prime\prime}(\gamma_0)}}\frac{1}{\sqrt{t}}+\cdots
\ee
The average position of the rightmost particle should
be equal to $X(t)$ since the FKPP equation describes the time
evolution of its probability distribution.
The mismatch between $X(t)$ and $X_{R,\text{cutoff}}(t)$ turns out to be exactly due
to the tip fluctuations neglected in the moving boundary mean-field model!
\end{beyond}

\clearpage


\section{\label{sec:applicationsQCD}Applications to QCD}

\begin{intro}
After our general analysis of branching random walks and of the properties
of the solutions to the FKPP equation and its avatars, we are now ready to address
the peculiar case of QCD. We shall first briefly 
recall the formulation of deep-inelastic scattering
in QCD at high energy,
then explain how branching random walks appear, and eventually
take over our knowledge of general branching random walks to QCD scattering amplitudes
in order to arrive at predictions and models which may be compared to experimental measurements.
\end{intro}

\subsection{QCD evolution at very high energies}

\begin{figure}[h]
\begin{center}
\includegraphics[width=.4\textwidth,angle=0]{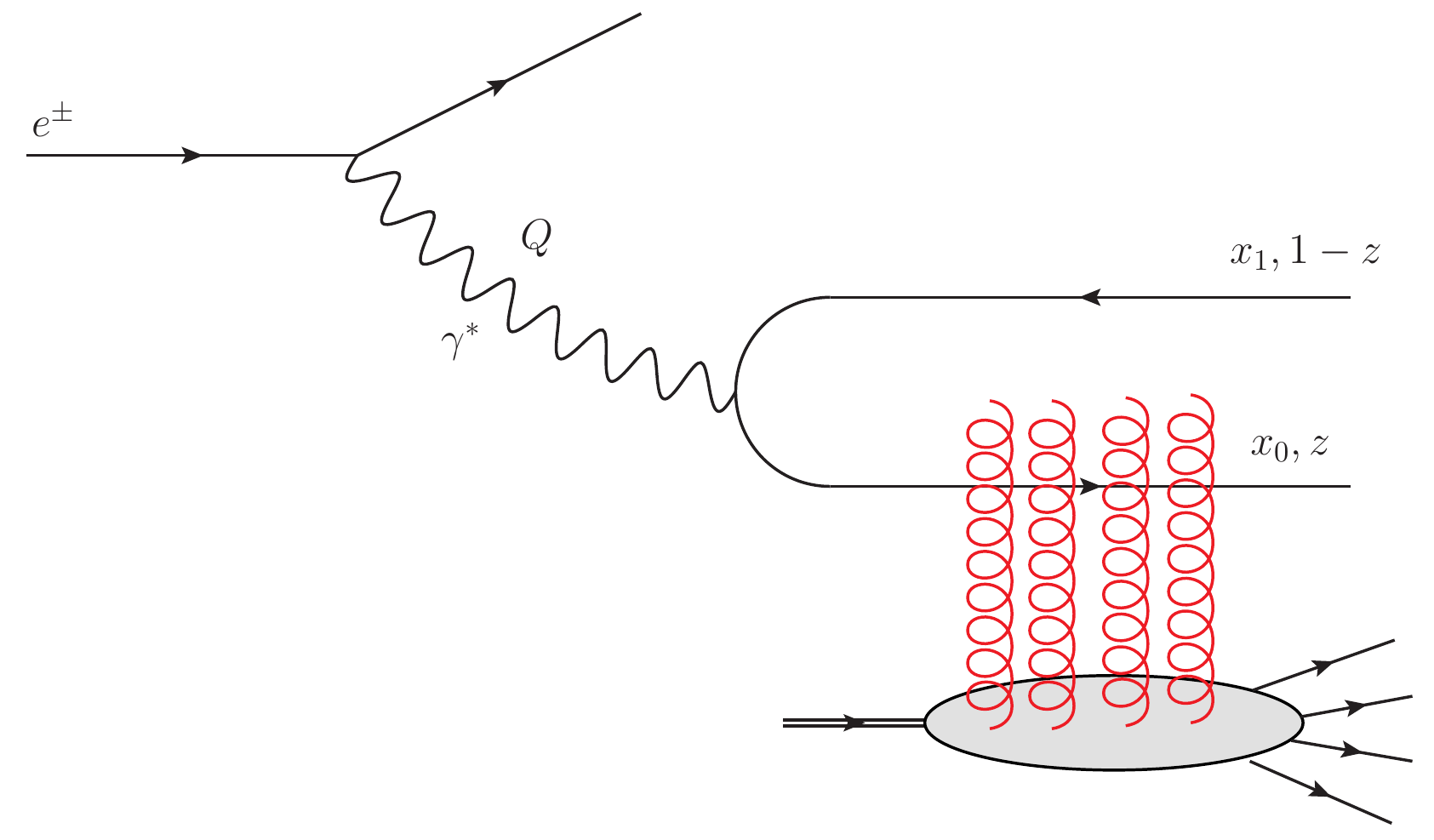}
\end{center}
\caption{\label{fig:dis}
Example of graph contributing to deep-inelastic scattering of an electron/positron
off a proton/nucleus target at high energy without quantum evolution.
The electron emits a virtual photon which interacts with the target
through a quark-antiquark fluctuation.
$x_0$ and $x_1$ are the coordinates of the
quark and of the antiquark respectively in the (two-dimensional) transverse plane, $z$ 
is the 
fraction of the momentum of the photon carried by the quark.
}
\end{figure}

We shall consider deep-inelastic scattering of an electron off some target proton
or nucleus in the dipole frame, namely the restframe of the target (see Fig.~\ref{fig:dis}).

On the target side, we know from general principles that the most probable 
states of the proton/nucleus 
at very high energies are dense states of gluons.
On the electron side, the electron can be seen as a Weizs\"aker-Williams cloud of 
virtual photons of virtuality say $Q^2=-q^2$, where $q$ is the four-momentum of the photon. 
Since the latter cannot interact directly
with gluons, it splits into a quark-antiquark pair. This pair, being globally color neutral, is a
color dipole. At lowest order in the coupling constants, 
the photon-target cross section reads
\be
\sigma^{\gamma^* p/A}(Q^2,y)=\int d^2{x_{01}} dz
|\psi^Q(x_{01},z)|^2 \int d^2b\, 2\,\text{Re}(1-S(x_{01},b,y))
\label{eq:sigmaDIS}
\ee
where $S(x_{01},b,y)$ is the $S$-matrix element for the elastic scattering of a dipole
of size $x_{01}\equiv x_0-x_1$ (a two-dimensional vector) 
at rapidity $y$ 
and impact parameter $b$ off the target proton/nucleus.
The total cross section is obtained from the optical theorem, which explains the presence
of the ``real part'' operator.
In our further discussion, we will drop the impact-parameter dependence almost throughout.

\begin{figure}[h]
\begin{center}
\includegraphics[width=.5\textwidth,angle=0]{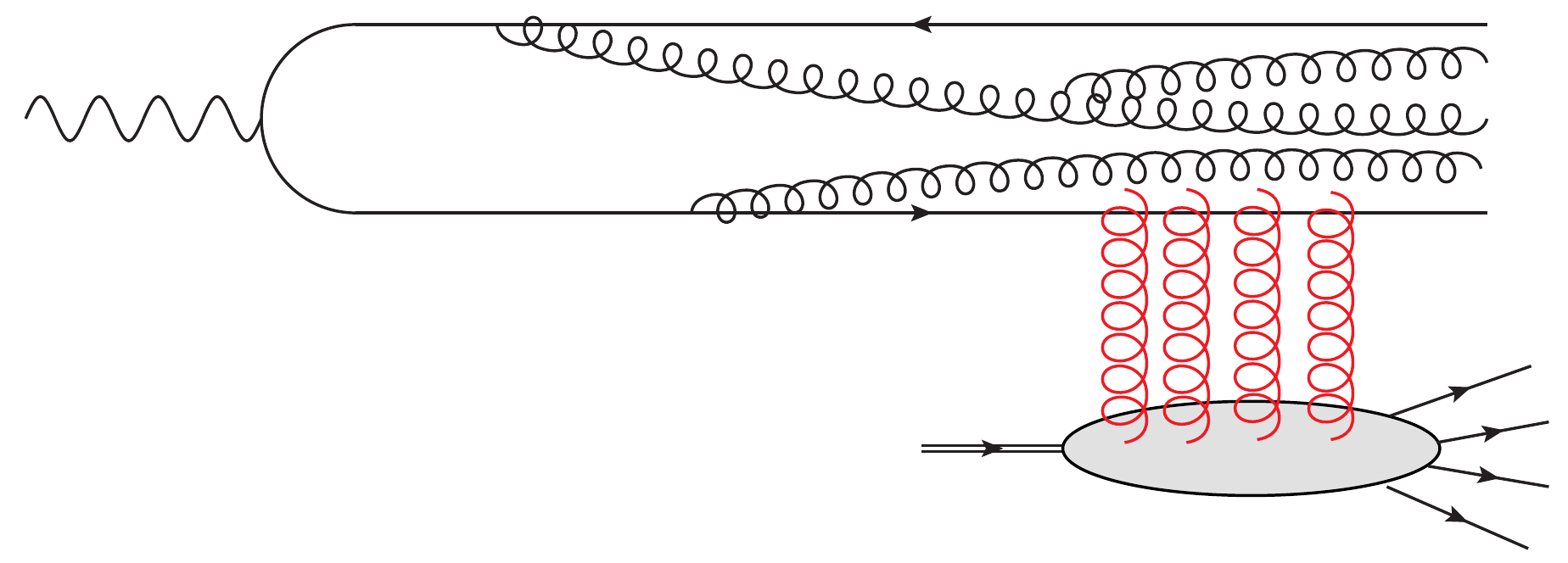}
\end{center}
\caption{\label{fig:dis2}
Example of graph contributing to the quantum evolution
of the deep-inelastic scattering process at very high energy.
}
\end{figure}

Now in quantum field theory, the states which actually interact 
are fluctuations of the initial objects (dipole or proton in their fundamental states;
see Fig.~\ref{fig:dis2}).
So we need to compute the probability distribution of the different
states (resulting from further fluctuations)
at the time of the interaction.

At high energies, as already mentioned, the dominant fluctuations are dense states of
soft gluons. To compute their probabilities at leading order when the coupling constant
is small and the rapidity large, it is enough to consider the successive emissions of
softer and softer gluons: Eventually, it is the process 
$\text{gluon}\rightarrow \text{gluon}+\text{gluon}$
which gives the main contribution.
Already at this stage, we see that this is a branching process, whose realizations are
``trees'' of gluons. There is also diffusion since the gluons which result from
the branching do not have the same (transverse) momentum as the parent gluon.
We need to perform a calculation in the framework of QCD
to make this statement precise and useful in practice.

\begin{figure}[h]
\begin{center}
\begin{tabular}{m{0.2\textwidth}m{2em}m{0.2\textwidth}}
\includegraphics[width=.2\textwidth,angle=0]{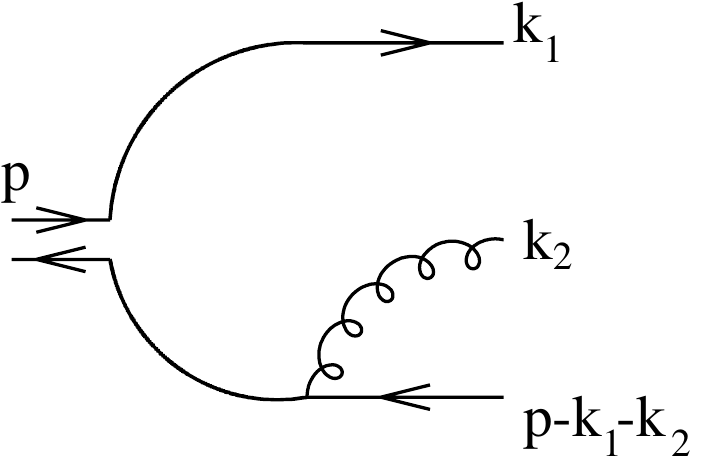}
&$+$&\includegraphics[width=.2\textwidth,angle=0]{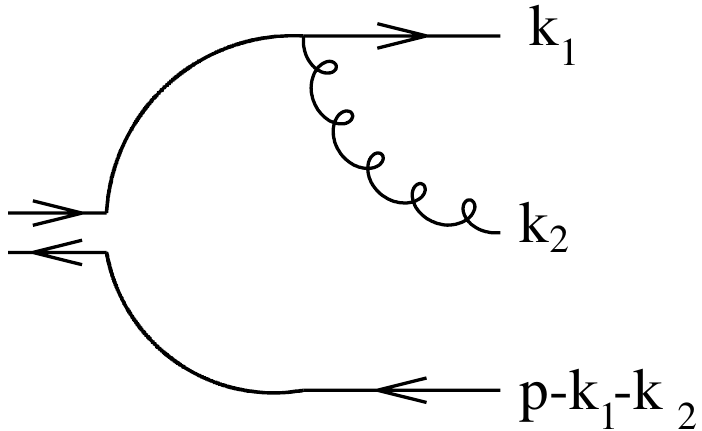}\\
\end{tabular}
\end{center}
\caption{\label{fig:dipoles}
Lightcone perturbation theory graphs 
contributing to the amplitude for the emission of a gluon by a quark-antiquark
pair.
}
\end{figure}

We start with one quark-antiquark color neutral pair (i.e. one color dipole), 
at rapidity 0: In DIS, this is the $q\bar q$ component of the
photon wave function.
The lowest-order fluctuation is a $q\bar q+\text{gluon}$ state.
We shall compute first
the probability to observe such a state at the time of the interaction
starting at lightcone time $\tau=-\infty$ with a bare $q\bar q$ pair.
Let us denote the quark momentum by $k_1$, and the antiquark momentum by
$p-k_1$.
The two graphs contributing to the probability
amplitude are represented in Fig.~\ref{fig:dipoles}.
The emitted gluon possesses the momentum $k_2$.

It turns out that switching from transverse momentum to transverse
coordinates simplifies a lot the problem. Physically, this stems
from the fact that high energies, the coordinates of a fast particle
are not altered by the interaction nor by the radiation of a soft gluon.
Hence we shall go to two-dimensional Fourier space, and introduce
the coordinates $x_0$ for the quark, $x_1$ for the antiquark, $x_2$ for the gluon.

After taking the modulus squared of the graphs shown in Fig.~\ref{fig:dipoles}
and summing over the polarization and the color
of the emitted gluon, the result reads \cite{Mueller:1993rr}
\be
dP(x_{01}\rightarrow x_{02},x_{12})
=\frac{2\alpha_s C_F}{\pi} dy
\frac{d^2 x_2}{2\pi}
\frac{x_{01}^2}{x_{02}^2 x_{12}^2},
\label{eq:dipolesplittingprob}
\ee
where $y$ is the rapidity of the gluon, $y=\ln k_{2+}$ and thus 
$dy=dk_{2+}/{k_{2+}}$.

Let us comment on this formula.
First, the emission probability is of course proportional to the strong
coupling constant $\alpha_s$, and to the $SU(N_c)$ Casimir of the
fundamental representation $C_F$ since we have summed over
the colors of the emitted gluon.
The differential element comes
from the phase space of the gluon.
The probability exhibits the two types of singularities present in gauge
theories: the soft singularity, which gives a logarithmic divergence to
the probability integrated over the ``+'' component of the gluon momentum $k_{2+}$,
and the collinear singularity in the last factor, which enhances the weight of the
configurations in which
the gluon is emitted collinearly either to the quark or to the antiquark.

\begin{figure}[h]
\begin{center}
\begin{tabular}{m{.14\textwidth}m{.01\textwidth}m{.14\textwidth}
m{.01\textwidth}m{.14\textwidth}m{.01\textwidth}m{.14\textwidth}
m{.01\textwidth}m{.14\textwidth}}
\includegraphics[width=.14\textwidth,angle=0]{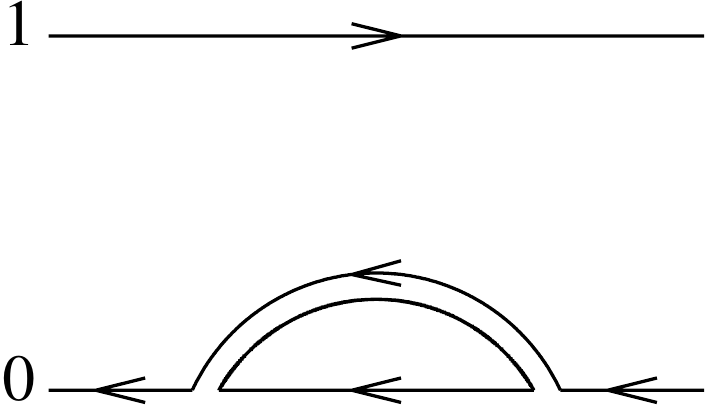}&$+$
&\includegraphics[width=.14\textwidth,angle=0]{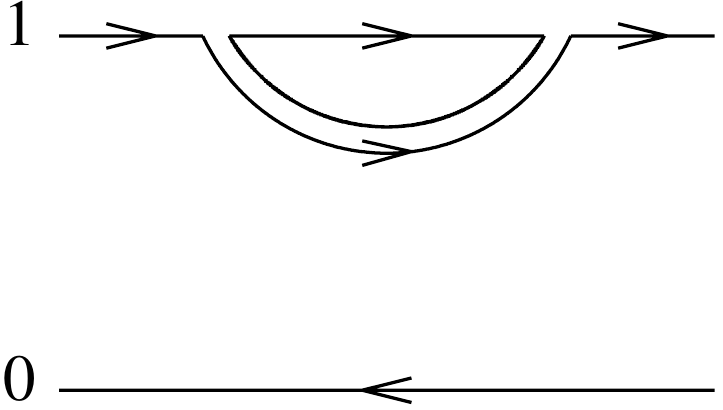}&$+$
&\includegraphics[width=.14\textwidth,angle=0]{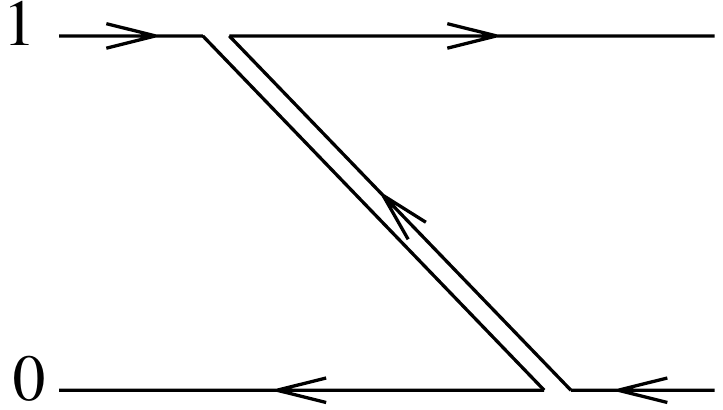}&$+$
&\includegraphics[width=.14\textwidth,angle=0]{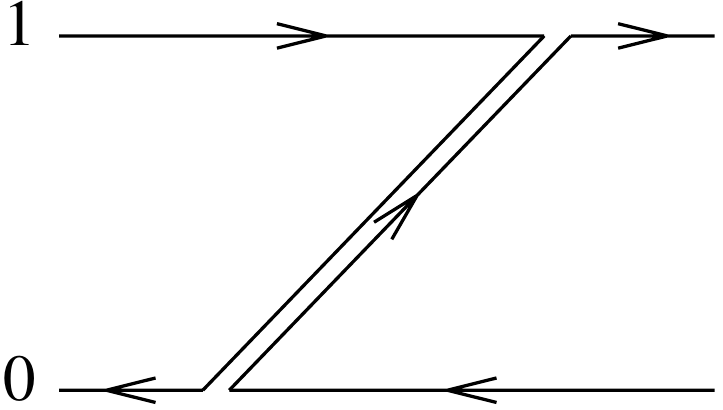}&$\equiv$
&\includegraphics[width=.14\textwidth,angle=0]{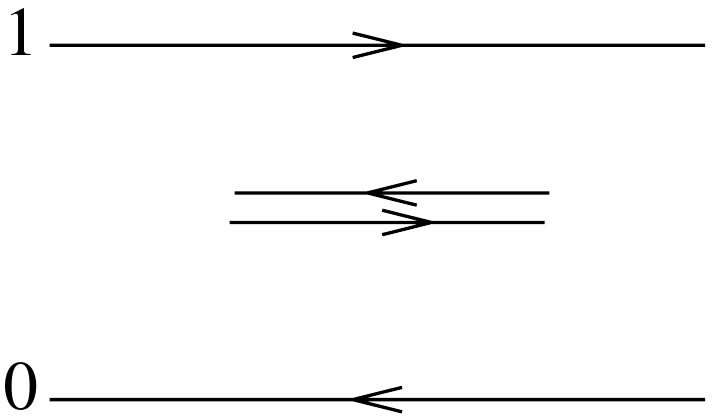}
\end{tabular}
\end{center}
\caption{\label{fig:grandnc}
Graphs contributing to the modulus squared of the amplitude for the emission of 
a gluon by a quark-antiquark pair in the large-$N_c$ limit.
The rightmost graph is a convenient representation for the sum of all possible
graphs which result in the dipole splitting probability~(\ref{eq:dipolesplittingprob}).
}
\end{figure}

It is convenient to go to the large-number-of-color limit (see Fig.~\ref{fig:grandnc}), 
since this limit suppresses
the planar diagrams and enables one to interpret gluons as zero-size quark-antiquark pairs.
(Moreover, in this limit, $C_F\rightarrow N_c/2$).
Under these simplifying (but systematic) assumptions, one obtains the color-dipole model
\cite{Mueller:1993rr}.

Indeed, the large-$N_c$ limit enables one to interpret the emission of the gluon as
the splitting of the initial dipole into two new dipoles. So $dP/dy$
can be interpreted as the rate at which a dipole of size $x_{01}$ splits
to two dipoles of respective sizes $x_{02}$, $x_{12}$ when the rapidity
is increased by the small quantity $dy$.

\begin{figure}[h]
\begin{center}
\begin{tabular}{m{.16\textwidth} 
>{\centering\arraybackslash}m{1cm}
>{\centering\arraybackslash} 
m{.16\textwidth}@{\hskip 0.5cm}|@{\hskip 0.5cm}
m{.16\textwidth} 
>{\centering\arraybackslash}m{1cm}
>{\centering\arraybackslash} 
m{.16\textwidth}
}
\includegraphics[width=.16\textwidth,angle=0]{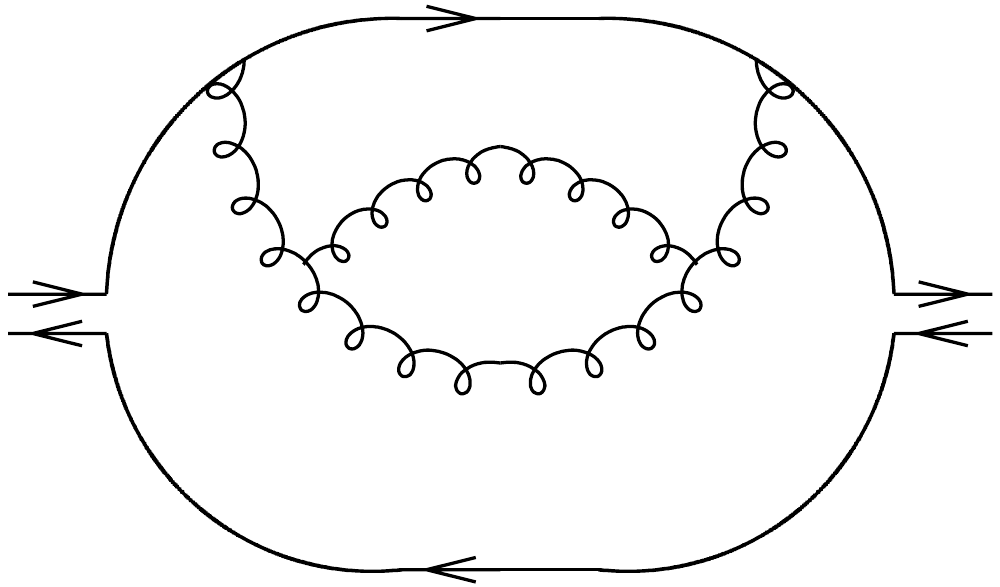}
&$\underset{\text{large $N_c$}}{\simeq}$&
\includegraphics[width=.16\textwidth,angle=0]{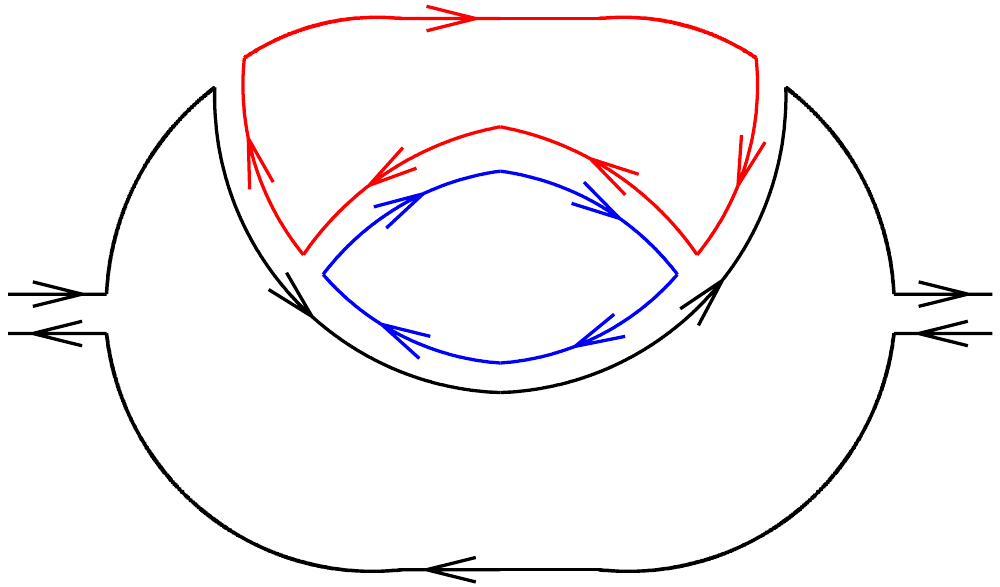}
& \includegraphics[width=.16\textwidth,angle=0]{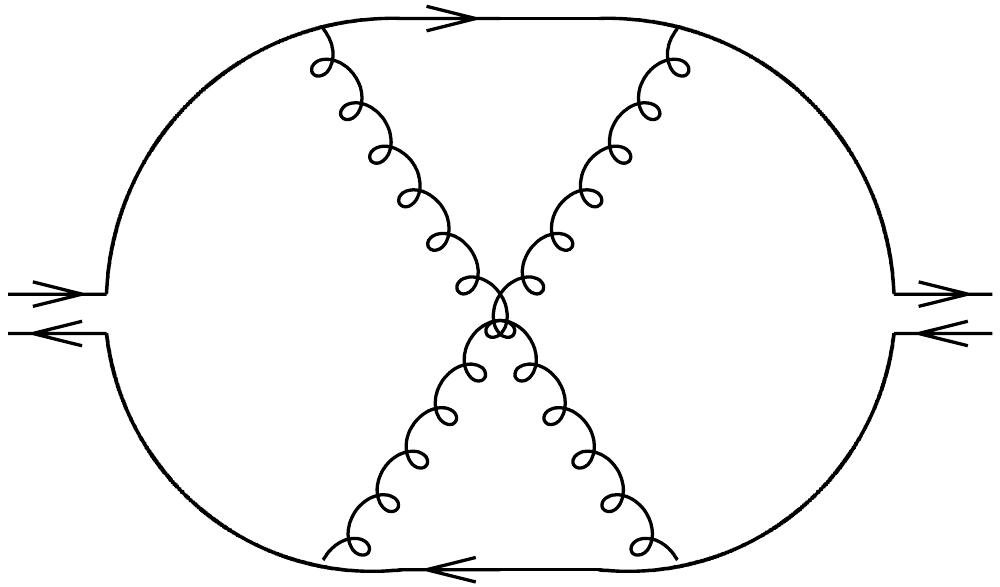}
&$\underset{\text{large $N_c$}}{\simeq}$&
\includegraphics[width=.16\textwidth,angle=0]{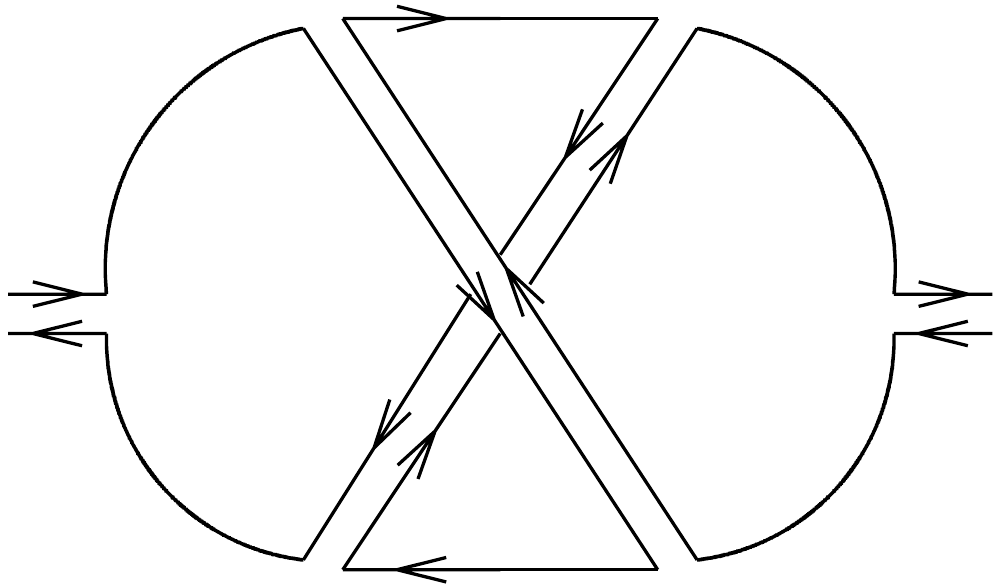}\\
\multicolumn{3}{c}{(a)} &
\multicolumn{3}{c}{(b)}
\end{tabular}
\end{center}
\caption{\label{fig:dipolesnot}
Example of a planar graph which contributes (a) and a nonplanar graph which does 
not contribute (b)
to the modulus squared of the amplitude for the emission of 
a gluon by a quark-antiquark pair in the large-$N_c$ limit.
}
\end{figure}

One may then iterate this process (see Fig.~\ref{fig:dipolesnot}): 
Thanks again to the large-$N_c$
limit in which nonplanar graphs are subdominant, the two dipoles, once emitted, evolve
independently of  each other. Upon rapidity evolution, we get a cascade of dipoles
through dipole branching.

We are now going to establish the Balitsky-Kovchegov (BK) equation in the
same way as we established the FKPP equation.


\subsection{QCD evolution as a branching random walk}

Let us compute the $S$-matrix element $S(x_{01},y)$ 
for the elastic scattering of a dipole
of size $x_{01}$ after an evolution over $y$ units of rapidity.

$S(x_{01},y)$ is the probability amplitude 
(in the sense that $S^2$ is the actual probability)
that there be no interaction between
the initial dipole of size $x_{01}$ and the target, after an evolution over $y$ units
of rapidity. At the time of the interaction, the projectile dipole has been replaced
by a random collection of $N(y)$ dipoles of sizes $\{r_i(y)\}$.
Since these dipoles are assumed independent, the probability amplitude that in a
particular configuration
none of them interact is the product of the $S$-matrix elements (at zero rapidity)
of each of them. The average over the different dipole configurations has eventually
to be taken. Hence we write
\be
S(x_{01},y)=\left\langle
\prod_{i=1}^{N(y)}S(r_i(y),y=0)
\right\rangle.
\label{eq:Stargetrest}
\ee
$S(r,y=0)$, which we may also denote by $S_\text{el}(r)$, 
represents the elementary interaction of a dipole of size $r$ with the target,
without any quantum evolution.

Now it is useful to recall that we proved that
\begin{equation}
F(x,t)\equiv\left\langle\prod_{i=1}^{N(t)}f(x-x_i(t))\right\rangle
\tag{\ref{eq:F}$'$}
\end{equation}
obeys the FKPP equation for any function $f$ when $\{x_i(t)\}$ is the set of the
positions of the $N(t)$ particles generated after a branching random walk starting
with a single particle at $x=0$ and running
over $t$ units of time.
The similarity between the last
two equations, Eq.~(\ref{eq:F}) and~(\ref{eq:Stargetrest}) is obvious:
It is enough to identify the functions $S$ to $F$, $S_\text{el}$ to $f$, the variables
$y$ to $t$, and as we will discover later on, $\ln x_{01}^2/r_i^2$ to $x_i$.

However, dipole splitting is not exactly the simple branching random walk which is
at the origin of the FKPP equation. It is a more sophisticated stochastic process,
and therefore, we shall establish the equivalent of the FKPP equation from scratch.

In order to establish such evolution laws, we try and express $S$ at rapidity
$y+dy$ as a function of $S$ at rapidity $y$. To do this, we consider what happens in
the small rapidity interval $[0,dy]$:
Either the dipole does not split, in which case, for this
particular event, the $S$-matrix element
would be given by $S_{\text{event}}(x_{01},y+dy)=S(x_{01},y)$, 
or it splits into two dipoles of respective sizes $x_{02}$ and $x_{12}$,
in which case $S_\text{event}(x_{01},y+dy)=S(x_{02},y)\times S(x_{12},y)$.
The fundamental assumption here is the complete 
independence of the evolution of the dipoles, leading to the latter factorization.
$S(x_{01},y+dy)$ is obtained by averaging $S_\text{event}(x_{01},y+dy)$ over
these two possible cases. We arrive at a sum of $S$ at rapidity $y$ weighted by
the dipole splitting probability in Eq.~(\ref{eq:dipolesplittingprob}):
\begin{multline}
S(x_{01},y+dy)=
\left\langle
S_\text{event}(x_{01},y+dy)
\right\rangle
=S(x_{01},y)\left[
1-dy\int \frac{dP}{dy}(x_{01}\rightarrow x_{02},x_{12})
\right]\\
+dy\int \frac{dP}{dy}(x_{01}\rightarrow x_{02},x_{12})
S(x_{02},y)S(x_{12},y).
\end{multline}
Letting $dy\rightarrow 0$, we obtain the following integro-differential equation:
\be
\partial_y S(x_{01},y)=\bar\alpha\int\frac{d^2 x_2}{2\pi}\frac{x_{01}^2}{x_{02}^2x_{12}^2}
\left[
S(x_{02},y)S(x_{12},y)-S(x_{01},y)
\right].
\label{eq:BKS}
\ee
This is the Balitsky-Kovchegov (BK) equation.
Introducing the scattering amplitude $T=1-S$, we get
\be
\partial_y T(x_{01},y)=\bar\alpha\int\frac{d^2 x_2}{2\pi}\frac{x_{01}^2}{x_{02}^2x_{12}^2}
\left[
T(x_{02},y)+T(x_{12},y)-T(x_{01},y)-T(x_{02},y)T(x_{12},y)
\right].
\label{eq:BKT}
\ee
Let us comment on this equation.
It is clear that the nonlinear term, of the form $-TT$, is important only when $T\sim 1$,
i.e. when typically more than one dipole interacts with the target (since in this case
the probability that there be no interaction $S^2=(1-T)^2$ tends to 0).
The BK equation boils then down to
\be
\partial_y T(x_{01},y)=\bar\alpha\int\frac{d^2 x_2}{2\pi}\frac{x_{01}^2}{x_{02}^2x_{12}^2}
\left[
T(x_{02},y)+T(x_{12},y)-T(x_{01},y)
\right],
\label{eq:BFKL}
\ee
which is nothing but the Balitsky-Fadin-Kuraev-Lipatov (BFKL) 
equation written in coordinate space.

In the following, we shall claim that the BK equation~(\ref{eq:BKT})
is {\it in the universality class} of the FKPP equation~(\ref{eq:FKPPu}).
To make this more precise, we shall first put the BK equation (or rather,
an appropriate approximation of it) in a form in which it is manifestly
equivalent to the FKPP equation.


\subsection{Mapping the Balitsky-Kovchegov equation to the FKPP equation }

Let us first analyze the linear limit of the BK equation for $S\rightarrow 1$, 
which is the BFKL equation.
To this aim, we need to find the eigenfunctions and the corresponding
eigenvalues of the BFKL equation.

\subsubsection{Calculation of the eigenvalues of the BFKL kernel}

We will need a few formulae of complex analysis.
First, the Euler gamma function is defined as
\be
\Gamma(x)=\int_0^{+\infty}dt\, e^{-t} t^{x-1},
\label{eq:defgamma}
\ee
whose main property is $x\Gamma(x)=\Gamma(1+x)$.
We will also need the Taylor expansion around $\varepsilon=0$ of the ratio
\be
\frac{\Gamma(a+n\varepsilon)}{\Gamma(a+m\varepsilon)}
=1+(n-m)\varepsilon \psi(a)+o\left(\varepsilon^2\right),
\label{eq:expansionGamma}
\ee
where $\psi(x)\equiv\Gamma^\prime(x)/\Gamma(x)$ and $n$ and $m$ are
two finite numbers.
The Euler Beta function is a combination of $\Gamma$ functions
and is the result of the following integration (see Appendix~\ref{sec:compint}):
\be
\int_0^1 dx\,x^{\alpha-1}(1-x)^{\beta-1}=
\frac{\Gamma(\alpha)\Gamma(\beta)}{\Gamma(\alpha+\beta)}\equiv
B(\alpha,\beta).
\ee
The main steps of our calculation will rely on
a similar-looking formula, but where the
integration extends over the whole complex plane:
\be
\int\frac{dz d\bar z}{2i\pi}|z|^{2\alpha-2}|1-z|^{2\beta-2}
=\frac{\Gamma(\alpha)\Gamma(\beta)}{\Gamma(\alpha+\beta)}
\frac{\Gamma(1-\alpha-\beta)}{\Gamma(1-\alpha)\Gamma(1-\beta)}.
\label{eq:conformal}
\ee
This formula is classical in the context of 2-dimensional conformal field
theory. For completeness, we propose a derivation in Appendix~\ref{sec:compint}.

The action of the BFKL kernel on a function $T$ of the 2-dimensional
vector $x_{01}$ reads
\be
K\otimes T(x_{01})=\bar\alpha \int\frac{d^2 x_2}{2\pi}
\frac{x_{01}^2}{x_{02}^2 x_{12}^2}\left[
T(x_{02})+T(x_{21})-T(x_{01})
\right].
\label{eq:BFKLkernel}
\ee
Note that the first two terms give identical contributions.
We restric ourselves to azimuthally symmetric solutions.
It is then natural to look for eigenfunctions of the form
\be
T(x)=|x|^{2\gamma}.
\label{eq:eigenpower}
\ee
(A general solution will be a linear combination of these power functions).
We insert Eq.~(\ref{eq:eigenpower}) into Eq.~(\ref{eq:BFKLkernel}), and go to complex
variables by defining
\be
z\equiv\frac{x_{02}^{(1)}+ix_{02}^{(2)}}{x_{01}^{(1)}+ix_{01}^{(2)}}
\ee
where the superscripts $(1)$ and $(2)$ label the components
of the vector.
Then
\be
|z|^2=\frac{x_{02}^2}{x_{01}^2},\ \
|1-z|^2=\frac{x_{21}^2}{x_{01}^2},\ \
d^2 x_2=\frac{dz d\bar z}{2i}|x_{01}|^2,
\ee
and the action of the kernel on the test functions~(\ref{eq:eigenpower})
reads
\be
K\otimes T(x_{01})=T(x_{01})\times\bar\alpha
\chi(\gamma),
\ee
where $\chi(\gamma)$ is defined by the following integral:
\be
\chi(\gamma)\equiv 
\int\frac{dzd\bar z}{4i\pi}
\frac{1}{|z|^2|1-z|^2}\left(|z|^{2\gamma}+|1-z|^{2\gamma}-1\right).
\ee
One easily sees that the power functions are indead eigenfunctions,
with eigenvalues $\bar\alpha\chi(\gamma)$.

Let us discuss the convergence of the integral defining $\chi(\gamma)$.
All terms converge at $z=+\infty$ when $0<\text{Re}(\gamma)<1$.
The first term converges also at $z=0$
but diverges at $z=1$.
As for the second term, it converges $z=1$ but diverges at
$z=0$. The last term diverges both at $z=0$ and $z=1$.
Hence one needs to regularize these integrals: We choose to slightly modify
the powers of the factors in the kernel: 
\be
\chi(\gamma)=\lim_{\varepsilon\rightarrow 0} 
\int\frac{dzd\bar z}{4i\pi}
\frac{1}{|z|^{2-2\varepsilon}|1-z|^{2-2\varepsilon}}\left(|z|^{2\gamma}+|1-z|^{2\gamma}-1\right)
\equiv I_\gamma-\frac12 I_0,
\ee
where
\be
I_\gamma\equiv\int\frac{dz d\bar z}{2i\pi}|z|^{2(\varepsilon+\gamma)-2}|1-z|^{2\varepsilon-2}.
\ee
It is straightforward to apply Eq.~(\ref{eq:conformal}) to $I_\gamma$, with
$\alpha=\varepsilon+\gamma$ and $\beta=\varepsilon$,
\be
\begin{split}
I_\gamma&=\frac{\Gamma(\varepsilon+\gamma)\Gamma(\varepsilon)}
{\Gamma(2\varepsilon+\gamma)}
\frac{\Gamma(1-2\varepsilon-\gamma)}
{\Gamma(1-\varepsilon-\gamma)\Gamma(1-\varepsilon)}\\
&=\frac{1}{\varepsilon}
\frac{\Gamma(1+\varepsilon)}{\Gamma(1-\varepsilon)}
\frac{\Gamma(\gamma+\varepsilon)}{\Gamma(\gamma+2\varepsilon)}
\frac{\Gamma(1-\gamma-2\varepsilon)}{\Gamma(1-\gamma-\varepsilon)}\\
&=\frac{1}{\varepsilon}
\left[
1+\varepsilon\left(2\psi(1)-\psi(\gamma)-\psi(1-\gamma)\right)
\right]
+{\cal O}(\varepsilon).
\end{split}
\label{eq:Igamma}
\ee
(Going from the first line to the second one makes use of the elementary identity
$\Gamma(1+x)=x\Gamma(x)$, while the expansion for small $\varepsilon$ is
obtained from Eq.~(\ref{eq:expansionGamma})).
As expected, Eq.~(\ref{eq:Igamma}) diverges  when $\varepsilon\rightarrow 0$.
The calculation of $I_0$ goes along the same lines. After expanding for
small $\varepsilon$, we find
\be
I_0=\frac{2}{\varepsilon}+{\cal O}(\varepsilon).
\ee
In the difference $I_\gamma-\frac12 I_0$, the divergence cancels
(This is expected, because $I_0$ actually corresponds to the renormalization of
the dipole wavefunction).
A finite term remains, which reads
\be
I_\gamma-\frac12 I_0=2\psi(1)-\psi(\gamma)-\psi(1-\gamma)=\chi(\gamma).
\ee
Hence the eigenfunctions of the BFKL kernel are the powers $|x_{01}|^{2\gamma}$,
and the corresponding eigenvalues are $\bar\alpha\chi(\gamma)$.

\begin{figure}
\begin{center}
\includegraphics[width=.8\textwidth,angle=0]{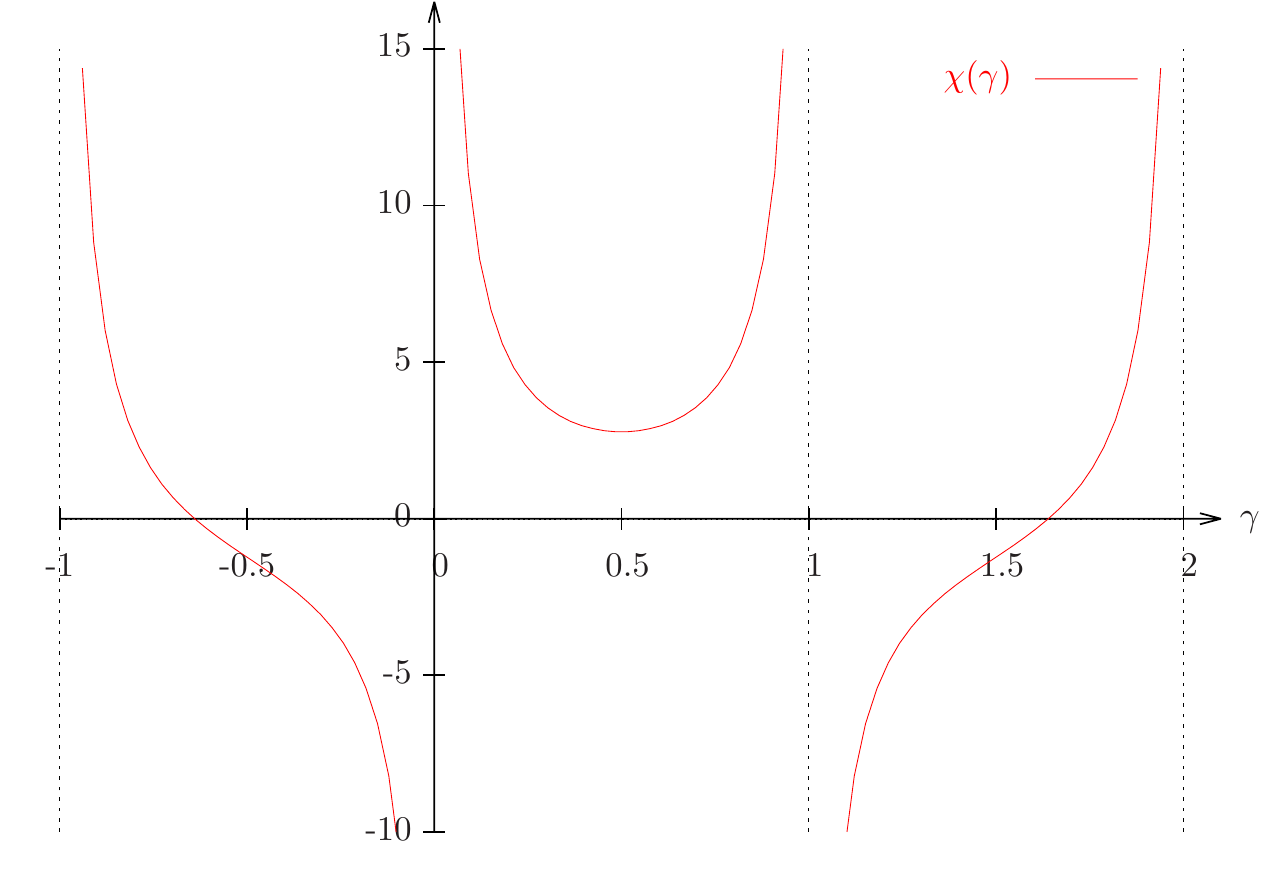}
\end{center}
\caption{\label{fig:plotchi}
The function $\chi(\gamma)$ to which the eigenvalues of the BFKL kernel are
proportional as a function of $\gamma$ for real $\gamma\in[-1,2]$.
There are simple poles at all integer values of $\gamma$.
}
\end{figure}
The structure of the function $\chi$ is shown in Fig.~\ref{fig:plotchi}.
It has poles for all integer values of $\gamma$.
The branch which gives the main contribution to the solution
of the BFKL equation
is $\gamma\in]0,1[$.
In the complex plane, $\chi$ has actually a saddle-point at $\gamma=\frac12$,
around which the solution may be expanded for large rapidities.

\subsubsection{Compact expression for the BFKL and BK equations}

We now aim at writing the BFKL and eventually the BK equations in a more compact
way.
We use the fact that a linear operator $K_x$ acting on some function $u(x)$
may be represented by its eigenvalues. Indeed,
\be
K_x\cdot u(x)=K_x\cdot\int\frac{d\gamma}{2i\pi}e^{\gamma x}\tilde u(\gamma)
=\int\frac{d\gamma}{2i\pi}\tilde u(\gamma)K_x\cdot e^{\gamma x}
\ee
and if $\chi(\gamma)$ is the eigenvalue of $K_x$ which corresponds to the eigenfunction
$e^{\gamma x}$, we arrive at a formal expression for the action of $K_x$ on the function $u(x)$:
\be
K_x\cdot u(x)=\int\frac{d\gamma}{2i\pi}\tilde u(\gamma)\chi(\gamma)\cdot e^{\gamma x}
=\chi(\partial_x)\int\frac{d\gamma}{2i\pi}\tilde u(\gamma)\cdot e^{\gamma x}
=\chi(\partial_x)u(x).
\ee
Applying this procedure to the BFKL equation,
\be
\partial_y T(x_{01},y)=\bar\alpha\chi\left(\partial_{\ln |x_{01}|^2}\right)T(x_{01},y).
\ee
To further analyze the BFKL and BK equations, it proves simpler to go to momentum space by
defining
\be
\tilde T(k,y)=\int \frac{d^2 x_{01}}{2\pi x_{01}^2}e^{i k x_{01}}T(x_{01},y).
\ee
The form of the BFKL equation remains essentially unchanged since the Fourier transform
is just a change of basis:
\be
\partial_y \tilde T(k,y)=
\bar\alpha\chi\left(-\partial_{\ln |k|^2}\right)\tilde T(k,y).
\ee
The nonlinear term turns out to drastically simplify in momentum space.
Its Fourier transform reads
\be
\int\frac{d^2 x_{01}}{2\pi x_{01}^2}e^{i x_{01}k}
\left[
\bar\alpha
\int\frac{d^2 x_2}{2\pi}\frac{x_{01}^2}{x_{02}^2 x_{12}^2}
T(x_{02},y)T(x_{12},y)
\right].
\ee
One can perform the change of variables in the integrals
$(x_{01},x_2)\rightarrow (x_{21},x_{02})$ to get
\begin{multline}
\bar\alpha
\int\frac{d^2 x_{21}}{2\pi x_{12}^2}
\frac{d^2 x_{02}}{2\pi x_{02}^2}
e^{i k(x_{02}+x_{21})}
 T(x_{02},y)T(x_{12},y)\\
=\bar\alpha\int\frac{d^2 x_{21}}{2\pi x_{12}^2}e^{i kx_{21}}T(x_{12},y)
\int\frac{d^2 x_{20}}{2\pi x_{02}^2}e^{i kx_{02}}T(x_{02},y).
\end{multline}
The two factors are just equal to $\tilde T(k,y)$.

All in all, the BK equation reads, in $k$ space
\be
\partial_y \tilde T(k,y)=
\bar\alpha\chi\left(-\partial_{\ln |k|^2}\right)\tilde T(k,y)-\bar\alpha
\left[
\tilde T(k,y)
\right]^2.
\ee
In this form, we see that this equation looks very much like
the FKPP equation~(\ref{eq:FKPPu}), except for the linear part which is not
a second-order differential operator, but an integral operator (up to changes
of variables).

However, this integral operator may be expanded.
Indeed, let us perform a Taylor expansion of the kernel eigenvalue
$\chi(\gamma)$ around some $\gamma_0$:
\be
\chi(\gamma)=\chi(\gamma_0)+(\gamma-\gamma_0)\chi^\prime(\gamma_0)+\frac12(\gamma-\gamma_0)^2
\chi^{\prime\prime}(\gamma_0).
\ee
The expanded kernel is obtained by replacing $\gamma$ by the differential
operator $-\partial_{\ln|k|^2}$, and thus the BFKL equation becomes a second-order
partial differential equation.
This is called the ``diffusive approximation''. 
We shall digress on this approximation
in connection to the solution to the linear BFKL equation.

\subsubsection{Diffusive approximation}

If $\gamma_0=\frac12$, then the diffusive approximation 
is equivalent to the saddle-point
approximation for the solution of the BFKL equation.
Recall that the full solution reads
\be
T(x_{01},y)=\int\frac{d\gamma}{2i\pi}\tilde T(\gamma)|x_{01}|^{2\gamma}e^{\bar\alpha\chi(\gamma)y}.
\ee
$\tilde T(\gamma)$ is the initial condition for the evolution, namely the scattering amplitude
at zero rapidity. If $y$ is very large, then the integral is dominated by the saddle point.
which is determined by the equation $\left(\bar\alpha\chi(\gamma)\right)^\prime=0$.
The latter is solved by $\gamma=\gamma_0=\frac12$.
One then expands the argument of the exponential to second order around $\gamma_0$.
After some trivial simplifications, one gets
\be
T(x_{01},y)=e^{\bar\alpha\chi(\frac12)y}
\int_{\frac12-i\infty}^{\frac12+i\infty}
\frac{d\gamma}{2i\pi}
\tilde T(\gamma)
e^{\frac{\bar\alpha}{2}(\gamma-\frac12)^2\chi^{\prime\prime}(\frac12)y}.
\ee
Changing integration variable by writing $\gamma=\frac12+i\nu$, 
\be
T(x_{01},y)=\tilde T({\scriptstyle\frac12})|x_{01}|e^{\bar\alpha\chi(\frac12)y}
\int_{-\infty}^{+\infty}
\frac{d\nu}{2\pi}
e^{-\frac{\bar\alpha}{2}\chi^{\prime\prime}(\frac12)y\nu^2+i\nu\ln|x_{01}|^2}.
\ee
The remaining integral over $\nu$ is a Gaussian integral.
Taking into account the special values of $\chi$
\be
\chi({\scriptstyle\frac12})=4\ln 2,\ \
\chi^{\prime\prime}({\scriptstyle\frac12})=28\zeta(3),
\ee
where $\zeta(x)=\sum_{n=1}^{\infty}\frac{1}{n^x}$ is the Riemann zeta function,
the final result reads
\be
T(x_{01},y)=\tilde T\left({\scriptstyle\frac12}\right)|x_{01}|
e^{4\ln 2 \bar\alpha y}
\frac{
\exp\left(
-\frac{\ln^2|x_{01}|^2}{56\zeta(3)\bar\alpha y}
\right)
}
{\sqrt{56\zeta(3)\pi\bar\alpha y}}.
\ee
When writing down this formula, we implicitly assume that the transverse distances are expressed
in units of the size of the target.

In the case in which $|x_{01}|$ is not too different from 1, namely if one scatters
a dipole whose size obeys $|\ln |x_{01}|^2|\ll \sqrt{y}$, then the
Gaussian factor tends to 1 and $T$ exhibits an exponential growth with the rapidity.

\begin{figure}
\begin{center}
\includegraphics[width=.8\textwidth,angle=0]{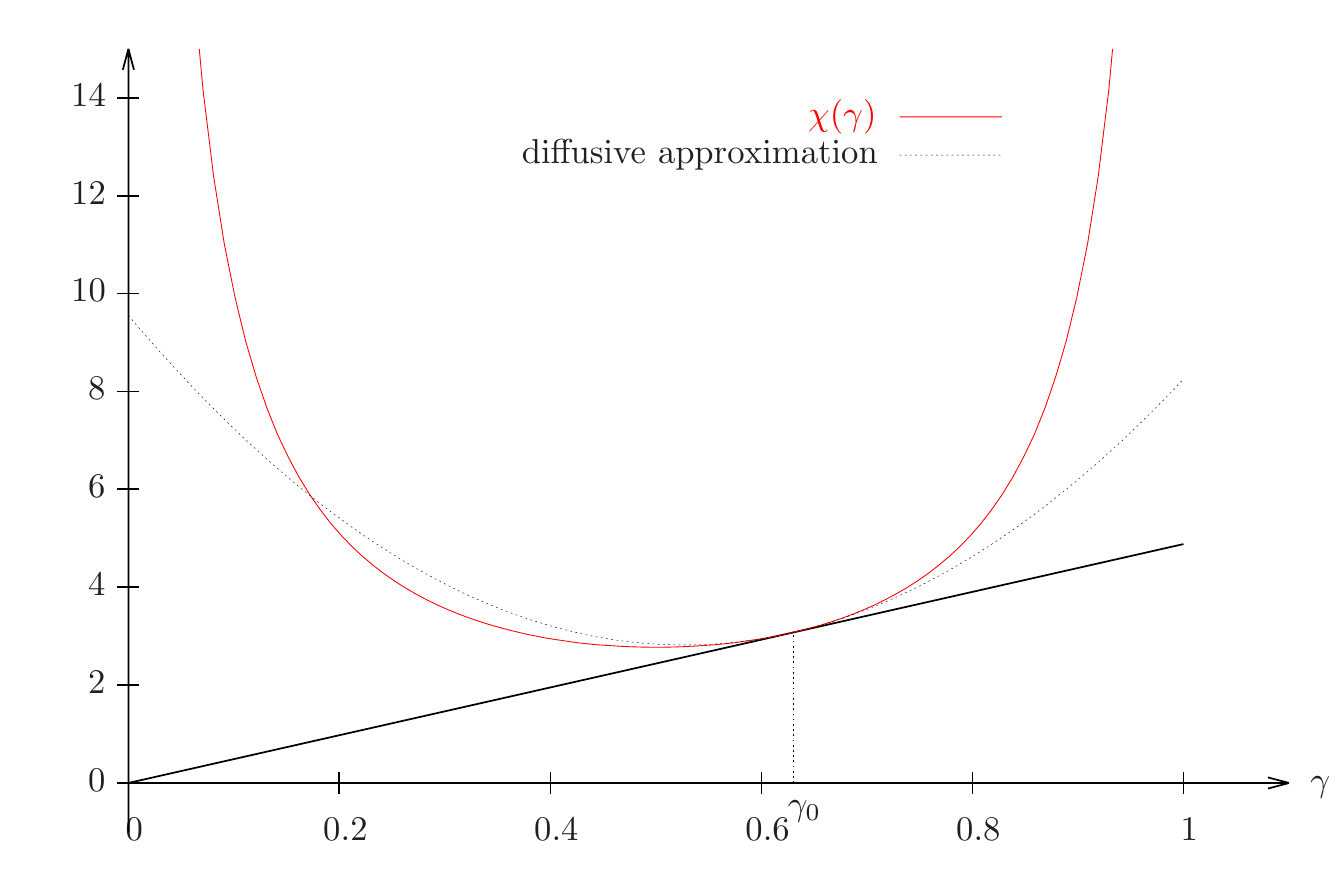}
\end{center}
\caption{\label{fig:plotchi2}
Principal branch of the function $\chi(\gamma)$ (red line),
graphical solution of the equation $\chi^\prime(\gamma_0)=\chi(\gamma_0)/\gamma_0$
which defines $\gamma_0$ (solid black line), and quadratic approximation
of $\chi(\gamma)$ around $\gamma=\gamma_0$ (dotted black line).
}
\end{figure}

Note that at very large rapidities, $T$ eventually tends to infinity.
On the other hand, since $T$ is related to a probability, it should be bounded.
The unitarity of $T$ is actually preserved by the BK equation, thanks to the
nonlinear term therein.
This forces us to consider constant values of $T$, namely to go to a frame which is
moving with the rapidity, instead
of fixing the dipole size.
In this case, the relevant eigenvalue is not $\chi(\frac12)$ but, as seen before,
$\chi(\gamma_0)$ where $\gamma_0$ solves $\chi(\gamma_0)/\gamma_0=\chi^\prime(\gamma_0)$
(see Fig.~\ref{fig:plotchi2}).

\subsubsection{BK in the diffusive approximation and FKPP}

We are now in a position to exhibit a rigorous mapping between BK and FKPP.
The BK equation reads, in the diffusive approximation
\be
\partial_y \tilde T(k,y)=
\bar\alpha\left[
-\chi^\prime(\gamma_0)\partial_{\ln |k|^2}\tilde T(k,y)
+\frac12\chi^{\prime\prime}(\gamma_0)(\partial_{\ln |k|^2}+\gamma_0)^2\tilde T(k,y)
\right]
-\bar\alpha
\left[
\tilde T(k,y)
\right]^2.
\label{eq:BKdiffusive}
\ee
One has to perform a mere change of variables in order to get the FKPP equation.
We leave the details as an exercise for the reader.

\begin{ex}
Show that the BK equation for $\tilde T$ in the diffusive approximation~(\ref{eq:BKdiffusive})
maps exactly to the FKPP equation~(\ref{eq:FKPPu}) for $u$ through the change of variables
\be
\bar\alpha y=\frac{2t}{\gamma_0^2\chi^{\prime\prime}(\gamma_0)},\ \
\ln |k|^2=\frac{x}{\gamma_0}-\frac{2}{\gamma_0}
\left(
1-\frac{\chi^{\prime}(\gamma_0)}{\gamma_0\chi^{\prime\prime}(\gamma_0)}
\right)t,\ \ 
\tilde T(k,y)=\frac{\gamma_0^2\chi^{\prime\prime}(\gamma_0)}{2}u(x,t).
\ee
\end{ex}

Strictly speaking, this mapping holds in momentum space.
But it is clear that the physics of the branching random walk is the same in
coordinate space.
On the practical side, coordinate space 
is maybe more convenient for model building:
$\sigma^{\gamma^* p/A}$ is a convolution involving the dipole cross section
in coordinate space.
On the theoretical side, it is in coordinate space that the unitarity bound can
be formulated for $T$.

We expect that all results obtained for $u$ solving the FKPP equation to go over also
to the QCD amplitudes in coordinate space, up to the substitution
$\ln |k|^2\rightarrow \ln 1/|x_{01}|^2$.


\subsection{Generalization: full BK and FKPP universality class}

We have exhibited a rigorous mapping between FKPP and an approximate 
(diffusive) form of BK.
But with some acquaintance with the physics of branching random walks,
once a given problem has been identified from general considerations to belong
to the class of branching random walks,
it is enough to identify the correct space and time variables
and the branching-diffusion kernel eigenvalue
function in order to be able
to conjecture the quantitative asymptotics of the solutions 
to the equivalent FKPP equation.
As for BK, the correspondence is given in Tab.~\ref{tab:mapping}.
\begin{table}
\begin{center}
\begin{tabular}{rcl}
\hline
FKPP & & BK\\
\hline
\begin{minipage}{0.35\textwidth}\raggedleft
{evolution variable $t$}
\end{minipage} &
$\longrightarrow$ &
\begin{minipage}{0.35\textwidth}
{rapidity $y$}
\end{minipage}\\
\\
\begin{minipage}{0.35\textwidth}\raggedleft
{spatial variable $x$}
\end{minipage} &
$\longrightarrow$ &
\begin{minipage}{0.35\textwidth}
{log of the dipole size  $\ln 1/|x_{01}|^2$
or of the transverse momentum
$\ln |k|^2$}
\end{minipage}\\
\\
\begin{minipage}{0.35\textwidth}\raggedleft
branching-diffusion kernel eigenvalues ($\gamma^2+1$ for FKPP)
\end{minipage} &
$\longrightarrow$ &
\begin{minipage}{0.35\textwidth}
{BFKL eigenvalues $\bar\alpha\chi(\gamma)$}
\end{minipage}\\
\\
\begin{minipage}{0.35\textwidth}\raggedleft
Position of the wave front $X(t)$
\end{minipage} &
$\longrightarrow$ &
\begin{minipage}{0.35\textwidth}
{log of the saturation scale $\ln Q_s^2(y)$}
\end{minipage}\\

\hline
\end{tabular}
\end{center}
\caption{\label{tab:mapping}Dictionary between FKPP and BK.}
\end{table}

We shall conjecture that the full BK equation is in the universality class
of the FKPP equation. By this we mean that asymptotic results
such as~(\ref{eq:gensolFKPPu}) may be applied to BK
just changing variables/functions as in Tab.~\ref{tab:mapping}.
However, before we can take over these results obtained for the
FKPP universality class, we should examine the shape of the initial
condition to understand whether the solution is determined by
the shape of the initial condition or if it is solely determined by
the dynamics.

\paragraph{Initial condition.}

We know that the properties of the solutions to the FKPP equation at large time
depend on the shape of the initial condition.
We saw that these properties are determined by the dynamics of the branching diffusion
process if the initial condition is ``steep enough''.

\begin{figure}
\begin{center}
\includegraphics[width=.3\textwidth,angle=0]{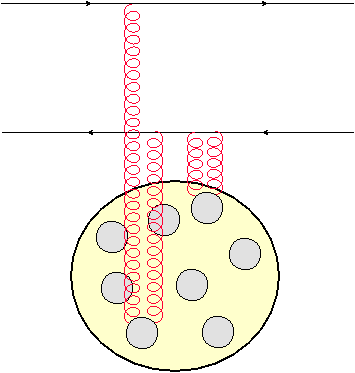}
\end{center}
\caption{\label{fig:MV}
One graph contributing to the McLerran-Venugopalan dipole-nucleus 
amplitude~(\ref{eq:SMV}).
The nucleus is assumed to consist in a large number of independent
nucleons, in such a way that the two-gluon exchanges be independent.
}
\end{figure}

A reasonable ansatz for the scattering of a dipole off a large nucleus
at zero rapidity is given by the McLerran-Venugopalan model, which basically assumes an
arbitrary number of independent two-gluon exchanges between the dipole and the various nucleons
inside the nucleus (see Fig.~\ref{fig:MV}).

The elastic $S$-matrix element in the McLerran-Venugopalan 
model~\cite{McLerran:1993ni,McLerran:1993ka}
reads\footnote{%
We keep only the main term in the exponential (strictly
speaking, there would be a correction proportional
to $x_{01}^2|\ln x_{01}^2|$).
}
\be
S_\text{MV}(x_{01})=e^{-\frac{x_{01}^2\bar Q_s^2}{4}}
\label{eq:SMV}
\ee
where $\bar Q_s$ is a momentum scale, the saturation scale of the nucleus.
It depends on the gluon density in the nucleons and on the atomic mass number of the nucleus.
Expressed in logarithmic scale for the transverse distances and expanded for $|x_{01}|$ small,
\be
S_\text{MV}(x_{01})=\exp\left(-e^{-\ln \frac{4}{\bar Q_s^2|x_{01}|^2}}
\right)
\simeq 1-e^{-\ln\frac{4}{\bar Q_s^2|x_{01}|^2}}
\ee
$T_\text{MV}=1-S_\text{MV}$ has the form $\text{const}\times e^{-\beta x}$, with
$x=\ln 1/|x_{01}|^2$ and $\beta=1>0.63=\gamma_0$.
Hence this initial condition is indeed steep enough so that we are
in the ``pulled front'' case.

Note that this feature is more general than the McLerran-Venugopalan model:
The fact that the QCD scattering amplitude of a color-neutral object of size $x_{01}$
vanishes as $x_{01}^2$ is a fundamental property of QCD known as {\it color transparency}.


\subsection{Properties of the solutions to the BK equation and models for DIS}

\subsubsection{Traveling wave property and geometric scaling}

Since the BK equation is in the universality class of the FKPP equation,
with an initial condition which is ``steep enough'', we can take over the
results obtained from the FKPP equation to QCD.

The main question we need to address is: What are the QCD traveling waves?
Are there phenomenological consequences of their existence?
We will then try and use the knowledge on the solutions to the BK equation
to build models and fit the data.

We recall that a traveling wave is a solution such that
\be
{u}(x,t)\underset{t\rightarrow +\infty}{\sim}{\cal U}(x-X^{(\beta)}(t)),
\tag{\ref{eq:travelingwaveproperty}$'$}
\ee
where $X(t)$ is the position of the wave front.
In general, starting from a given initial condition, the traveling wave
appears asymptotically for large $t$.

We now know that $T$ obeys an equation similar to the FKPP equation, with
the spatial variable being the logarithm of the dipole size 
$x\rightarrow \ln|x_{01}|^2$, and the evolution variable the rapidity:
$t\rightarrow y$. Let us introduce a rapidity-dependent distance $R_s(y)$, and the
associate momentum $Q_s(y)=1/R_s(y)$ that we shall call ``saturation momentum''.

\begin{figure}
\begin{center}
\begin{tabular}{c c}
\includegraphics[width=.45\textwidth,angle=0]{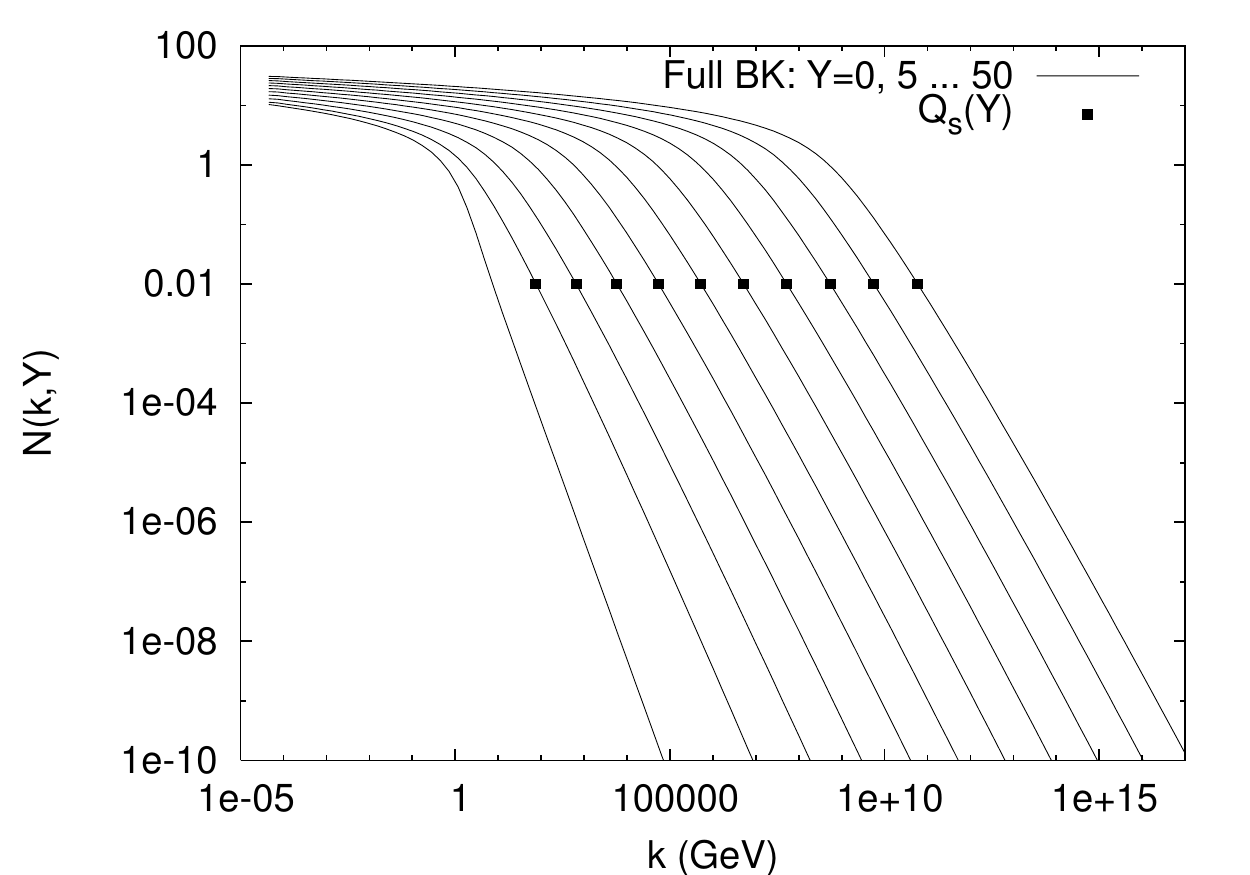}
&\includegraphics[width=.45\textwidth,angle=0]{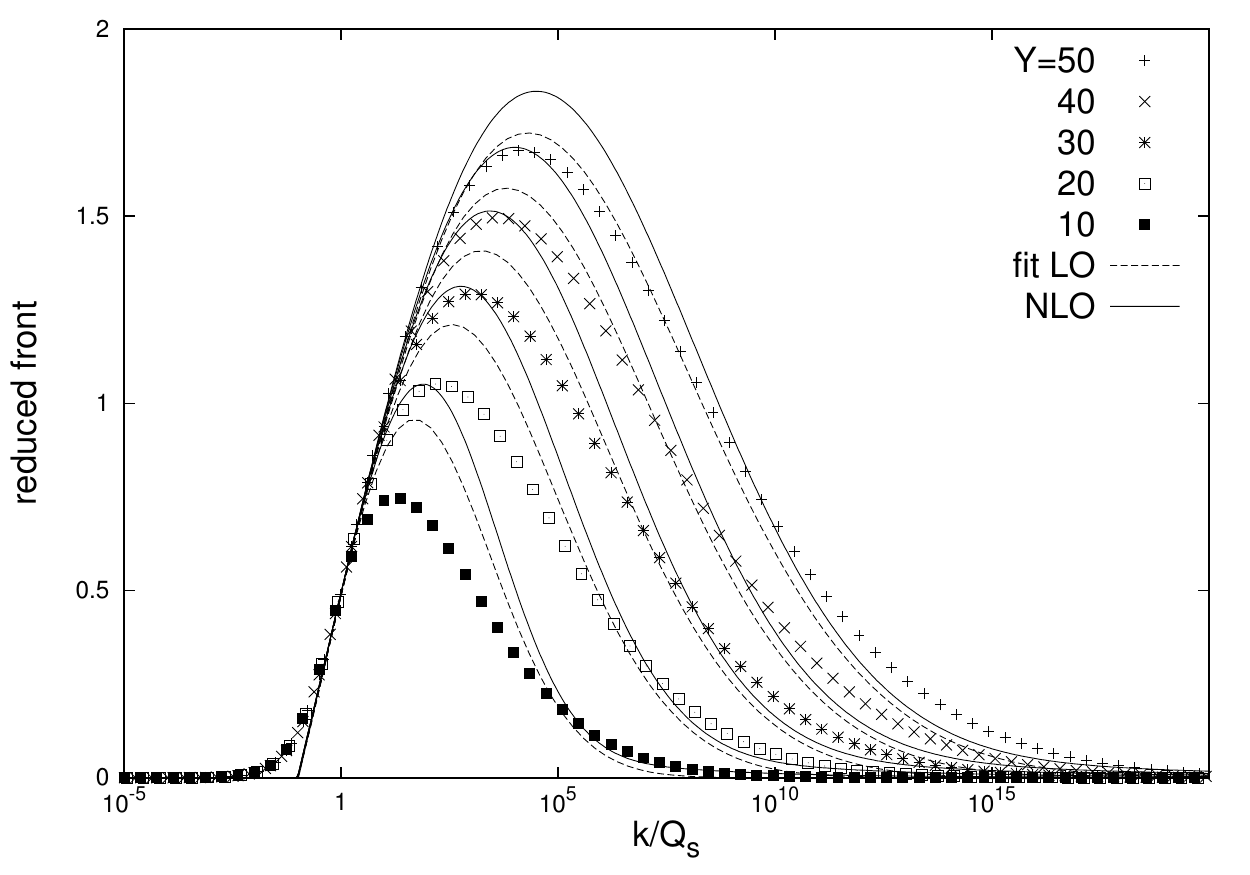}\\
(a) & (b)
\end{tabular}
\end{center}
\caption{\label{fig:bknumsol}
Numerical solution of the BK equation in momentum space,
starting from a McLerran-Venugopalan initial condition.
(a) Front shape (logarithmic scale) for different rapidities. (b) Reduced front shape
in the frame of the wave (logarithmic scale on the $x$-axis).
[Plots from Ref.~\cite{Enberg:2005cb}].
}
\end{figure}

Then the traveling wave property for the QCD amplitude $T$ reads
\be
T(x_{01},y)=T(\ln |x_{01}|^2-\ln R_s^2(y)),
\label{eq:scalingT}
\ee
which means that for $y$ very large, $T$ only depends on the product
$|x_{01}|^2\times Q_s^2(y)$.
This scaling property can be checked for the BK equation in a numerical
simulation\footnote{%
There are several numerical implementations of the BK equation.
We used the one described in Ref.~\cite{Enberg:2005cb}, ``BKsolver'',
which can be
downloaded from R.~Enberg page at:
{\tt http://rikardenberg.wordpress.com/bksolver/}.} 
and is indeed well verified, see Fig.~\ref{fig:bknumsol}.

The scaling~(\ref{eq:scalingT}) which should hold for
the abstract dipole scattering amplitude 
actually goes (approximately) over to the DIS cross section,
at least when assuming that the quarks are massless.
Indeed, taking into account the relation $S=1-T$ and
the fact that $S$ is essentially real at high energies, 
and assuming furthermore that the impact-parameter dependence 
is constant over a disk of radius $R$ (namely replacing
$\int d^2b\rightarrow \pi R^2$),
we rewrite Eq.~(\ref{eq:sigmaDIS}) as
\be
\sigma^{\gamma^* p/A}(Q^2,y)=2\pi R^2\int d^2{x_{01}} dz
|\psi^Q(x_{01},z)|^2 T(x_{01},y)
\label{eq:sigmaDISbis}
\ee
The explicit expression for the photon wave function in quarks reads...

Thus we find that $\sigma^{\gamma^* p/A}(Q^2,y)$ is actually a function
of a single variable, namely $Q^2/Q_s^2(y)$.
This is called geometric scaling: It is the statement that all DIS data
(at small enough $x$) should fall in the same curve when plotted against the
scaling variable. This is a prediction which can be tested against the data.

Now the question is what is the $y$-dependence of the saturation scale.
We just need to notice that $\ln Q_s^2(y)$ is the position of the traveling
wave. Hence
\be
\ln Q_s^2(y)/\bar Q_s^2=\bar\alpha\chi^\prime(\gamma_0)y-\frac{3}{2\gamma_0}\ln y+\text{const}
\label{eq:Qsy}
\ee
where we introduce the natural scale for $Q_s$, namely the saturation scale of the nucleus
$\bar Q_{s}$. Exponentiating the above equation,
\be
Q_s^2(y)=\bar Q_s^2\frac{e^{\bar\alpha\chi^\prime(\gamma_0)y}}{y^{{3}/{2\gamma_0}}}
\simeq \bar Q_s^2 e^{\lambda y}
\label{eq:expQsy}
\ee
Thus we see that the saturation momentum grows exponentially with the rapidity
at the rate $\lambda=\bar\alpha\chi^\prime(\gamma_0)$,
up to power corrections.
This means that as energy increases, the nucleus becomes less and less transparent also
to small dipoles, or, said in another way, absorbs dipoles of smaller and smaller sizes.

\begin{figure}
\begin{center}
\includegraphics[width=.6\textwidth,angle=0]{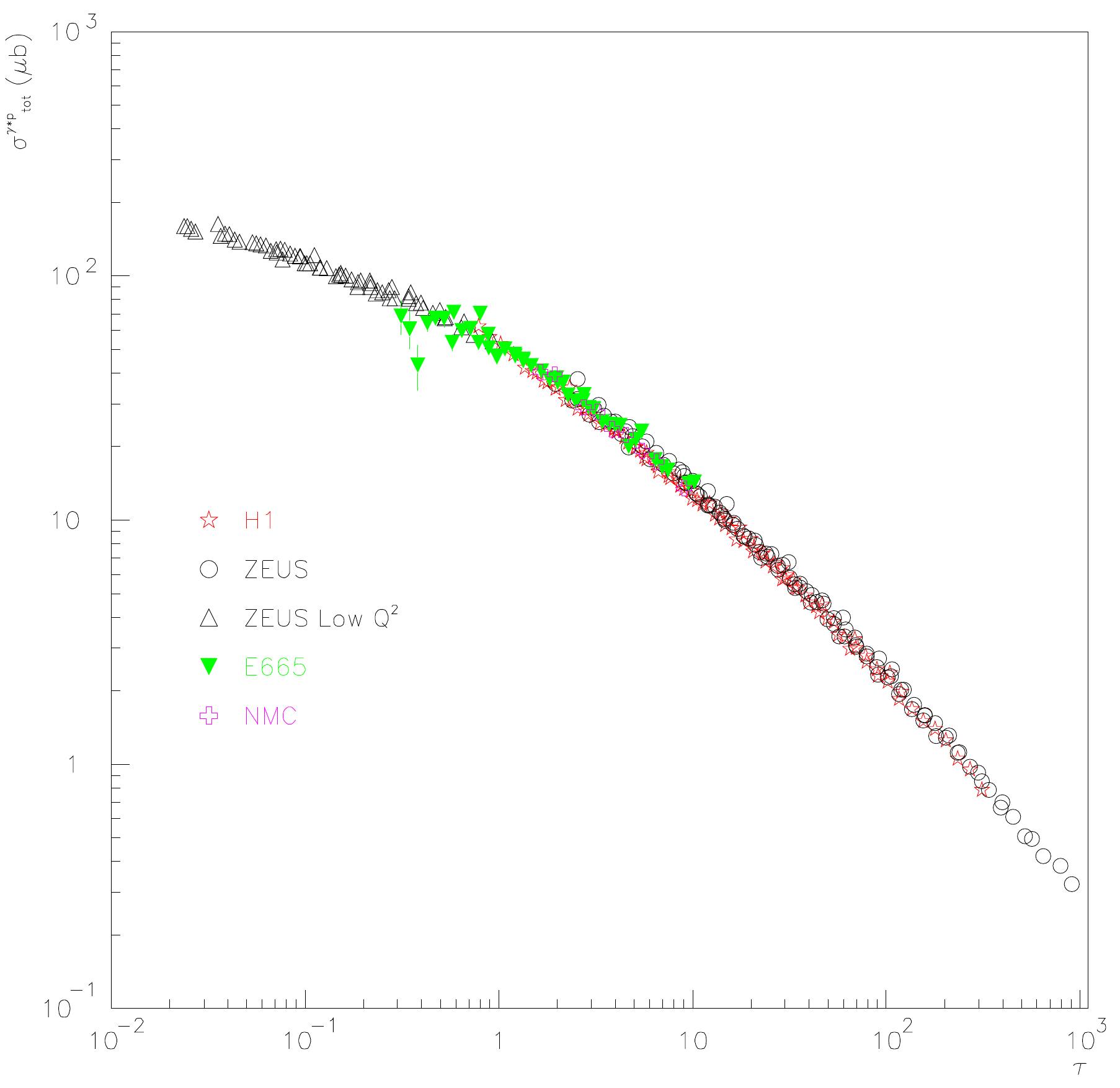}
\end{center}
\caption{\label{fig:gs}Inclusive $\gamma^* p$ cross section extracted from
the deep-inelastic scattering data for $x_\text{Bj}<10^{-2}$, plotted against $Q^2/Q_s^2(y)$.
All the data fall on the same curve: This is geometric scaling, or equivalently,
which may be interpreted as
the traveling wave properties of the solution to the BK equation.
[Figure from Ref.~\cite{Marquet:2006jb}; original plot in Ref.~\cite{Stasto:2000er}].}
\end{figure}

As seen in Fig.~\ref{fig:gs}, geometric scaling is a very striking
feature of the deep-inelastic scattering data\footnote{Actually,
geometric scaling was found in the data before it was understood that it is
actually a property of solutions to the BK equation. We shall review the
history of this field
later on, see Sec.~\ref{sec:history}.} at small $x$.

We can also predict the shape of the dipole amplitude from the solutions
to (generalized) FKPP equations. From these elements (saturation scale and
shape of the amplitude), one can try and build
models for the dipole cross section, and apply them to a description
of the scattering amplitudes measured in deep-inelastic scattering experiments.

\subsubsection{Towards a model for deep-inelastic scattering}

In order to arrive at predictions for the DIS cross section,
all we need is a model for the $S$-matrix element (see Eq.~(\ref{eq:sigmaDIS})
or, equivalently, for the amplitude $T$, see Eq.~(\ref{eq:sigmaDISbis}))
for the forward elastic interaction of a dipole with the target proton or 
nucleus.

\paragraph{The Golec-Biernat and W\"ustoff model 
\cite{GolecBiernat:1998js,GolecBiernat:1999qd}.}
The simplest model consists in promoting the saturation scale in the McLerran-Venugopalan
model~(\ref{eq:SMV}) to a function of the rapidity
\be
\bar Q_s^2\rightarrow Q_s^2(y)=\bar Q_s^2 e^{\lambda y}
\label{eq:QsGBW}
\ee
and let $\bar Q_s$ and $\lambda$ be free parameters.
Equation~(\ref{eq:QsGBW}) 
is the leading-order approximation
if $\lambda=\bar\alpha\chi^{\prime}(\gamma_0)$ but this expression for $\lambda$
is too crude since it is based on the leading-order BK equation, which is not
accurate enough for phenomenology.

Using Eq.~(\ref{eq:sigmaDISbis}) with $S$ given by Eq.~(\ref{eq:SMV}) and
the saturation scale therein being replaced by Eq.~(\ref{eq:QsGBW}), we obtain
the famous Golec-Biernat and W\"ustoff model.
It has only three free parameters ($R$, $\bar Q_s$ and $\lambda$), and it turns out
that it is able to describe reasonably well\footnote{%
Dipole models (improved versions of the Golec-Biernat and W\"usthoff model)
seem however to do less well with the most precise HERA data, see e.g. 
Ref.~\cite{Luszczak:2013rxa}.
} all inclusive 
(and also diffractive)
data for deep-inelastic
scattering of electron/positron off protons at small $x_{\text{Bj}}\equiv Q^2/s$.
(By ``small'' is usually meant $x_\text{Bj}\leq 10^{-2}$).

There are several refinements of this model one may think of.
The first one would be to build an amplitude whose shape is closer to the
shape of the BK traveling waves, as we shall see in the next paragraph.
The second one, which we will not discuss in detail here, would be to
introduce a more elaborate impact-parameter profile.

\paragraph{Refinements.}
Going back to Eq.~(\ref{eq:gensolFKPPu}), we perform the changes of variables
according to Tab.~\ref{tab:mapping}. The dipole scattering amplitude reads,
for dipole sizes smaller than the inverse saturation scale
\be
T(y,x_{01})\propto \left|\ln\left(|x_{01}|^2 Q_s^2(y)\right)\right|
\left(|x_{01}|^2 Q_s^2(y)\right)^{\gamma_0}
e^{-\frac{\ln^2\left(|x_{01}|^2 Q_s^2(y)\right)}
{2\bar\alpha\chi^{\prime\prime}(\gamma_0)y}}.
\label{eq:Tdilute}
\ee
The first two factors exhibit geometric scaling, while the last one,
which is different from one whenever 
$\left|{\ln\left(|x_{01}|^2 Q_s^2(y)\right)}\right|$ is not
small with respect to
$\sqrt{2\bar\alpha\chi^{\prime\prime}(\gamma_0)y}$, 
encodes geometric scaling violations
at finite rapidity $y$ since it exhibits an explicit $y$ dependence.
As already mentioned, this formula is valid for small enough dipole sizes,
hence in the ``dilute'' regime where $T\ll 1$.
To construct the full amplitude, we also
need to understand the properties of $T$ 
in the saturation regime, where $T\sim 1$ or $S\ll 1$
To this aim, we go back to the BK equation for $S$ given in 
Eq.~(\ref{eq:BKS}),
and rewrite it in this limit, in which the nonlinear term $S(y,x_{02})S(y,x_{12})$
can be neglected:
\be
\partial_y S(y,x_{01})=\bar\alpha
\int_{1/Q_s^2(y)}\frac{d^2 x_2}{2\pi}\frac{x_{01}^2}{x_{02}^2 x_{12}^2}
\left[-S(y,x_{01})\right].
\ee
The lower bound on the integral means that both $|x_{02}|$ and $|x_{12}|$
have to be larger than $1/Q_s(y)$, which is the condition for
the equation to linearize.
Now the $S$ factor goes out of the integral since it has no $x_2$ dependence, 
and the dominant contribution to the latter
comes from the two collinear regions $|x_{02}|,|x_{12}|\ll |x_{01}|$
\be
\int_{1/Q_s^2(y)}\frac{d^2 x_2}{2\pi}\frac{x_{01}^2}{x_{02}^2 x_{12}^2}
\simeq
2\int_{1/Q_s^2(y)}^{x_{01}^2}\frac{d^2 x_2}{2\pi}\frac{1}{x_{02}^2}
=\ln (|x_{01}|^2 Q_s^2(y)).
\ee
We may then integrate the differential equation from some initial rapidity $y_0$ to $y$:
\be
S(y,x_{01})=S(y_0,x_{02})e^{-\bar\alpha\int_{y_0}^y
dy^\prime\,\ln(|x_{01}|^2Q_s^2(y^\prime))}.
\label{eq:Ssat}
\ee
We leave the final integration as an exercise.
\begin{ex}
Knowing the expression for $Q_s(y)$ (keep only the leading term),
complete the calculation by integrating over $y^\prime$.\\
{\normalfont The final result is known as the ``Levin-Tuchin law''~\cite{Levin:1999mw}.
A refined version of this calculation has come out very recently, see
\cite{Contreras:2014oba}.
}
\end{ex}

Now we may try and match the two expressions~(\ref{eq:Tdilute}) and~(\ref{eq:Ssat}):
We get the IIM model~\cite{Iancu:2003ge}, which successfully describes the HERA data.

\begin{recap}
Parton evolution in the high-energy of QCD is a peculiar branching random walk.
The Balitsky-Kovchegov equation~(\ref{eq:BKT}), which drives the energy evolution
of QCD scattering amplitudes, is in the universality class of the FKPP equation~(\ref{eq:FKPPu}).
It turns out that there is an exact mapping in the diffusive approximation 
for the kernel of the BK equation, but more
generally, one can argue that
the asymptotics of the two equations should be identical up to the identification
of the relevant variables.
Properties of the solutions to the BK equation can be inferred from what is known 
on the solutions to the FKPP equation, 
leading to the expression~(\ref{eq:Tdilute}) for
the scattering amplitude, and~(\ref{eq:expQsy}) for the saturation scale.
The traveling wave property corresponds to geometric scaling, which
was found in the deep-inelastic scattering data (Fig.~\ref{fig:gs}).
\end{recap}

\begin{beyond}
One limit which was taken to arrive
the BK evolution equation is the large-number of color ($N_c$) limit.
An equation which takes into account finite-$N_c$ corrections is known:
It is the so-called Jalilian-Marian-Iancu-McLerrran-Leonidov-Kovner (JIMWLK)
equation, see Ref.~\cite{Iancu:2000hn,Ferreiro:2001qy} and references therein.
Numerical solutions of the latter \cite{Rummukainen:2003ns} seem to show that its solutions are
very similar to the solutions to the BK equation.
But we shall cautiously deem that the question whether the JIMWLK equation
contains the same physics as the BK equation is still an open question.

\noindent
High-energy QCD, the dipole model and the BK equation were introduced 
in Prof.~Mueller's lectures at Wuhan's school.
For more,  see the 
original paper~\cite{Mueller:1993rr}
and the recent 
textbook of 
Ref.~\cite{kovchegov2012quantum}.
\end{beyond}

\clearpage


\section{Beyond the simple branching random walk -- 
Beyond the Balitsky-Kovchegov equation}

\begin{intro}
So far, we have considered that gluons evolve through 
$\text{gluon}\rightarrow \text{gluon}+\text{gluon}$
splittings only, and we have neglected the correlations of
multiple gluons in the course of partonic evolution.
However, QCD would a priori also allow for gluon recombinations,
and further color charge interactions. One expects these processes
to play a role when the density becomes large enough.
While this has not been properly formulated in QCD, 
we may introduce such recombination/saturation processes
in branching random walks, and obtain modified evolution equations for
e.g. the particle density.
It turns out that the solutions to these equations have
universal features which are independent of the details of the recombination
processes. This is what we will be after in this section.
\end{intro}

\subsection{Motivation}

Deep-inelastic scattering at high energies may be seen as a dipole-target interaction
process, once the virtual photon wave function in $q\bar q$ pairs has been factorized.
The simplest model for each of the nucleons in the target is a color dipole.

\begin{figure}[ht]
\begin{center}
\begin{tabular}{m{.12\textwidth}m{1em}
m{.1\textwidth}m{1em}
m{.1\textwidth}m{1em}
m{.1\textwidth}m{1.5cm}
m{.1\textwidth}
}
\includegraphics[width=.12\textwidth,angle=0]{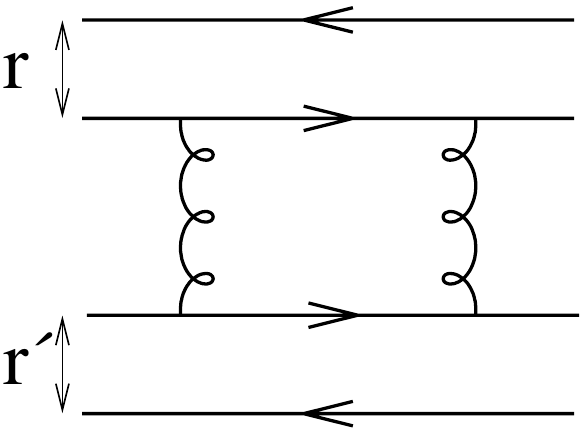} &$+$&
\includegraphics[width=.1\textwidth,angle=0]{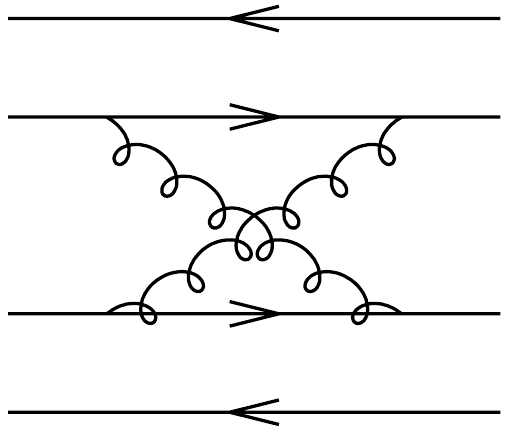} &$+$&
\includegraphics[width=.1\textwidth,angle=0]{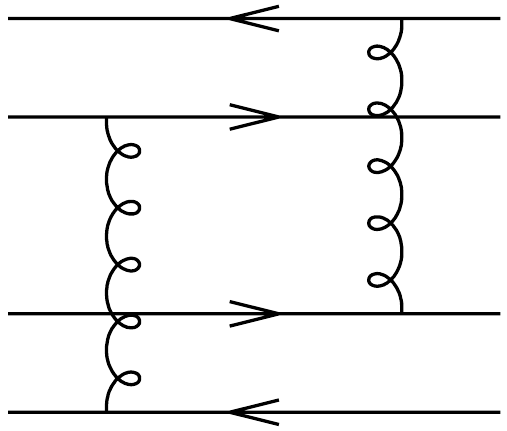} &$+$&
\includegraphics[width=.1\textwidth,angle=0]{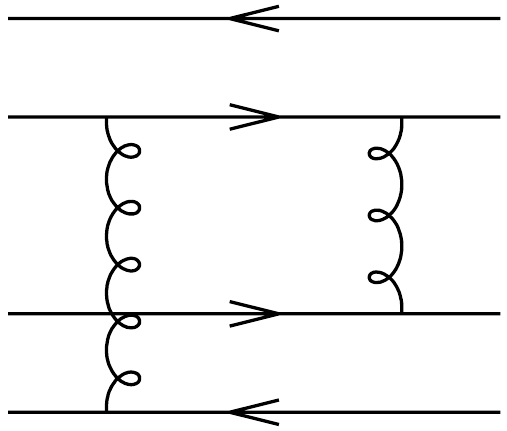} &$+\ \text{[sym.]}\equiv$&
\includegraphics[width=.1\textwidth,angle=0]{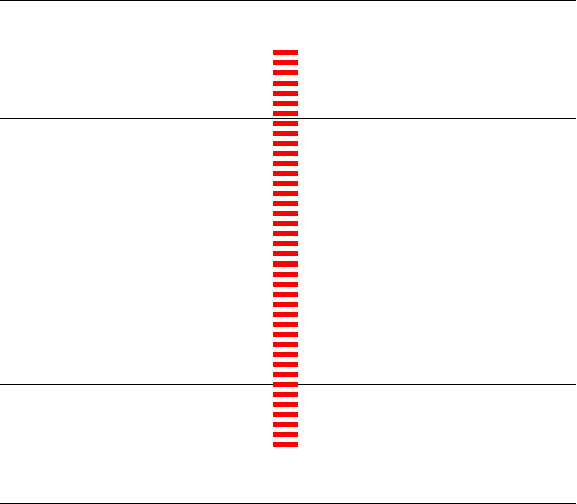}
\end{tabular}
\end{center}
\caption{\label{fig:dipoledipole}
Graphs contributing to the scattering of two color dipoles 
(at zero rapidity; see Eq.~(\ref{eq:Telsimple}))
and the graphical representation of their sum that we shall use later.
}
\end{figure}

The lowest order contribution to the dipole-dipole elastic scattering amplitude
is given by the two-gluon exchange graphs (see Fig.~\ref{fig:dipoledipole}).
The interaction is local in impact parameter, in the sense that
dipoles which have no geometric overlap, namely whose centers do not sit
within a distance smaller than the size of say the largest dipole, interact very weekly.
The amplitude for the scattering of dipoles of respective sizes $r$ and $r_0$ sitting
on top of each other in transverse space
approximately reads
\be
T_\text{el}(r,r_0)\simeq\alpha_s^2\frac{r_<^2}{r_>^2},
\ee
where $r_<=\min(r,r_0)$ and $r_>=\max(r,r_0)$.
In logarithmic coordinates, we can consider that this is a local interaction
also in the dipole sizes and replace $T_\text{el}$ by
\be
T_\text{el}(r,r_0)\simeq \alpha_s^2\delta(\ln r^2/r_0^2).
\label{eq:Telsimple}
\ee

\begin{figure}[ht]
\begin{center}
\begin{tabular}{c@{\hskip 3cm}c}
\includegraphics[width=.16\textwidth,angle=0]{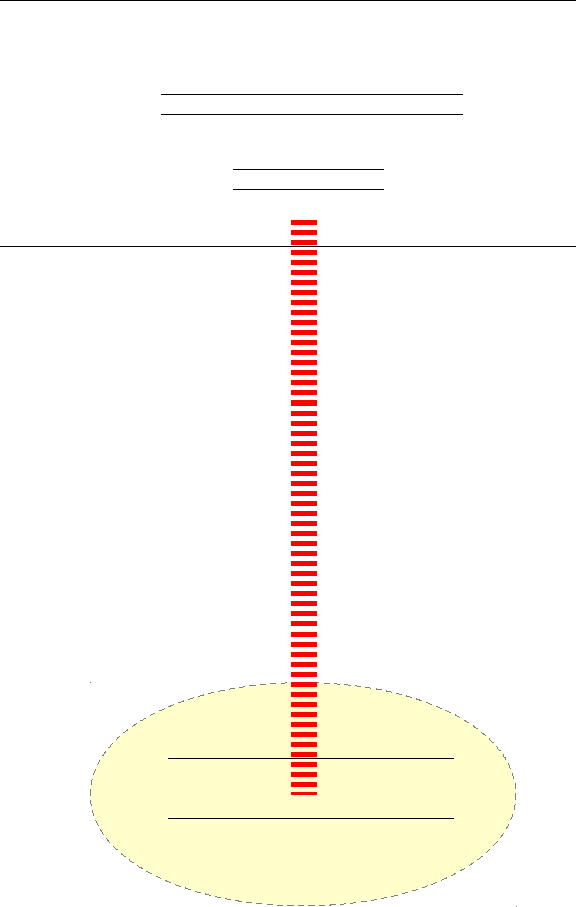} &
\includegraphics[width=.16\textwidth,angle=0]{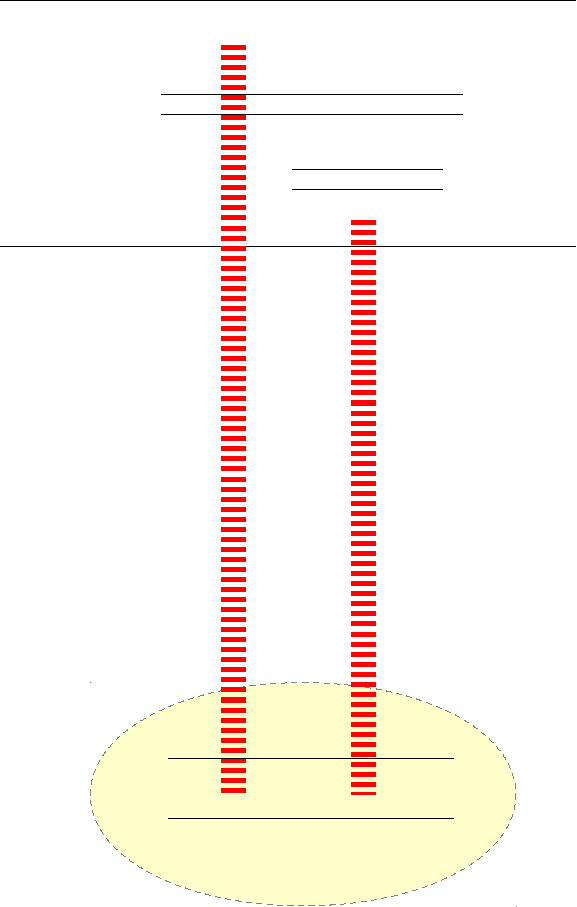}\\
\\
(a) & (b)
\end{tabular}
\end{center}
\caption{\label{fig:dipoledipoleinteraction}
(a) Example of dipole graph contributing to dipole-dipole scattering in the restframe
of the lower dipole (the so-called ``target''), and
whose sum is represented by the BFKL equation. There is a single two-gluon exchange,
represented by the red vertical line.
(b) At higher rapidity, multiple two-gluon exchanges may occur and, if the target
is a single dipole, the interactions are no longer independent.
This case has not been formulated in a satisfactory way yet.
}
\end{figure}

QCD evolution replaces an initial dipole of size $x_{01}$ by a density
$n(r,y|x_{01})$ of dipoles of size $r$ at rapidity $y$ (see Fig.~\ref{fig:dipoledipoleinteraction}).
Let us assume for a moment that the target is a single dipole of size $r_0$.
Then the amplitude reads
\be
T(x_{01},y)=\int \frac{dr^2}{r^2} T_\text{el}(r,r_0)\int dP[n] n(r,y|x_{01}),
\ee
where $dP[n]$ is a formal notation for the probability of a dipole configuration of
density $n$. The last integral just gives the mean density of dipoles of size $r$.
Using Eq.~(\ref{eq:Telsimple}) to perform the integral over $r^2$, we arrive at the formula
\be
T({x_{01},y})\simeq\alpha_s^2\langle n(r_0,y|x_{01})\rangle.
\label{eq:TBFKL}
\ee
This formula says that the scattering amplitude is proportional to the average number
of dipoles of size which matches the size $r_0$ of the target after evolution
of the projectile of initial size $x_{01}$
over $y$ units of rapidity.
We know that the mean density of dipoles grows like the exponential of the rapidity:
$\langle n(r_0,y)\rangle\sim e^{\bar\alpha y}$. Hence when
$\bar\alpha y\geq \bar\alpha y_\text{sat}\equiv\ln(1/\alpha_s^2)$, $T$ becomes larger than 1.
But $T$ can be interpreted as the probability that a dipole in the Fock state of the
projectile interact with the target, and thus should be less than unity throughout the evolution.
Actually, this means that the approximation in which there is only one elementary interaction
(in which the BFKL equation is justified)
breaks down at $y\sim y_\text{sat}$, and one should take into account multiple exchanges
(see Fig.~\ref{fig:dipoledipoleinteraction}b).

\begin{figure}[ht]
\begin{center}
\begin{tabular}{c@{\hskip 3cm}c}
\includegraphics[width=.16\textwidth,angle=0]{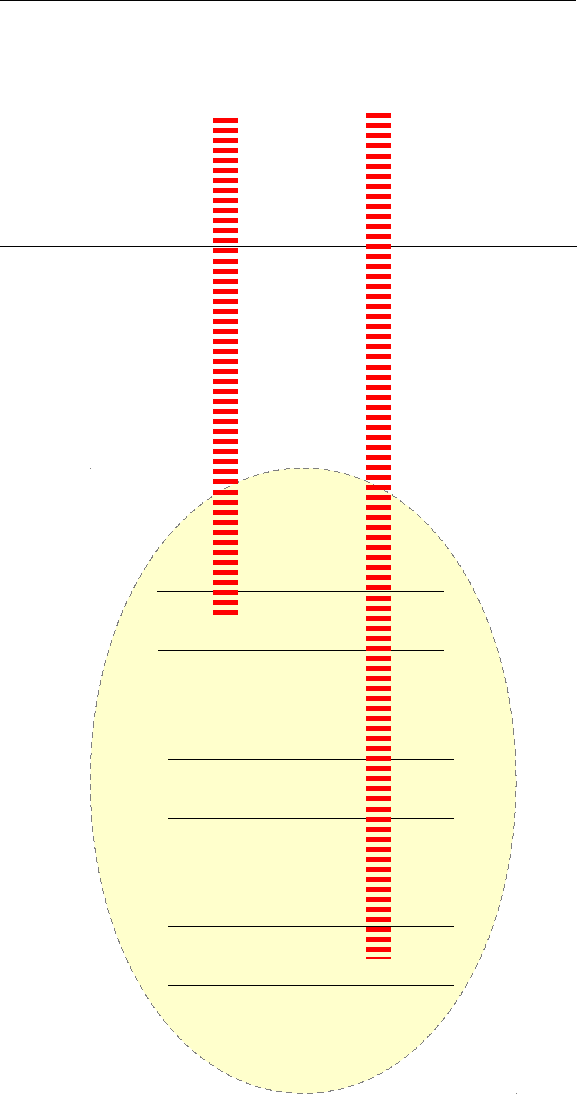} &
\includegraphics[width=.16\textwidth,angle=0]{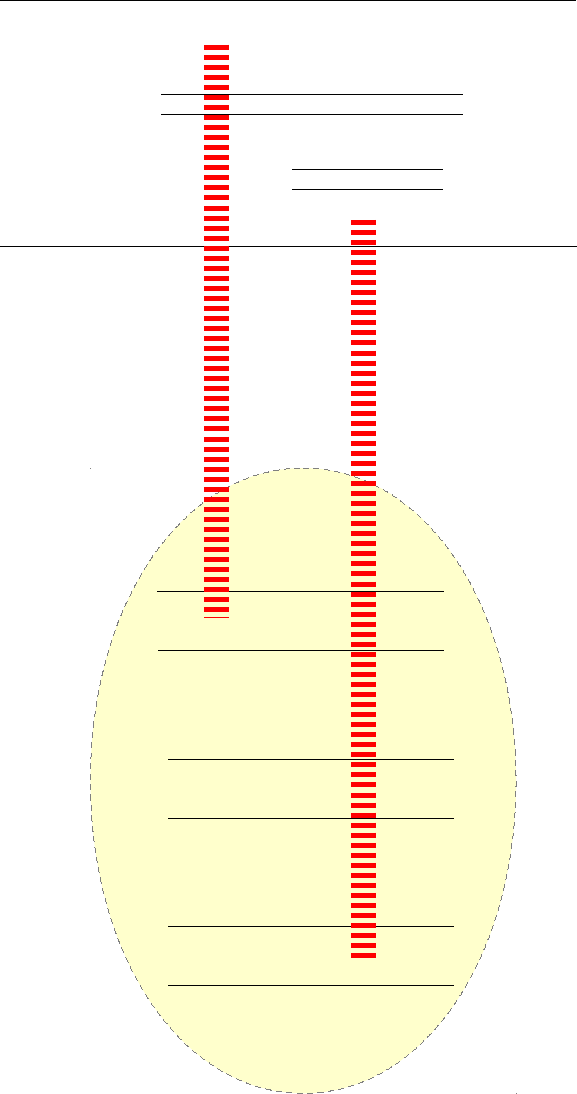}\\
\\
(a) & (b)
\end{tabular}
\end{center}
\caption{\label{fig:dipolenucleusinteraction}
(a) Example of dipole graph contributing to dipole-nucleus scattering at zero
rapidity.
Such graphs are resummed by the McLerran-Venugopalan formula.
(b) When the upper dipole is boosted, 
it interacts through quantum fluctuations. The corresponding graphs
are resummed by the BK equation.
}
\end{figure}

If the target is a very large nucleus instead of a single dipole,
then these interactions are all independent (see Fig.~\ref{fig:dipolenucleusinteraction}), 
since combinatorially, the probability that two interactions occur with the same nucleon
is small. In this case, one should simply replace Eq.~(\ref{eq:TBFKL}) by the BK equation
with the appropriate initial condition representing the scattering of an elementary dipole
of size $x_{01}$ with a set of dipoles of size $r_0$. This is the McLerran-Venugopalan model.

If however the target is a single dipole, then the interactions are necessarily correlated
and the BK equation is a priori not justified (although, as we will see later, its solution
may represent correctly the physics of dipole-dipole scattering for low enough rapidity),
as seen in Fig.~\ref{fig:dipoledipoleinteraction}b.

Let us go back to the BFKL evolution in order to estimate roughly at which rapidity the
BFKL description is expected to break down in dipole-dipole scattering.
\begin{figure}[ht]
\begin{center}
\begin{tabular}{m{.16\textwidth}m{1cm}
|@{\hskip 1cm}m{0.16\textwidth}m{1cm}
|@{\hskip 1cm}m{0.16\textwidth}m{1cm}}
\includegraphics[width=.16\textwidth,angle=0]{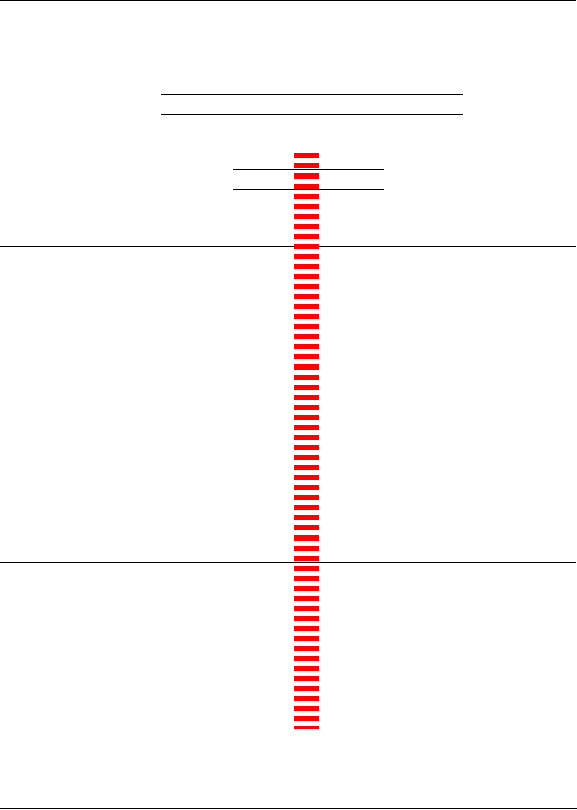} &
\begin{tabular}{c}
$n(y)$
\\\\\\\\\\
1
\end{tabular} &
\includegraphics[width=.16\textwidth,angle=0]{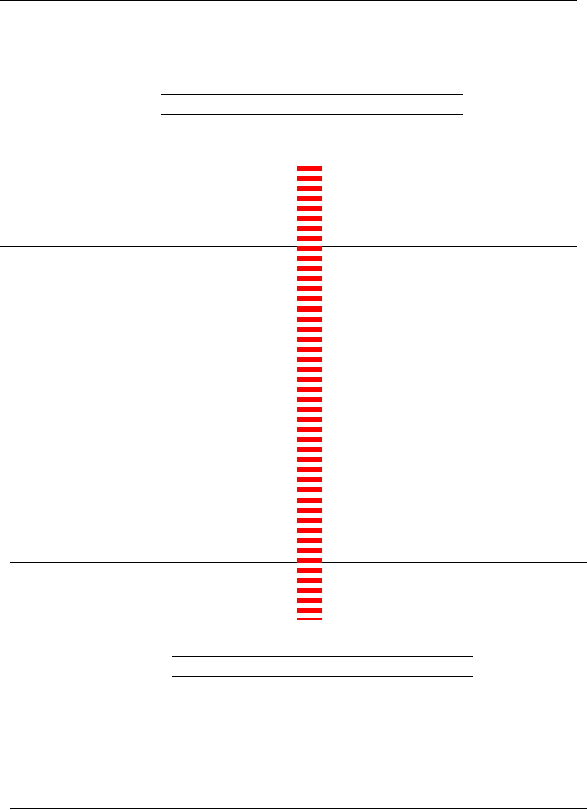} &
\begin{tabular}{c}
$n({\scriptstyle \frac{y}{2}})$
\\\\\\\\\\
$n({\scriptstyle \frac{y}{2}})$
\end{tabular} &
\includegraphics[width=.16\textwidth,angle=0]{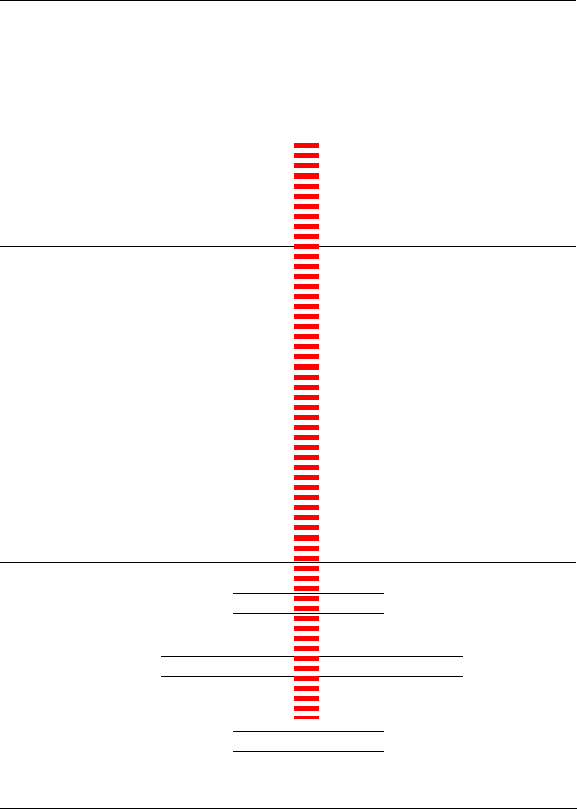} &
\begin{tabular}{c}
$1$
\\\\\\\\\\
$n(y)$
\end{tabular}\\
\\
\multicolumn{2}{c}{(a)}& \multicolumn{2}{c}{(b)} & \multicolumn{2}{c}{(c)}
\end{tabular}
\end{center}
\caption{\label{fig:bfklframes}
BFKL evolution viewed in different frames.
(a) Restframe of the lower dipole,
(b) center-of-mass frame where the rapidity is equally shared between
the dipoles,
(c) resframe of the upper dipole.
The average number of dipoles in each of the objects depends on the rapidity
and on the frame,
and is indicated in the figure.
}
\end{figure}
The answer actually depends on the frame (see Fig.~\ref{fig:bfklframes}).
In the restframe of one or of the other dipole, the scattering amplitude is roughly given by
Eq.~(\ref{eq:TBFKL}), namely $T(y)\sim\alpha_s^2 \langle n(y)\rangle$ (we kept only the
$y$ dependence in this equation). 
$T$ should be less than one for the BFKL equation to apply, which means that
the number of dipoles should be effectively 
less than $1/\alpha_s^2$ for the dipole evolution
being linear, namely for it being an ordinary branching random walk.

We already recalled that the dipole number grows
exponentially with $\bar\alpha y$.
This gives a maximum rapidity for the dipole evolution to be linear in
the laboratory frame equal to
\be
Y_\text{lab}\sim y_\text{sat}=\frac{1}{\bar\alpha}\ln\frac{1}{\alpha_s^2}.
\ee
We dropped uninteresting constants.

\begin{figure}[ht]
\begin{center}
\begin{tabular}{c@{\hskip 2cm}c}
\includegraphics[width=.25\textwidth,angle=0]{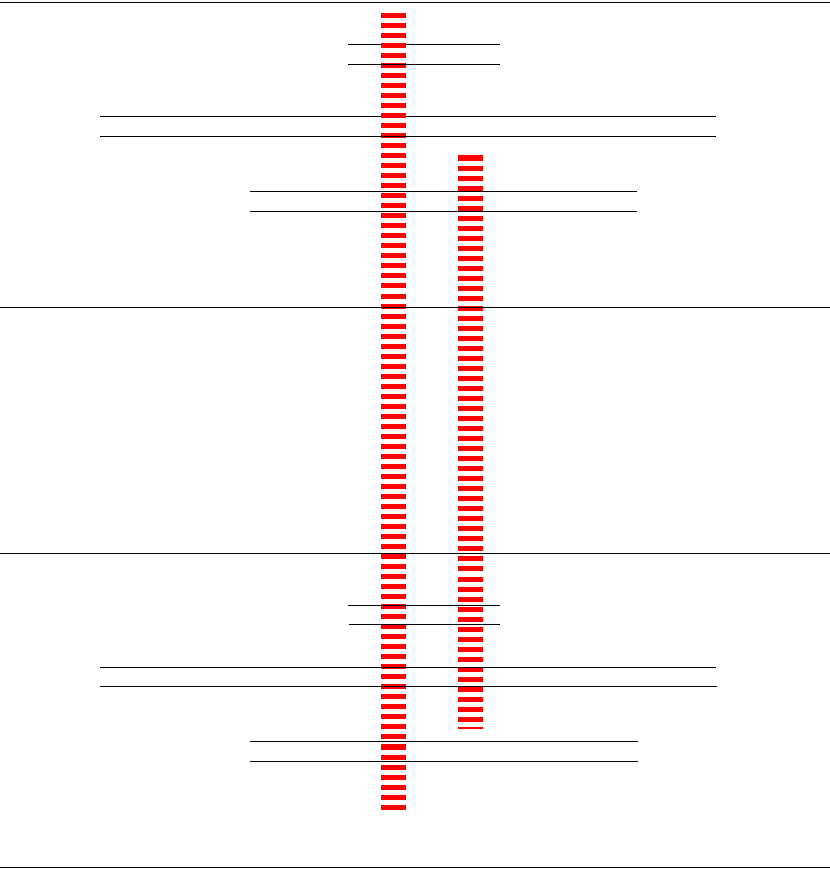} &
\includegraphics[width=.375\textwidth,angle=0]{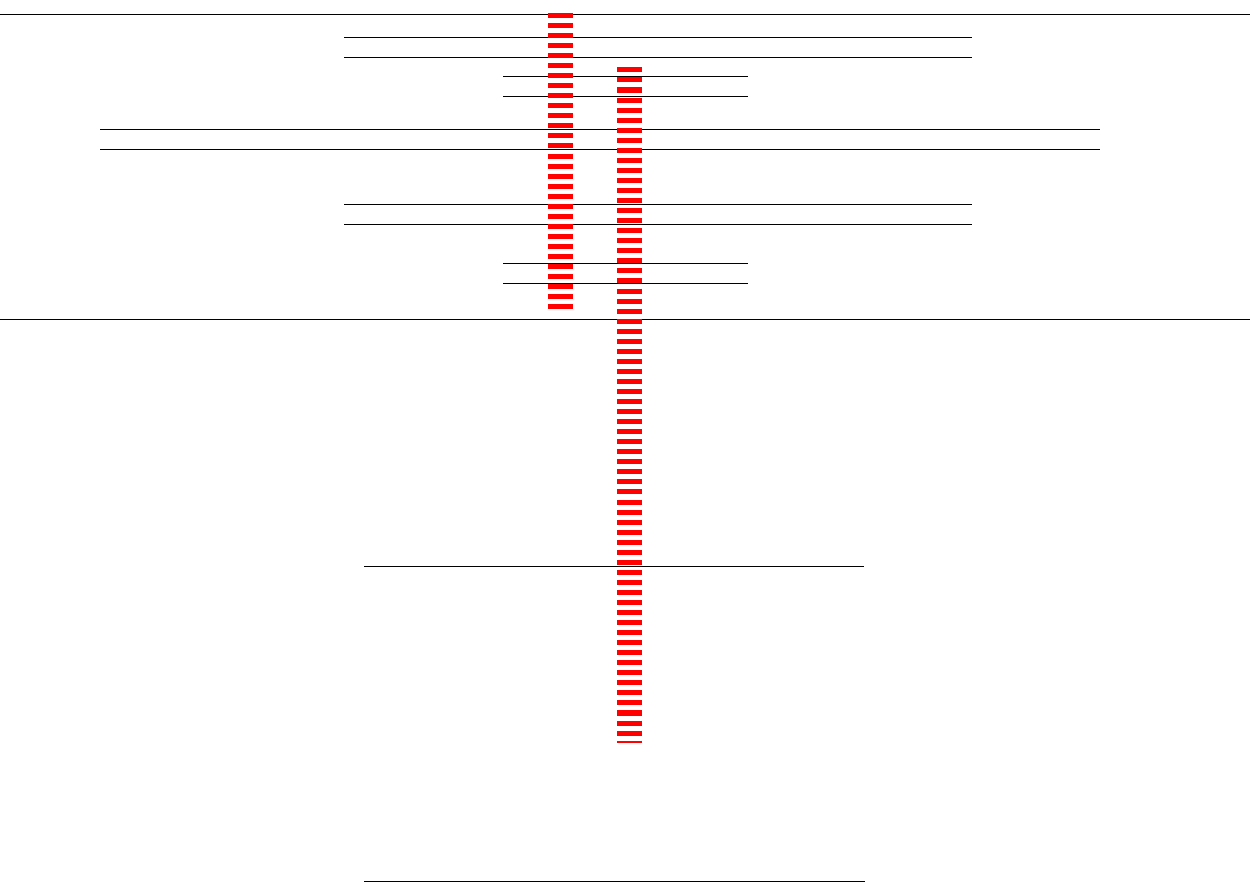}\\
\\
(a) & (b)
\end{tabular}
\end{center}
\caption{\label{fig:multiple}
(a) Example of dipole graph contributing to dipole-dipole scattering at
rapidity $\frac{1}{\bar\alpha}\ln\frac{1}{\alpha_s^2}<y<\frac{2}{\bar\alpha}\ln\frac{1}{\alpha_s^2}$
in the center-of-mass frame. In this regime, the dipole wave functions evolve according
to linear dipole evolution, but the total amplitude is in the saturation regime
and multiple exchanges are needed in order to unitarize the amplitude.
(b) The same boosted to the lab frame (restframe of the target dipole).
Now one expects nonlinear interactions in the course of the evolution of the
projectile dipole.
}
\end{figure}

In the center-of-mass frame instead, the amplitude reads
$T(y)\sim\alpha_s^2 \langle n({\scriptstyle\frac{y}{2}})\rangle
\langle n({\scriptstyle\frac{y}{2}})\rangle$, leading to a different expression for
the maximum rapidity for the dipole evolution being linear:
\be
Y_\text{com}\sim \frac{2}{\bar\alpha}\ln\frac{1}{\alpha_s^2}=2 Y_\text{lab}.
\ee
But of course, for $y=Y_\text{com}$, the overall amplitude
$T(y)$ would be larger than one, and so multiple scatterings
must occur (see Fig.~\ref{fig:multiple}a). Since the dipoles are correlated, the evolution of the
amplitude cannot be described by the BK equation.

Hence a proper formulation of dipole-dipole scattering seems to require
the introduction of a nonlinear mechanism {\it in the evolution itself}
which would effectively limit the density of dipoles to $\sim 1/\alpha_s^2$.
How to do this is not known yet.
However, we may try and understand the effects
of these nonlinearities
starting with a simple branching random walk supplemented with recombinations,
and then figure out what is universal and thus what may be taken over to QCD.


\subsection{BRW with selection/recombination: stochastic traveling waves}

\subsubsection{A simple model with stochastic traveling wave for Darwinian population evolution}

We have already introduced a model for population evolution in Sec.~\ref{sec:gen}
as an example of branching-diffusion process.
We had a population of individuals, each characterized by the ``fitness'' $x$, a real
number.
The time evolution of the population 
was defined by the following rule (see Fig.~\ref{fig:offspring}): 
\begin{figure}[h]
\begin{center}
\includegraphics[width=.25\textwidth,angle=0]{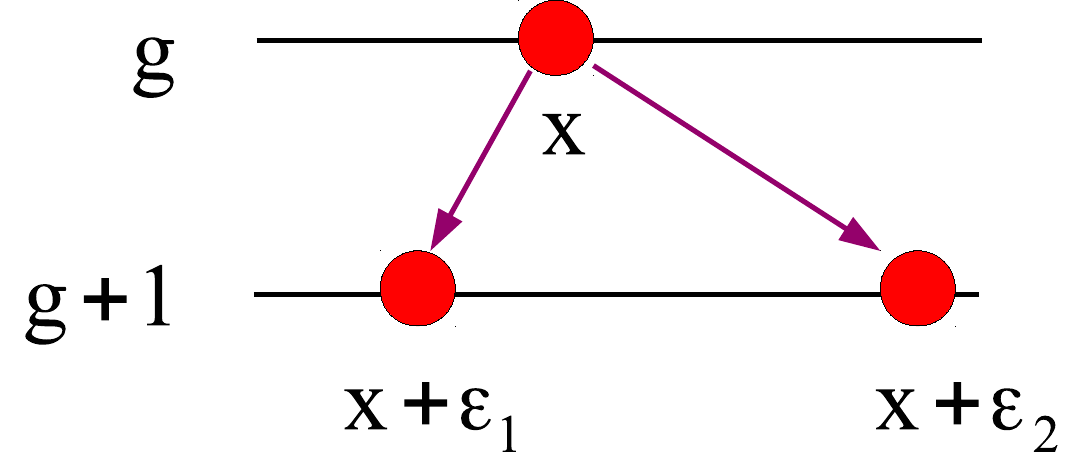}
\end{center}
\caption{\label{fig:offspring}
Elementary processe defining the evolution of each individual in the population
from generation $g$ to $g+1$.
}
\end{figure}
Each individual with fitness $x$ present in the population at generation number $g$ is replaced
at the next generation $g+1$ 
by two offspring, which have respective fitnesses $x_1$ and $x_2$ such that
\be
x_1=x+\varepsilon_1,\ x_2=x+\varepsilon_2,
\tag{\ref{eq:mutations}'}
\ee
where $\varepsilon_1,\varepsilon_2$ are random numbers distributed according 
to a probability
distribution $\rho(\varepsilon)$.
One now adds another rule for the evolution: Whenever the total population reaches some integer $N$,
for the further evolution, one removes from the population the least ``fit'' individuals
in such a way as to always keep the population size constant and equal to $N$.
This is a selection mechanism, and our model is now a simple model for Darwinian population
evolution. Indeed, the fitness is inherited by the offspring, up to
stochastic variations which represent the mutations. The selection mechanism enforces
the fact that only the fittest survive.

\begin{figure}
\begin{center}
\begin{tabular}{m{0.45\textwidth}|m{0.45\textwidth}}
\includegraphics[width=.45\textwidth,angle=0]{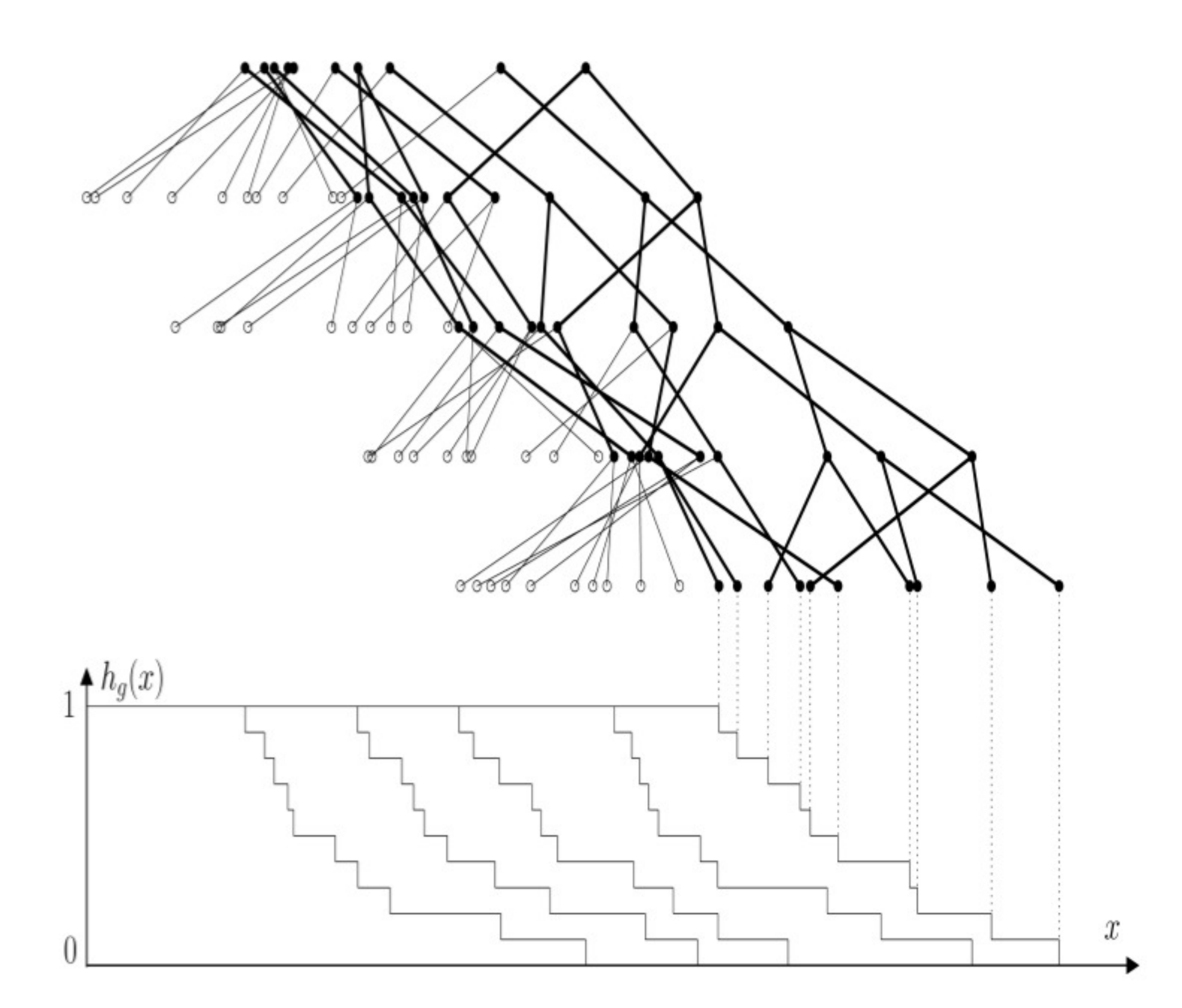}&
\includegraphics[width=.45\textwidth,angle=0]{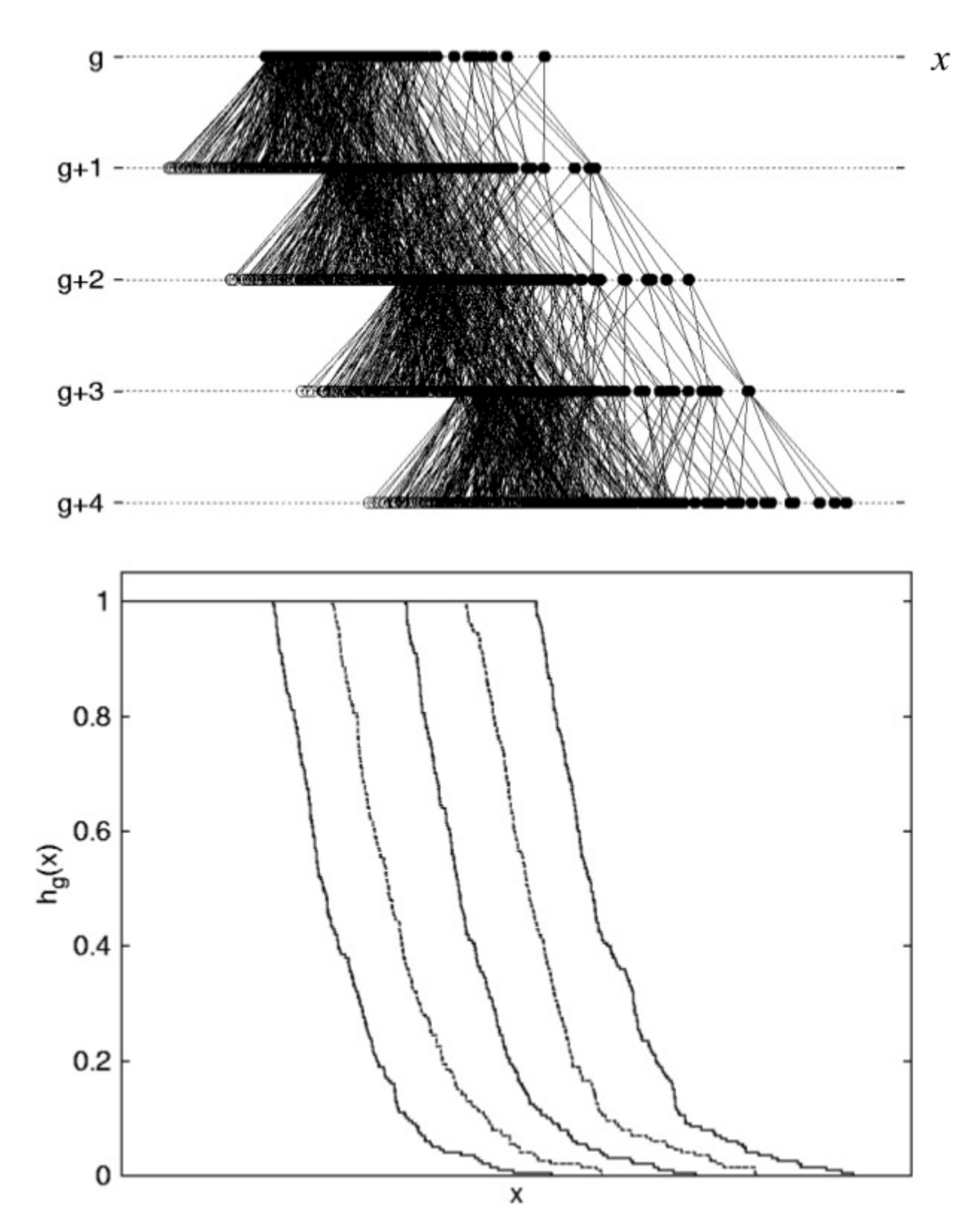}\\
\centerline{(a)} & \centerline{(b)}
\end{tabular}
\end{center}
\caption{\label{fig:popevol}
One realization of the population evolution model, for different
population sizes: (a) $N=10$, (b) $N=200$. In each case, the fitnesses of the individuals
are represented for 5 generations. The links indicate the genealogy.
The curves at the bottom of the figure represent
the fraction $h_g(x)$
of individuals which have fitness larger than $x$ at generation $g$.
(Arbitrary scale on the $x$-axis.)
}
\end{figure}

Realizations of such a model are represented in Fig.~\ref{fig:popevol},
in the case of a small population ($N=10$,  Fig.~\ref{fig:popevol}a) and also
for a larger population ($N=200$, Fig.~\ref{fig:popevol}b).
A function which exhibits traveling wave properties is $h_g(x)$, the fraction 
of individuals which have a fitness larger than $x$ at generation $g$.\footnote{%
Another interesting ``observable'' to study with these models
is the properties of the genealogies: Consider $k$
individuals chosen randomly at generation $g$, what are the statistical
properties of their most recent common ancestor?
This problem turns out to be intimately related to the propagation
of stochastic traveling waves, see Ref.~\cite{Brunet:2006zn}.
However, while it is an interesting problem in a biological context,
we have not found any application of genealogies to the QCD context so far.
}

We can make a few remarks looking at Fig.~\ref{fig:popevol}.
First, we see that 
the dispersion in fitness of the population remains of the same order
of magnitude throughout the evolution and its mean increases. These features
are of course due to the selection mechanism.
We also see that when the population increases, the curves representing $h_g(x)$
look smoother: The noise gets averaged due to the large number of objects 
(see Fig.~\ref{fig:popevol}b).
But actually, the stochasticity always remains significant in the low-density tip of
the front.

\subsubsection{Reaction-diffusion model}

\begin{figure}[h]
\begin{center}
\begin{tabular}{cc|c|c}
\multicolumn{2}{c}{Ordinary Brownian motion} & 
\multicolumn{1}{c}{Branching} &
\multicolumn{1}{c}{Recombination}
\\
\\
\includegraphics[width=.2\textwidth,angle=0]{jumpleft-eps-converted-to.pdf}\ \ &
\includegraphics[width=.2\textwidth,angle=0]{jumpright-eps-converted-to.pdf}\ \ &
\includegraphics[width=.2\textwidth,angle=0]{split-eps-converted-to.pdf}\ \ &
\includegraphics[width=.2\textwidth,angle=0]{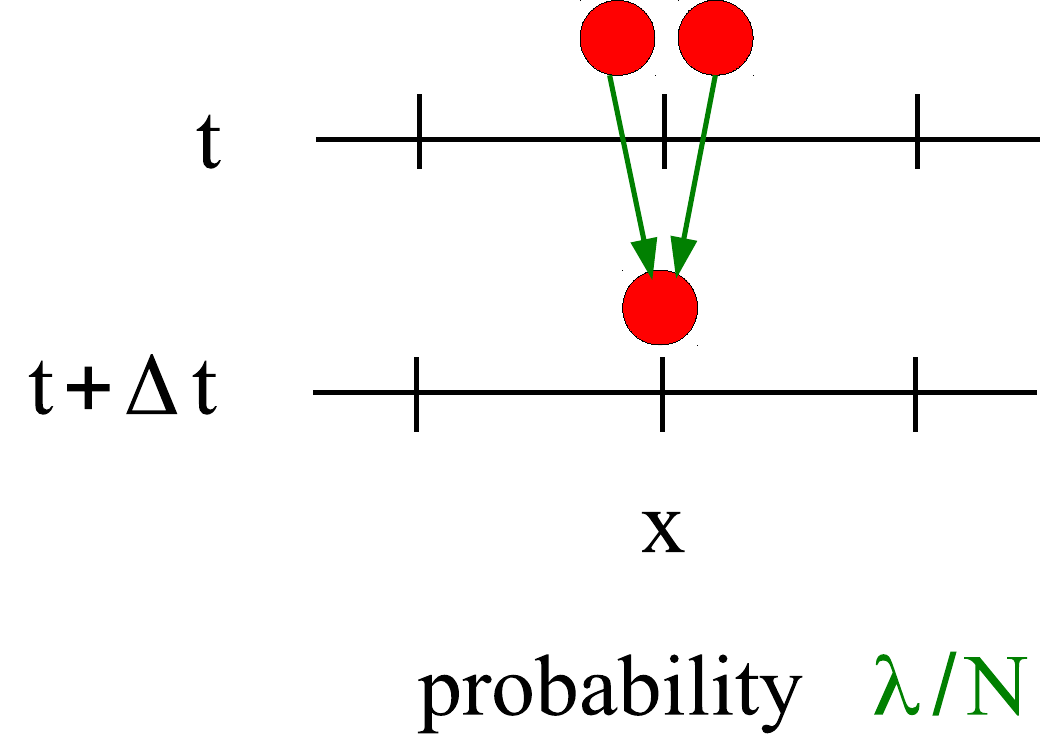}
\end{tabular}
\end{center}
\caption{\label{fig:elemRD}
Elementary processes defining the reaction-diffusion model.
}
\end{figure}

Let us come back to our branching random walk process on a lattice
introduced in Sec.~\ref{sec:introBRW}. We shall
just add a recombination process: Any pair of particles on site $x$ recombines
to one single particle with probability $\lambda/N$, where $N$ is a new parameter
(see Fig.~\ref{fig:elemRD}).
This is a reaction-diffusion model, which may apply to the context of chemical
reactions or of the spread of diseases.

The evolution equation for the average number of particles on site $x$ as time $t$
increases is easy to obtain. We assume a configuration $n(x,t)$ of particles at time $t$,
and write the equation for the average $\langle n\rangle$ at time $t+\Delta t$ knowning the configuration
at time $t$:
\begin{multline}
\langle n(x,t+\Delta t)\rangle_{[t,t+\Delta t]}=\left(1-2\mu+\lambda\right)n(x,t)
+\mu\left[
n(x+\Delta x,t)+n(x-\Delta x,t)
\right]\\
-\frac{\lambda}{N}n(x,t)\left[n(x,t)-1\right].
\end{multline}
The first term in the right-hand side accounts for the mean fraction $2\mu$ of particles
which leave the site $x$ due to diffusion and the mean fraction $\lambda$ which are added due to
particle splittings. The second term is a gain term due to diffusion from the nearby sites.
The last term is the mean number of particles which disappear due to recombination.

We now average over the full history of the stochastic process which leads to the configuration
$n(x,t)$
\begin{multline}
\langle n(x,t+\Delta t)\rangle-\langle n(x,t)\rangle
=\mu\left[
\langle n(x+\Delta x,t)\rangle+\langle n(x-\Delta x,t)\rangle-2\langle n(x,t)\rangle
\right]\\
+\lambda \langle n(x,t)\rangle
-\frac{\lambda}{N}\langle n(x,t)[n(x,t)-1]\rangle.
\end{multline}
In order to get a partial differential equation, we take the continuum limit
\be
\Delta t\rightarrow 0,\ \ \Delta x\rightarrow 0\ \ \text{with}\ \
\mu\frac{(\Delta x)^2}{\Delta t}=1,\ \ \lambda=\Delta t,
\ee
and we arrive at
\be
\partial_t \langle n\rangle=\partial_x^2 \langle n\rangle+\langle n\rangle-\frac{1}{N}
\langle n(n-1)\rangle.
\label{eq:reacdiff}
\ee
We observe that this is not a closed equation since the right-hand side has
a term of the form $\langle n^2\rangle$.
The most strightforward way to arrive at a closed equation is to assume
the factorization of this correlator: $\langle n^2\rangle=\langle n\rangle^2$.
This is a mean-field approximation: It consists in neglecting the fluctuations.
It is expected to be a good approximation when the number of particles gets large.
The above equation then boils down to
\be
\partial_t \langle n\rangle=\partial_x^2 \langle n\rangle+\langle n\rangle-\frac{1}{N}
\langle n\rangle^2,
\ee
where we have also neglected\footnote{%
Actually, replacing directly $\langle n(n-1)\rangle$ 
in Eq.~(\ref{eq:reacdiff})
by $\langle n\rangle^2$
is the so-called ``Poissonian approximation''.
} a term of order $1/N$. 

Defining the rescaled mean particle number
$u\equiv \langle n\rangle/N$, we arrive at
\be
\partial_t u=\partial_x^2 u+u-u^2,
\tag{\ref{eq:FKPPu}'}
\ee
which is of course again the FKPP equation.

\begin{beyond}
The full evolution equation for $n$ would be a stochastic partial
differential equation
with a complicated noise term.
There is an elegant formulation of reaction-diffusion processes
in terms of a partial differential equation with Gaussian multiplicative
noise (see e.g. Ref.~\cite{peliti1985}), 
but it requires the introduction of an abstract ``field''
$\phi(x,t)$ of coherent states.
(The moments of $\phi$ are related to the factorial moments of $n/N$).
$\phi$ solves an equation of the form
\be
\partial_t\phi=\partial_x^2\phi+\phi-\phi^2+\sqrt{\frac{1}{N}\phi(1-\phi)}\,\nu,
\label{eq:sFKPP}
\ee
where the field $\nu$ is a Gaussian white noise, defined by the correlators
\be
\langle\nu(x,t)\rangle=0,\ 
\langle\nu(x,t)\nu(x^\prime,t^\prime)\rangle=\delta(x-x^\prime)\delta(t-t^\prime).
\ee
One should specify that
Eq.~(\ref{eq:sFKPP}) has to be understood in the It$\bar{\text{o}}$ sense 
(see e.g. Ref.~\cite{gardiner2004handbook}).
\end{beyond}


\subsection{Properties of stochastic traveling waves}

Insight into stochastic traveling waves was developed in Refs.~\cite{PhysRevE.56.2597}
and~\cite{Brunet:2005bz}. Since our presentation here is
rather concise, we refer the reader to those papers for details and
to the review paper of Ref.~\cite{Munier:2009pc}.

\subsubsection{General considerations}

We first need to gain some intuition on stochastic traveling waves.
Thinking of the reaction-diffusion model on a lattice, it is clear that
in bins in which the number of particles is large, the evolution is
essentially deterministic, hence given by the corresponding equation in 
the FKPP universality class. 
We expect the noise to be important only in bins in which
the number of particles is of order unity.
So if $N$ is large, the mean-field approximation (i.e. the FKPP equation)
should have some validity, yet to be understood.

We observe that the main important 
property of stochastic fronts which is missed when going
to the infinite-$N$ limit is the fact that the number of particles on each lattice site
is not a continuous variable, but takes integer values, $0,1,2,\cdots$.
In particular, starting with a localized initial condition,
there must be a rightmost and a leftmost occupied site.
So the exponential shape of the front which solves asymptotically (for
large times) the FKPP equation, $u(x,t)\sim e^{-(x-X(t))}$,
cannot represent the (normalized) number of particles
in a given realization. The problem is most stringent 
in regions in which $u(x,t)<1/N$ (i.e. in which
the number of particles $n=N\times u$ would become a fraction 
of unity if it solved the deterministic FKPP equation).

From this remark, we may first try and guess the velocity of
the front describing the particle density in individual realizations, 
and eventually figure out a method of
taking into account discreteness.

We recall that in the (generalized) FKPP case,
the front velocity is tightly connected to its shape.
Starting from a localized initial condition, it reads
\be
\dot X(t)=\chi^\prime(\gamma_0)-\frac{3}{2\gamma_0 t},
\label{eq:vFKPP}
\ee
and this actually is the velocity of a front whose exponential 
shape $e^{-\gamma_0(x-X(t))}$ extends over a region of size 
$\sim\sqrt{\chi^{\prime\prime}(\gamma_0)t}$ (see the Gaussian factor in 
Eq.~(\ref{eq:gensolFKPPu})).
We have just argued that the exponential shape cannot be correct
when $u<1/N$. So the front must have a size $L$ which is such that
$u(X(t)+L,t)= 1/N$, which, taking into account
the fact that its shape is exponential,
gives
\be
L=\frac{1}{\gamma_0}\ln N.
\ee
Starting from a steep initial condition, the time $t_\text{diffusion}$ at which 
the exponential shape extends over the full allowed range $L$
is of order $L^2/\chi^{\prime\prime}(\gamma_0)$ (see again Eq.~(\ref{eq:gensolFKPPu})),
and at that time, from Eq.~(\ref{eq:vFKPP}), the front velocity reads
\be
\chi^\prime(\gamma_0)-\text{const}\times
\frac{3\chi^{\prime\prime}(\gamma_0)}{2\gamma_0 L^2}.
\ee
After this time, the front cannot extend any further, and so this should also be,
on the average, the asymptotic front velocity at large time.
The constant cannot be determined from this naive estimate, but
the parametric form should be correct.

Note that with respect to the asymptotic velocity of the deterministic FKPP front
($\chi^\prime(\gamma_0)$), the correction scales like $1/\ln^2 N$.
Naively, one would have expected a correction of the order of $1/N$
since taking into account discreteness amounts to cutting 
off a fraction $1/N$ of the tail of the front.
The correction we have just argued is much larger!

\begin{figure}
\begin{center}
\includegraphics[width=.95\textwidth,angle=0]{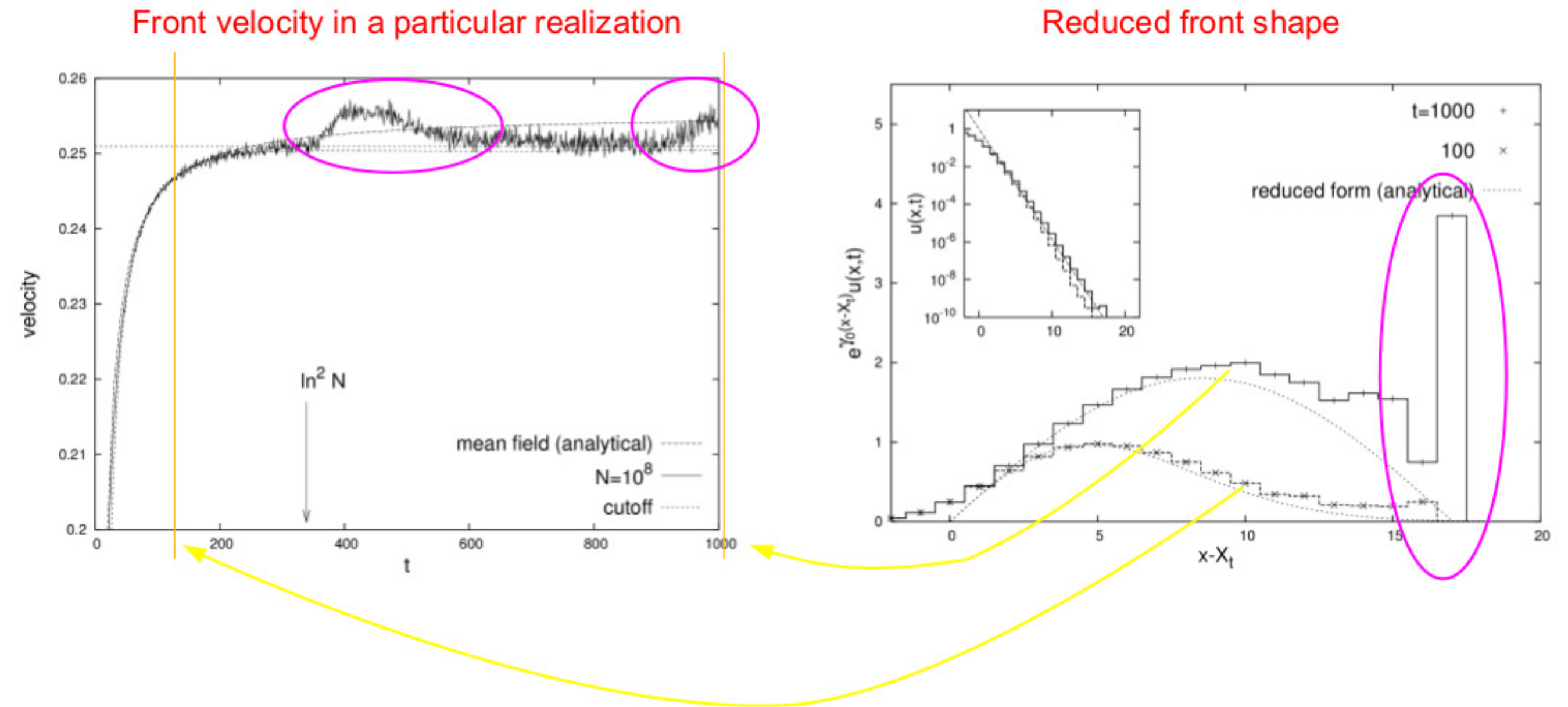}
\end{center}
\caption{\label{fig:beyondbknum1}
{\it Left:} Front velocity as a function of time for a particular realization of the
reaction-diffusion model.
{\it Right:} Reduced front shape in the frame of the wave for the same realization,
at two different times.
For $t\ll \ln^2 N$, we see that the reduced front shape is consistent with e.g.
Eq.~(\ref{eq:shapeFKPP}) while at larger $t$, the shape starts to look more
symmetric and large fluctuations occur at the tip.
}
\end{figure}

Support for the scenario just outlined can be found in numerical simulations.
Figure~\ref{fig:beyondbknum1} (left) represents the velocity of the front
in one particular realization of the simulation of a reaction-diffusion 
model (for all details, see Ref.~\cite{Enberg:2005cb}). 
We see that indeed, at some time $t\sim \ln^2 N\sim t_\text{diffusion}$, 
the increase of the front velocity
seems to stop, and except for small-amplitude short-term noise 
and for large but rare upward jumps,
the velocity becomes constant.
It also seems that when one reaches this ``constant'' velocity,
the shape of the reduced front (namely the front divided by the
exponential $e^{-\gamma_0 x}$, Fig.~\ref{fig:beyondbknum1} (right)) departs
from the shape predicted by the FKPP equation, see 
Eq.~(\ref{eq:shapeFKPP}).

We are going first to set up a precise calculation of the average front
velocity, and then come back to the large positive fluctuations in the velocity.

\subsubsection{Accounting for saturation and discreteness}

The simplest way to account for the fact that $u$ cannot exhibit the exponential
shape $e^{-\gamma_0 x}$ in regions in which $u<1/N$ is to put an appropriate cutoff, 
namely an absorptive boundary, in the tail.

To solve the FKPP equation, we already had an absorptive boundary instead of
the nonlinearity. Now we need a second cutoff to represent discreteness.
Since the asymptotic front has a length $L$, the second cutoff
sits at a fixed distance $L$ of the first one. 
In between, the evolution equation is linear and deterministic.

Hence we want to solve the linear equation
\be
\partial_t u=\partial_x^2 u+u
\ee
with the two boundary conditions
\be
u(X(t),t)=0,\ \ u(X(t)+L,t)=0,
\label{eq:boundaryu}
\ee
and some localized initial condition.
(Its precise form is not relevant since it turns out that it will be
``forgotten'' after a sufficiently large time).
An appropriate ansatz is
\be
u(x,t)=e^{-\xi}h\left(\frac{\xi}{L},\frac{t}{L^2}\right)
\ee
where $\xi=x-X(t)$.
Indeed, we know that the dominant shape of the front is a decreasing exponential,
that is why we factorized $e^{-\xi}$. Furthermore, the natural distance scale
in the problem is $L$, and the natural time scale is the diffusion time over such
a distance, namely $L^2$.
Let us introduce the dimensionless variables $\rho\equiv \xi/L$ and $\tau\equiv t/L^2$.
Then, in terms of these new variables,
\be
\begin{split}
\partial_t u&=\left[\dot X\left(h-\frac{1}{L}\partial_\rho h\right)
+\frac{1}{L^2}\partial_\tau h\right]e^{-\xi}\\
\partial_x u&=\left[-h+\frac{1}{L}\partial_\rho h\right]e^{-\xi}\\
\partial_x^2 u&=\left[h+\frac{1}{L^2}\partial_\rho^2 h-\frac{2}{L}\partial_\rho h\right]e^{-\xi},
\end{split}
\ee
hence
\be
\partial_\tau h=\partial_\rho^2 h+L(\dot X-2)\partial_\rho h+L^2(2-\dot X)h.
\label{eq:evolh}
\ee
In order to have a nontrivial stationary solution ($\partial_\tau h=0$), we need to make sure that 
the terms proportional to $h$ and $\partial_\rho^2 h$ have a coefficient of order one.
Recall that $L$ is a large parameter:
We must thus set $\dot X=2-\frac{c^2}{L^2}$ where $c$ is an undetermined constant so far.
Then, the term proportional to $\partial_\rho h$ becomes negligible.
Equation~(\ref{eq:evolh}) boils down to
\be
\partial_\tau h=\partial_\rho^2 h+c^2 h.
\ee
A stationary solution obviously solves the second-order differential equation
\be
\partial_\rho^2 h+c^2 h=0,
\ee
whose solution reads (for $c>0$)
\be
h(\rho,\tau)=A\sin c\rho+B\cos c\rho,
\ee
where $A$ and $B$ are arbitrary constants.
Compatibility with the boundary conditions~(\ref{eq:boundaryu}) requires $c=\pi$ and $B=0$.

Going back to the initial variables, we have thus found
\be
u(x,t)\propto e^{-(x-X(t))}\ln N
\sin
\left[
\frac{\pi}{\ln N}\left(x-X(t)\right)
\right],
\ \ \text{where}\ \
X(t)=\left(2-\frac{\pi^2}{\ln^2 N}\right)t.
\label{eq:solbranchingdiffusion}
\ee
Now for a generic model whose branching diffusion kernel is characterized by
the eigenvalue $\chi(\gamma)$:
\be
u(x,t)\propto  e^{-\gamma_0(x-X(t))}\frac{\ln N}{\gamma_0}
\sin
\left[
\frac{\pi\gamma_0}{\ln N}\left(x-X(t)\right)
\right],\
\text{where}\ \ 
X(t)=V_{\text{BD}} t,
\ee
and we introduced the front velocity
\be
V_{\text{BD}}\equiv 
\chi^\prime(\gamma_0)-\frac{\pi^2\chi^{\prime\prime}(\gamma_0)}
{2\gamma_0 L^2},\ \text{where}\ L=\frac{1}{\gamma_0}\ln N.
\label{eq:VBD}
\ee
(The subscript ``BD'' stands for ``Brunet-Derrida'').
As before, $L$ is the size of the front, namely in the calculation, 
up to an additive constant,
the distance between the two absorptive boundaries.

\subsubsection{Beyond the deterministic equations: modeling noise at the tip}

So far, we have replaced the stochastic evolution equation by a deterministic
equation with two cutoffs: one for unitarity, ensuring that $n\leq N$, the other
one for discreteness, ensuring, in some sense that $n\geq 1$, or more precisely,
that $n$ represents indeed a number of particles.
Now the full problem is stochastic. Finite-$N$ corrections should reflect
more precisely the stochasticity.
The question we shall address in this paragraph is how to go beyond
the Brunet-Derrida cutoff.

We know that stochasticity plays a role in the tail, 
where the density of particles is low.
Our basic assumption is that the first correction beyond the cutoff,
which in some way enforces discreteness,
is well represented by
a single particle randomly sent a distance $\delta$ 
ahead of the deterministic tip of the front, at a rate $p(\delta) d\delta$.
Except for this stochastic process,
all evolution is assumed
to be deterministic. In particular, once this particle
is randomly produced, its further time evolution
is purely deterministic.

Let us imagine that a fluctuation occurs at time $t$. The position of the
tip at $t$ is
\be
X_\delta(t)=X(t)+\delta,
\ee
and at later time $t+\Delta t$, after the fluctuation has evolved into a front,
\be
X_\delta(t+\Delta t)=X_\delta(t)+V_\text{BD}\Delta t-3\ln L.
\ee
The last negative term is a ``delay'' induced by the formation of the front,
and due to the fact that until times of the order of $\Delta t\sim L^2$, the
front velocity differs from $V_\text{BD}$ by $-3/(2\Delta t)$.
The front without fluctuation has just translated by $V_\text{BD}\Delta t$:
\be
X_\delta(t+\Delta t)=X(t)+V_\text{BD}\Delta t.
\label{eq:frontwithoutfluct}
\ee
Now the shape of the front in the forward part is essentially exponential.
The front after the fluctuation has relaxed (at time such that $\Delta t\gg L^2$)
is the sum of the front without fluctuation translated at time $t+\Delta t$
at the constant velocity $V_\text{BD}$,
and the front originated from the fluctuation:
\be
\begin{split}
e^{-(x-X_\text{tot}(t+\Delta t))}&=e^{-(x-X(t+\Delta t))}+C e^{-(x-X_\delta(t+\Delta t))}\\
&=e^{-(x-X(t)-V_\text{BD}\Delta t)}\left(1+C e^{\delta-3\ln L}\right).
\end{split}
\ee
We find that the position of the front with the fluctuation reads
\be
X_\text{tot}(t+\Delta t)=X(t)+V_\text{BD}\Delta t
+R(\delta),\ \ \text{where}\ \ R(\delta)\equiv\ln\left(1+C\frac{e^\delta}{L^3}\right).
\label{eq:XtotRdelta}
\ee
$R(\delta)$ just introduced is the additional shift of the front position induced by
a forward fluctuation.

In order to be able to compute the effect of these fluctuations, we still need
to know the rate at which the forward fluctuations occur, and their distribution in $\delta$.
It is natural to assume that the latter is exponential $e^{-\delta}$, 
and thus we shall conjecture
the rate
\be
p(\delta)d\delta=C_1 e^{-\delta}d\delta,
\ee
where $C_1$ is an unknown constant.

With these elements, we can write the following effective theory for the evolution of $X(t)$:
\be
X(t+\Delta t)=
\begin{cases}
X(t)+V_\text{BD}\Delta t & \text{probability}\ \ 1-\Delta t\int_0^{+\infty}d\delta\,p(\delta)\\
X(t)+V_\text{BD}\Delta t+R(\delta) & \text{probability}\ \ \Delta t\, d\delta\, p(\delta).
\end{cases}
\label{eq:rules}
\ee
The generating function for the cumulants of $X(t)$ is defined by
\be
G(\lambda,t)=\ln\left\langle e^{\lambda X(t)}\right\rangle.
\ee
Let us write its time evolution:
\be
\begin{split}
G(\lambda,t+\Delta t)&=\ln\left\langle e^{\lambda X(t+\Delta t)}\right\rangle\\
&=\begin{multlined}[t][10cm]\ln\left\langle
\left[
\Delta t\int_0^{+\infty}d\delta\,p(\delta)e^{\lambda \left(X(t)+V_\text{BD}\Delta t+R(\delta)\right)}\right.\right.\\
\left.\left.+\left(
1-\Delta t\int_0^{+\infty}d\delta\,p(\delta)
\right)
e^{\lambda \left(X(t)+V_\text{BD}\Delta t\right)}
\right]
\right\rangle
\end{multlined}
\\
&=\ln\left\langle
e^{\lambda X(t)}
\right\rangle
+\lambda V_\text{BD}\Delta t
+\ln\left[1+\Delta t\int_0^{+\infty}d\delta\,p(\delta)
\left(
e^{\lambda R(\delta)}-1
\right)
\right].
\end{split}
\ee
The average over the processes occurring in the time interval $[t,t+\Delta t]$ is
done using the rules~(\ref{eq:rules}), applied to go from the first to the second line.
The remaining brackets $\langle\cdot\rangle$ represent the average over the time
interval $[0,t]$.

The first term in the right-hand side of the previous equation is, by definition, 
nothing but $G(\lambda,t)$.
Now we take the $\Delta t\rightarrow 0$ limit,\footnote{%
The reader may see a contradiction in $\Delta t$ being
infinitesimal here, while we said earlier that $\Delta t\gg L^2\gg 1$ 
(see after Eq.~(\ref{eq:frontwithoutfluct})).
Actually, this is justified because the fluctuations which contribute to the
shift of $X(t)$ turn out to occur every $L^3$ steps of time, and $L^3\gg L^2$, since
$L\equiv\ln N$ is assumed a large number.
} in which we can expand the logarithm in the
last term and write $G(\lambda,t+\Delta t)=G(\lambda,t)+\Delta t\,\partial_t G(\lambda,t)$.
We arrive at the equation
\be
\frac{\partial G(\lambda,t)}{\partial t}
=\lambda V_\text{BD}+\int_0^{+\infty}d\delta\,p(\delta)
\left(
e^{\lambda R(\delta)}-1
\right),
\ee
which is trivial to integrate.
One expands the result in powers of $\lambda$ to get the cumulants of $X(t)$.
For large $t$ (which enables us to neglect the unknown integration constant),
\be
\begin{split}
\frac{
\left\langle
\left[X(t)\right]^n
\right\rangle_c
}{t}
&=\delta_{n,1}V_\text{BD}+\int_0^{+\infty}d\delta\,p(\delta)\left[R(\delta)\right]^n\\
&=\delta_{n,1}V_\text{BD}+C_1\int_0^{+\infty}d\delta\,e^{-\delta}
\ln^n\left(1+C\frac{e^\delta}{L^3}\right).
\end{split}
\ee
The integral over $ \delta$ is performed by the change of variable 
$\delta\rightarrow x\equiv L^3 e^{-\delta}/C$:
\be
\begin{split}
C_1\int_0^{+\infty}d\delta\,e^{-\delta}\ln^n\left(1+C\frac{e^\delta}{L^3}\right)
&=\frac{CC_1}{L^3}\int_0^{L^3/C}dx\,\ln^n\left(1+\frac{1}{x}\right)\\
&=\frac{CC_1}{L^3}n!\zeta(n)+{\cal O}(1/L^6),
\end{split}
\ee
where $\zeta(n)$ is the Euler Zeta function.
Note that the large-$L$ expansion in the last line can be performed only for $n\geq 2$.

Keeping the leading term when $L\gg 1$ (which is obtained simply by setting the
upper bound of the integral over $x$ to $+\infty$), we get
\be
\frac{\left\langle
\left[X(t)\right]^n
\right\rangle_c}{t}
=CC_1 n!\zeta(n)\frac{1}{L^3}.
\label{eq:cumulants0}
\ee
$CC_1$ is an overall constant, the same for all cumulants.

We need to address the case $n=1$ (first moment of $X(t)$) separately, since in this
case, the integral over $x$ is logarithmic and thus the upper bound cannot be
sent to infinity:
\be
\frac{\langle X(t)\rangle}{t}=V_\text{BD}
+\frac{CC_1}{L^3}\int_0^{L^3/C}dx\,\ln\left(1+\frac{1}{x}\right).
\ee
The integral is easy to perform. In the limit of large $L$, the leading term
just reads $\ln (L^3)$, in such a way that
\be
\frac{\langle X(t)\rangle}{t}=V_\text{BD}
+{CC_1}\frac{3\ln L}{L^3}.
\label{eq:cc1invelocity}
\ee
The constant $CC_1$ appears also here.

We still need to determine this constant. We have not found a way to
compute it, but we can try and guess it.
For completeness, let us briefly sketch the argument.

We write the expression for the mean displacement rate of the front
due to fluctuations only:
\be
\int d\delta p(\delta)R(\delta)=
\int d\delta
e^{-\delta}\ln\left(1+C\frac{e^\delta}{L^3}\right).
\ee
We see that as long as $\delta\ll 3\ln L$, then 
the integrant may be approximated by $C/L^3$, which has no $\delta$-dependence,
while for $\delta\gg 3\ln L$, the integrant is cut off exponentially.
Hence is seems that effectively, the fluctuations extend the front
by $3\ln L$, and so the total effective size of the front reads
\be
L_{\text{eff}}=L+3\ln L.
\ee
As we already commented, since $X(t)/t$ with $X(t)$ from 
Eq.~(\ref{eq:solbranchingdiffusion}) may be interpreted
as the velocity of a front of length $L$, if we replace $L$ by $L_\text{eff}$ therein,
we get
\be
V\equiv
\frac{X(t)}{t}=2-\frac{\pi^2}{L_\text{eff}^2}
\underset{L\gg 1}{\sim}
2-\frac{\pi^2}{L^2}+6\pi^2\frac{\ln L}{L^3}+\cdots
\ee
and identifying the result of the expansion to Eq.~(\ref{eq:cc1invelocity})
leads to the determination $CC_1=2\pi^2$, and thus of all cumulants, see
Eq.~(\ref{eq:cumulants0}).
For a generic model, $CC_1=\pi^2\chi^{\prime\prime}(\gamma_0)$, and thus
the front velocity and cumulants of its position read
\be
\begin{split}
V&=\chi^\prime(\gamma_0)-\frac{\pi^2\gamma_0\chi^{\prime\prime}(\gamma_0)}{2\ln^2 N}
+\pi^2\gamma_0\chi^{\prime\prime}(\gamma_0)\frac{3\ln \ln N}{\ln^3 N},\\
\frac{\text{[$n$-th cumulant]}}{t}&=\pi^2\gamma_0^2\chi^{\prime\prime}(\gamma_0)
\frac{n!\zeta(n)}{\gamma_0^n\ln^3 N}.
\end{split}
\label{eq:stochafrontfinal}
\ee
We notice that all cumulants are of order $t/\ln^3 N$, which means that they
are small for $t\ll \ln^3 N$: Up to times of this order of magnitude,
the traveling wave behaves ``deterministically''. (Only the velocity
differs from the FKPP velocity already for times $t>\ln^2 N$: Indeed,
at $t\sim \ln^2N$, the front velocity becomes constant and equal to $V_\text{BD}$).
Hence $\ln^3 N$ is a new time scale, generated by the fluctuations.

At such times, different realizations of the evolution have different front positions:
The dispersion is related to the second-order cumulant, namely it is of the order
of $\sqrt{t/\ln^3 N}$.
\begin{figure}
\begin{center}
\includegraphics[width=.7\textwidth,angle=0]{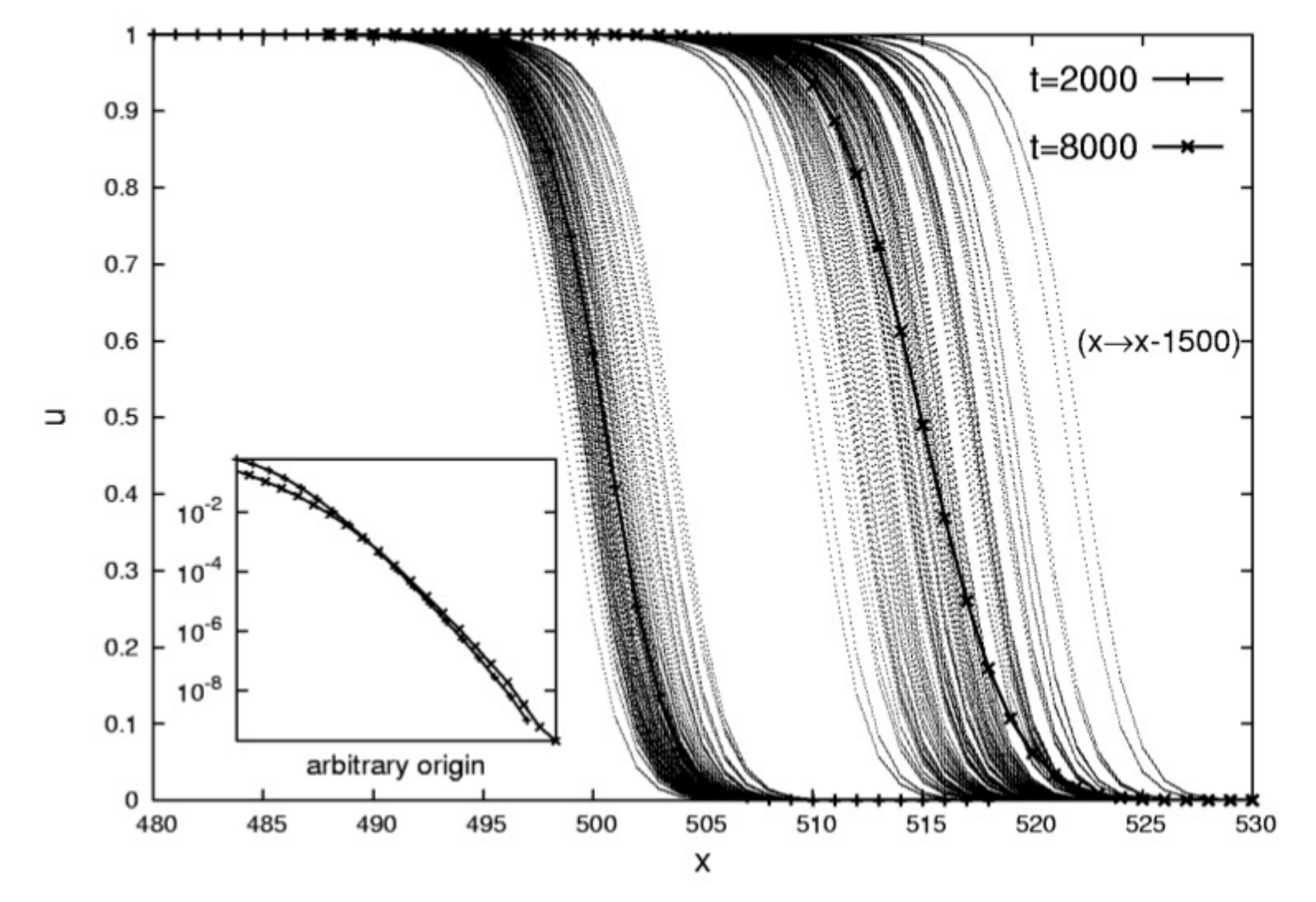}
\end{center}
\caption{\label{fig:beyondbknum2}
Front shape for several realizations of the reaction-diffusion model at two different times.
{\it Inset:} Comparison of the average front shapes at these different times, to illustrate
the property of diffusive scaling.
[Plot from Ref.~\cite{Enberg:2005cb}; see therein for all details on the
simulated model].
}
\end{figure}
This is very well seen in a numerical simulation of reaction-diffusion models, see
Fig.~\ref{fig:beyondbknum2}.
There is an interesting consequence of this fact:
The average of $u(x,t)$ over the noise does not depend on the geometric scaling variable
$x-\langle X(t)\rangle$, but on a different variable:
\be
\langle u(x,t)\rangle={\cal U}
\left(
\frac{x-\langle X(t)\rangle}{\sqrt{t/\ln^3 N}}
\right).
\label{eq:DS}
\ee
This is again seen in Fig.~\ref{fig:beyondbknum2}, see the inset.
This new scaling which replaces geometric scaling 
at very large times is sometimes called
``diffusive scaling''.
We leave the proof as an exercise for the reader.

\begin{ex}
Prove the diffusive scaling pattern~(\ref{eq:DS}) for $u(x,t)$
averaged over the realizations.
Remember that each realization of the stochastic evolution
looks like a deterministic traveling wave with shape which
may be approximated by
\be
u(x,t)=
\theta(x-X(t))e^{-\gamma_0 x}
+\theta(X(t)-x),
\ee
and that $X(t)$ is a stochastic variable whose distribution,
for the purpose of this calculation,
can be approximated by a Gaussian of width $\sim\sqrt{t/\ln^3 N}$.
\end{ex}


\subsection{Applications to QCD: Beyond BK}

We have argued that in the context of QCD, a nonlinear mechanism 
(something like gluon recombination?) should 
supplement the dipole model to tame the exponential growth of the dipole/gluon
density with the rapidity as soon as the .
The equations which would describe such effects are not yet known for sure.

We have just analyzed most generally reaction-diffusion models, which are
very similar to the dipole model for the linear part (diffusion and
exponential growth of the number of objects), supplemented with recombination
which, on the average, 
effectively limits the density of particles to~$N$.
The results we obtained for the front position and for the shape of the
particle density depend only on the branching-diffusion process
(through the eigenvalues $\chi(\gamma)$) and in $N$.

So it is enough to use the dictionary~(\ref{tab:mapping}) to which
one adds the QCD quantity corresponding to $N$, namely $1/\alpha_s^2$,
in order to be able to 
take over the results found above for the front position~(\ref{eq:stochafrontfinal})
and particle density profile~(\ref{eq:DS})
in reaction-diffusion processes
to scattering amplitudes in QCD.
We therefore arrive at predictions for the rapidity-dependence of
the saturation scale, and
we also predict a departure from geometric scaling at very high rapidities.

\begin{recap}
We have argued that there should be
some kind of nonlinearity which effectively limits
the density of gluons (or equivalently of dipoles) 
in the quantum evolution 
to $\sim 1/\alpha_s^2$.
The precise mechanism is not known in QCD, but one may get an idea of the effects
of such nonlinearities in simple branching-diffusion models with recombination
(reaction-diffusion, or population evolution).
We found that realizations of such models are {\it stochastic} traveling waves.
Their position is now a random variable. We were able to compute all its cumulants,
which depend only on a few parameters and not on the details of the recombination
mechanism: This universality enables one to take over the results obtained
in generic reaction-diffusion models to QCD, although the proper evolution 
equation has not been derived in QCD.
We found that the main physical consequence on the QCD amplitudes
is the substitution of geometric scaling by diffusive
scaling at ultrahigh rapidities.
\end{recap}

\clearpage


\section{Conclusion}

\subsection{Summary: the big picture}

Let us summarize the picture to which we have arrived in these lectures for
the rapidity evolution of scattering amplitudes. 

We were concerned essentially
with dipole-nucleus scattering for which the BK equation seems firmly established,
but in the last section, we turned also to dipole-dipole scattering.

We were able to identify three (well-)separated rapidity scales,
which delimitate 4 regimes:
\setlength{\unitlength}{1cm}
\begin{center}
\boxed{
\begin{picture}(14,3)
\put(13.2,1.7){$y$}
\put(1,2){\vector(1,0){12}}
\put(3,1.3){$y_\text{sat}=\frac{1}{\bar\alpha}\ln\frac{1}{\alpha_s^2}$}
\put(3.8,1.9){\line(0,1){0.2}}
\put(6,1.3){$y_\text{diff}=\frac{1}{\bar\alpha}\ln^2\frac{1}{\alpha_s^2}$}
\put(6.8,1.9){\line(0,1){0.2}}
\put(9,1.3){$\frac{1}{\bar\alpha}\ln^3\frac{1}{\alpha_s^2}$}
\put(9.5,1.9){\line(0,1){0.2}}
\put(1.5,.5){Region I}
\put(4.5,.5){Region II}
\put(7.25,.5){Region III}
\put(10.5,.5){Region IV}
\end{picture}
}
\end{center}
\begin{itemize}

\item {\it Region I: Low-density region.}
The BFKL equation is valid since the gluon density is low.
Of course, it applies both to dipole-dipole and dipole-nucleus scattering.

\item {\it Region II: High density.}
When the rapidity is higher than $y_\text{sat}$, nonlinear effects
set in. In the dipole-nucleus case, for $y_\text{sat}\ll y_\text{diff}$,
the latter correspond to {\it independent} multiple scatterings between
the evolved dipole and the target. They are described by the BK equation.
In the dipole-dipole case instead, the BK equation cannot be established since
the nonlinear effects are to be included in the evolution itself.
The right equation may be something like a ``stochastic BK equation''.
However, it seems that in this region, the scattering amplitude has the same properties as if it
were a solution of the BK equation.
In particular, it exhibits geometric scaling.

\item {\it Region III: Modified saturation scale.}
The BK equation breaks down at this point also in the diple-nucleus case.
The saturation scale becomes independent of the rapidity.
Geometric scaling still holds:
More precisely, it seems that the scaling variable is the same, but
the precise shape of the amplitude is different in the dipole-dipole
and dipole nucleus cases, see the recent work of ours, Ref.~\cite{Mueller:2014fba}.

\item{\it Region IV: Diffusive scaling.}
One enters a regime dominated by fluctuations, which manifest
themselves in the form of a new scaling form for the amplitude, ``diffusive
scaling''.
This holds both for the
dipole-dipole and the dipole-nucleus
amplitudes.

\end{itemize}

\subsection{\label{sec:history}Historical note}

Our presentation of these topics may lead one to think that geometric scaling
was predicted from the mathematics exposed here, and then found in the data.
Actually, the story went almost exactly the other way round.
Let us briefly sketch the main steps which led to the understanding of
QCD amplitudes at very high energies that we have explained here.
(Of course, we do not claim exhaustivity).

The Balitsky-Kovchegov equation was first established in 1996~\cite{Balitsky:1995ub}, 
and rederived
in 1999 in the context
of the dipole model~\cite{Kovchegov:1999yj,Kovchegov:1999ua}. 
Until year 2000, no one knew how to solve it.
In the meantime, Golec-Biernat and W\"usthoff proposed a saturation model
\cite{GolecBiernat:1999qd,GolecBiernat:1998js}
which described very well virtually all HERA data in the small-$x$ regime.
Geometric scaling was accidentally postulated in this model, a fact which was
noticed by Sta\'sto, Golec-Biernat and Kwieci\'nski and subsequently discovered
in the data~\cite{Stasto:2000er}.
Attemps to derive geometric scaling from QCD were made in the next few years,
first through numerical works, and then 
analytically~\cite{GolecBiernat:2001if,Iancu:2002tr,Mueller:2002zm}.
(Actually, the form of the rapidity dependence of the saturation scale, namely 
what we related in these lectures to the velocity of the traveling wave 
$\bar\alpha_s \chi(\gamma_0)/\gamma_0$, 
was known much before from the solution of the
first equation for saturation derived from physical arguments by Gribov,
Levin, Ryskin \cite{Gribov:1984tu}) (and in in the double-leading logarithmic
approximation by Mueller and Qiu~\cite{Mueller:1985wy}).
The interpretation of geometric scaling as FKPP traveling waves came
only after~\cite{Munier:2003vc,Munier:2003sj}.
The first attempt to go beyond the BK equation was achieved in
Ref.~\cite{Mueller:2004sea}, and the result obtained there was then
recognized to stem also from the stochastic FKPP equation and to be
related to the discreteness of quanta in Ref.~\cite{Iancu:2004es}.

\subsection{Concluding remarks and prospects}

We recognized that the BK equation,
which governs the rapidity/energy evolution of QCD amplitudes in
the high-energy limit, belongs to a large universality class,
whose simplest representative is the FKPP equation.

Essentially, this holds because parton evolution is a peculiar branching diffusion
process. This is likely to be a very general statement, beyond the particular
realization of parton evolution (namely the color dipole model) we have been focusing
on in these lectures.

This identification is useful because many of the main properties of traveling
waves are universal: They can be understood on simple toy models, and the obtained
results can then simply be taken over to QCD.

From the mathematical point of view, we are trying to understand the properties of solutions
(or better, realizations) of
nonlinear (stochastic) partial differential equations.
Since the latter appear in many different fields, any progress in this direction
may have numerous potential applications.

From the physical point of view, this link between QCD and more general mathematical problems
can help to understand the very essence of saturation in QCD, and also to
learn how to go beyond the BK equation.
It sets a general framework for understanding saturation effects,
which are conceptually interesting, and are likely to play an important
role for the phenomenology at the LHC.


\section*{Acknowledgements}

I warmly thank the organizers 
Prof.~Xin-Nian Wang, Prof.~Bo-Wen Xiao and Prof.~Guang-You Qin
for the support, for
the welcome in Wuhan, and for the perfect organization of the school,
as well as the students and colleagues who attended the lectures
for their interest and for their questions.
I also thank Dr. E.~Petreska for her reading of these notes.


\clearpage


\appendix

\section{\label{sec:compint}
Computation of the complex integral which appears in the BFKL eigenvalue problem}

In this section, we shall compute the integral
\begin{equation}
I=\int\frac{dzd\bar z}{2i}
z^{\alpha-1}
\bar z^{\tilde\alpha-1}
(1-z)^{\beta-1}
(1-\bar z)^{\tilde\beta-1}.
\end{equation}
Such integrals appear
in the context of 
various problems involving
conformal field theory, and the computation below may be
found in different places in the literature
(see e.g.~\cite{DiFrancesco:1997nk} for a textbook, or Ref.~\cite{Xiao:2007te} 
Appendix~A, or \cite{Ciafaloni:2010vh} or Ref.~\cite{Geronimo:2003:CIA}
for a more general integral of this type).

For $I$ to be well defined, the integrand must have trivial monodromies
around the singularities at $z=0,1$. This is the case
if
$\alpha-\tilde\alpha$, $\beta-\tilde\beta$
are integer numbers.
We shall restrict ourselves to real exponents, which
simplifies the discussion and is enough for the purposes of
this paper.
Furthermore, the integral converges only if
$\text{Re}(\alpha+\tilde\alpha)>0$, $\text{Re}(\beta+\tilde\beta)>0$
(at $z=0,1$) and 
$\text{Re}(\alpha+\tilde\alpha+\beta+\tilde\beta)<2$
(at $|z|\rightarrow\infty$).

Our calculation is a heuristic way to arrive at an expression for this
integral in terms of known functions. 

The first step is to write $I$ as a double integral over real variables.
Defining $z\equiv x+iy$, one gets
\be
I=\int_{-\infty}^{+\infty}dx
\int_{-\infty}^{+\infty}dy
(x+iy)^{\alpha-1}
(x-iy)^{\tilde\alpha-1}
(1-x-iy)^{\beta-1}
(1-x+iy)^{\tilde\beta-1}.
\ee
Then, one performs a Wick rotation 
$y\rightarrow e^{i(\pi/2-2\varepsilon)}y\simeq i(1-2i\varepsilon)y$,
where the term proportional to $\varepsilon$ 
hampers that the 
integration path go along the branch cuts.
We get
\begin{multline}
I=i\int_{-\infty}^{+\infty}dx
\int_{-\infty}^{+\infty}dy
(x-y+2i\varepsilon y)^{\alpha-1}
(x+y-2i\varepsilon y)^{\tilde\alpha-1}
(1-x+y-2i\varepsilon y)^{\beta-1}\\
\times(1-x-y+2i\varepsilon y)^{\tilde\beta-1}.
\end{multline}
Next, the change of variables
$X_+=x+y$,
$X_-=x-y$
casts the integral in the form
\begin{multline}
I=-\frac{i}{2}
\int_{-\infty}^{+\infty}
dX_+
\int_{-\infty}^{+\infty}
dX_-
[X_- + i\varepsilon(X_+ - X_-)]^{\alpha-1}
[X_+ - i\varepsilon(X_+ - X_-)]^{\tilde\alpha-1}\\
\times[1 - X_- - i\varepsilon(X_+ - X_-)]^{\beta-1}
[1 - X_+ + i\varepsilon(X_+ - X_-)]^{\tilde\beta-1}.
\end{multline}
The integration over $X_+$ may be written as a sum of contributions
from the integration domains
$]-\infty,0[$, $]0,1[$, $]1,+\infty[$.
The position of the branch points in the $X_-$ plane 
with respect to the integration contour is then specified:
\begin{equation}
X_-=-i\varepsilon X_+\ ,\ \
X_-=1-i\varepsilon (X_+ - 1).
\end{equation}
Note however that the $X_-$ contour crosses the cuts.
For example for $X_+\in]-\infty,0[$, there is a cut 
along the negative real axis in the $X_-$ plane, which
intersects the contour
at $X_-=X_+$:
The branch point at $0$ is in the upper-$X_-$ plane, but 
the cut then goes to the lower plane when $X_-<X_+$.
But as we shall see, this is not
a problem since the initial integral is well defined.
Let us write the contribution of the integration region
$(X_+,X_-)\in]-\infty,0[^2$ as
\begin{multline}
I_{11}=-\frac{i}{2}\int_{-\infty}^0 dX_+\bigg[
\int_{-\infty}^{X_+}dX_-(X_-+i\varepsilon)^{\alpha-1}
(X_+-i\varepsilon)^{\tilde\alpha-1}
f(X_+,X_-)\\
+\int_{X_+}^{0}dX_-(X_--i\varepsilon)^{\alpha-1}
(X_++i\varepsilon)^{\tilde\alpha-1}
f(X_+,X_-)
\bigg],
\end{multline}
where $f(X_+,X_-)$ gathers the remaining factors, which are real
on the contours of integration considered here.
Since
\begin{multline}
\int_{-\infty}^0 dX_+
\int_{-\infty}^{X_+}dX_- 
(X_-+i\varepsilon)^{\alpha-1}
(X_+-i\varepsilon)^{\tilde\alpha-1}
f(X_+,X_-)\\
=e^{-2i\pi(\alpha-\tilde\alpha)}
\int_{-\infty}^0 dX_+
\int_{-\infty}^{X_+}dX_- 
(X_--i\varepsilon)^{\alpha-1}
(X_++i\varepsilon)^{\tilde\alpha-1}
f(X_+,X_-),
\end{multline}
and since $\alpha-\tilde\alpha$ is an integer, 
the contributions of the discontinuity at $X_+=X_-$
cancel between the two integrations and
the cut may
safely be kept in the upper plane for all relevant
values of $X_+$ and $X_-$.
The same must be true also for the other cases.

\begin{figure}[h]
\begin{center}
\begin{tabular}{c|c|c}
\includegraphics[width=0.3\textwidth]{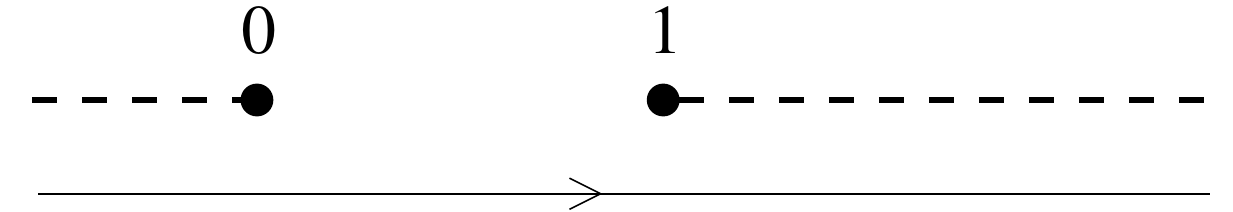}&
\includegraphics[width=0.3\textwidth]{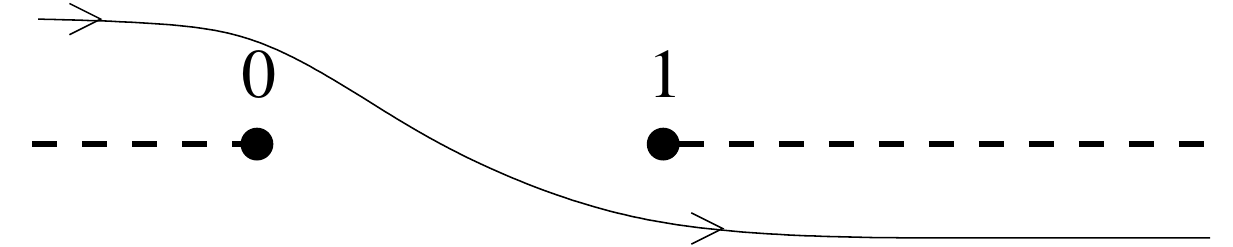}&
\includegraphics[width=0.3\textwidth]{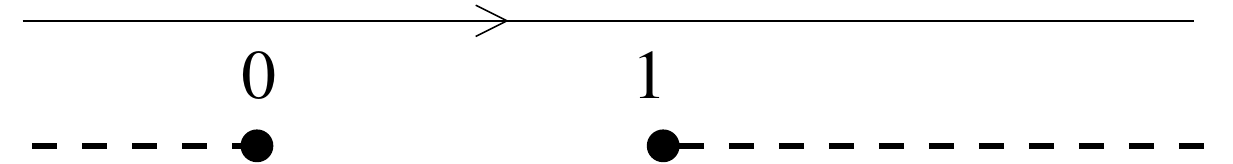}
\\
$X_+<0$&$0<X_+<1$&$X_+>1$
\end{tabular}
\end{center}
\caption{
\label{fig:contours}
Integration contours for the $X_-$ variable
corresponding to the different possible values of
the $X_+$ variable.}
\end{figure}

The contours in the $X_-$ plane are shown 
in Fig.~\ref{fig:contours} in the different
ranges of $X_+$.
We see that only the 
integral over $X_-$ 
in the domain $X_+\in]0,1[$ 
contributes to $I$, since in the other cases, the
contour in the $X_-$ plane may be shrunk to a point.
After appropriate contour deformations, one gets
\begin{equation}
I=\sin\pi\alpha \int_0^1 dX_+ X_+^{\tilde\alpha-1}
(1-X_+)^{\tilde\beta-1}
\times\int_{-\infty}^0 
dX_- (-X_-)^{\alpha-1}
(1-X_-)^{\beta-1}
\label{eq:I2}
\end{equation}
The sine factor comes from the discontinuity across the cut
$]-\infty,0]$.
The integral over $X_+$ clearly
is the beta function $B(\tilde\alpha,\tilde\beta)$. 
Indeed, the latter is defined by
\be
B(a,b)=\int_0^1 dx\,x^{a-1}(1-x)^{b-1},
\ee
and admits a representation in terms of $\Gamma$ functions:
\be
B(a,b)
=\frac{\Gamma(a)\Gamma(b)}{\Gamma(a+b)}.
\label{eq:defbeta}
\ee
{\footnotesize
This relation may be proven as follows.
Using the definitions of the $\Gamma$ function (Eq.~(\ref{eq:defgamma}))
and of the $B$ function (Eq.~(\ref{eq:defbeta}),
we write
\be
\Gamma(a+b)B(a,b)=\int_0^\infty dy\, y^{a+b-1}e^{-y}
\int_0^1 dx\,x^{a-1}(1-x)^{b-1},
\ee
and perform the change of variable $x=X/y$.
Then
\be
\Gamma(a+b)B(a,b)=\int_0^\infty dy\,e^{-y}
\int_0^y dX\,X^{a-1}(y-X)^{b-1}.
\ee
Next, we exchange the order of the integrations
and subsequently shift the $y$-variable by $X$:
\begin{multline}
\int_0^\infty dy\,e^{-y}
\int_0^y dX\,X^{a-1}(y-X)^{b-1}
=\int_0^\infty dX\,X^{a-1}\int_X^\infty dy\, (y-X)^{b-1}e^{-y}\\
=\int_0^\infty dX\, X^{a-1}e^{-X}
\int_0^\infty dy\, y^{b-1}e^{-y},
\end{multline}
which is simply the product $\Gamma(a)\Gamma(b)$.
}

The integral over $X_-$ in Eq.~(\ref{eq:I2})
reduces to $B({\alpha,1-\alpha-\beta})$
after the change of variable $X_-=x/(x-1)$ has been 
performed.
Using the identity
\begin{equation}
\frac{\pi}{\sin\pi x}=\Gamma(x)\Gamma(1-x),
\label{eq:singamma}
\end{equation}
one may rewrite $I$ in several equivalent ways.
Useful formulas are
\begin{equation}
I=\pi\frac{\Gamma(\tilde\alpha)\Gamma(\tilde\beta)}
{\Gamma(\tilde\alpha+\tilde\beta)}
\frac{\Gamma(1-\alpha-\beta)}
{\Gamma(1-\alpha)\Gamma(1-\beta)}
=
B({\alpha,\beta}) B({\tilde\alpha,\tilde\beta})
\frac{\sin\pi\alpha\sin\pi\beta}
{\sin\pi(\alpha+\beta)}.
\label{eq:res2sing}
\end{equation}

\clearpage


\addcontentsline{toc}{section}{References}

\end{document}